\documentclass[prodmode,acmtoms]{acmsmall}

\usepackage{times}

\usepackage{pifont}
\usepackage{epsfig}
\usepackage{tabularx}
\usepackage{amssymb}
\usepackage{latexsym}
\usepackage{amsmath}
\usepackage{delarray}
\usepackage{moreverb}
\usepackage{xspace}

\usepackage{lscape}
\usepackage{changebar}
\usepackage{graphicx}
\usepackage{graphics}
\usepackage{mflogo}
\usepackage{xspace}
\usepackage{texnames}
\usepackage{rotating}
\usepackage{alltt}

\usepackage{amsmath}
\usepackage{epsfig}
\usepackage{booktabs,paralist}

\newcommand{\SPr}[2]{{<}#1,#2{>}}

\newcommand{\h}{\psi}
\newcommand{\g}{\gamma}
\newcommand{\f}{\varphi}
\newcommand{\Z}{\mathbb{Z}}

\newcommand{\N}{\mathbb{N}}
\newcommand{\R}{\mathbb{R}}

\newcommand{\D}{\mathbb{D}}

\newcommand{\Vc}[1]{{\bf #1}}

\newcommand{\Transducer}[1] {{\mathbf #1 }}

\newcommand{\TableRef}[1]{Table \ref{#1}}

\newcommand{\SectionRef}[1]{Section \ref{#1}}

\newcommand{\mypar}[1]{{\bf #1.}}

\newcommand{\doublefigure}[5]{{\begin{figure}[ht] %
\centering %
\includegraphics[width=#1\linewidth]{#2}
\includegraphics[width=#1\linewidth]{#3} %
\caption{#4}%
\label{#5}%
\end{figure}}}

\newcommand{\Doublefigure}[5]{{\begin{figure*} %
\centering %
\includegraphics[width=#1\linewidth]{#2}
\includegraphics[width=#1\linewidth]{#3} %
\caption{#4}%
\label{#5}%
\end{figure*}}}

\newcommand{\singlefigure}[4]{{\begin{figure}[htp] %
\centering %
\includegraphics[width=#1\linewidth]{#2}\\
\caption{#3}%
\label{#4}%
\end{figure}}}

\graphicspath{{Figures/}}

\begin{document}

\markboth{P. D'Alberto et al.}{Non-Parametric N-Sample Series Comparison}

\title{Non-Parametric Methods Applied to the N-Sample Series Comparison}

\author{PAOLO D'ALBERTO
  \affil{FastMMW} 
  ALI DASDAN
  \affil{Knowledge Discovery Consulting,}
  CHRIS DROME
  \affil{Yahoo! Inc. }
}

\begin{abstract}
  Anomaly and similarity detection in multidimensional series
  have a long history and have found practical usage in many different
  fields such as medicine, networks, and finance. Anomaly detection is
  of great appeal for many different disciplines; for example,
  mathematicians searching for a unified mathematical formulation
  based on probability, statisticians searching for error bound
  estimates, and computer scientists who are trying to design fast
  algorithms, to name just a few.
  
  In summary, we have two contributions: First, we present a
  self-contained survey of the most promising methods being used in
  the fields of machine learning, statistics, and bio-informatics
  today. Included we present discussions about conformal prediction,
  kernels in the Hilbert space, Kolmogorov's information measure, and
  non-parametric cumulative distribution function comparison methods
  (NCDF). Second, building upon this foundation, we provide a powerful
  NCDF method for series with small dimensionality. Through a
  combination of data organization and statistical tests, we describe
  extensions that scale well with increased dimensionality.  
\end{abstract}




\acmformat{P. D'Alberto, A. Dasdan, and C. Dromen 2012. Non-Parametric
  Methods Applied to the N-Sample Series Comparison}

\begin{bottomstuff}
Author's addresses: P. D'Alberto paolo@FastMMW.com, A. Dasdan
ali\_dasdan@yahoo.com, and C. Drome cdrome@yahoo-inc.com
\end{bottomstuff}

\maketitle

\tableofcontents

\section{Introduction}
\label{sec:introduction} 
Let us begin by exploring what change detection in a multidimensional
series is through a couple of examples. If we are investigating
whether cheating is happening at a game of chance, we could try to
find a bias or pattern in a single game or a set of games, which would
constitute a single dimensional series and a multi-dimensional series
respectively. We may look for a statistically significant difference
from the expected win-loss pattern. As another example, an Internet
search company may collect hundreds of annotation features for a host,
then the search engine will monitor the feature changes over time in
order to identify potential service outages or other issues with that
host. In both cases, a variation from the norm can only be identified
and understood after it happens, which may prompt further
investigations into the cause.

The question of why one should be concerned about change detection
should be obvious even when considering what is at stake from the
two previous examples. Although it is true that we may be able to
identify a change only after it has already occurred, one important
goal of change detection is to minimize the amount of delay required
to identify such a change, and thereby, hopefully, minimize the impact
or effects of the change.

In this paper, we consider change detection, and consequently
similarity detection, as a data mining problem on a large volume of
data. In particular we are interested in optimizing the early
response, recall, cost, and sensibility with respect to a change:
\begin{itemize}

\item for the {\em early response}, we try to minimize the number of
  events or data points that pass by undetected before a change is
  identified;

\item for the {\em recall}, we try to increase the accuracy of
  identifying real changes versus perceived changes;

\item for the {\em cost}, we try to reduce the cost and increase the
  speed at which the required computations are performed;

\item for the {\em sensibility}, we try to minimize the number of data
  points required to declare a significant difference (early response
  and sensibility are related: we will need a fast response to have a
  sensible response, but we will need some confidence associated with
  the variation to improve the recall).

\end{itemize}

We organized this work into three parts: In the first part, we present
a review of a variety of existing methods using a unified and concise
notation. Each method is described in detail with an emphasis on
pointing out their relative advantages and disadvantages.  In the
second part, we propose a new method and compare it against the
previously described methods. In the third part, we present
experimental results for the various methods. To this end, we have
produced a self-contained work which can be understood by others who
do not have an in-depth knowledge of this field\footnote{A comparable
  understanding of the topics contained herein would require the
  reader to comprehend half a dozen papers covering just the
  statistical aspects. A similar amount of papers would be required to
  understand the implementation details, and another handful to cover
  experimental results. Furthermore, each paper would be presented in
  its own notation making comparison between methods challenging.}.

In summary, our goals for this work are as follows: The first goal is
to introduce the reader to six methods spanning four different
non-parametric method families. We do this in the context of a unified
framework which easily facilitates comparison between methods. This
framework is designed to be flexible enough that adding new methods is
easy. Our work is similar to the work by Siegle in 1959
\cite{Siegel1959}, in which he focused on explaining the power and
usage of non-parametric methods. Rather, our work focuses on the {\em
  computational} aspects of the methods in order to create a useful
statistical tool set.

A secondary goal, arising from the first, is to show that there is
no single best method; rather, the most appropriate method is a function
of the data. That said, each method does contribute insights into
understanding the data better. Hence, we propose that the methods should
be used in conjunction, thus enhancing the set of statistical tools
available to the reader.

The third goal is to build upon this knowledge to propose a new method
based on works of Bickel \cite{Bickel1969} and Friedman--Rafsky
\cite{FriedmanR1979}. We have extended these works in two original
ways, by sorting the data in topological order, and then applying
a robust, non-parametric test set based on empirical cumulative
distribution functions.

The fourth goal is to contribute an implementation of our proposed
framework, which includes codes of the methods described herein. The
library is implemented in C and optimized to reduce the computational
costs of each method. We supply wrappers to allow the use of the
library to be called from within such languages as Java, Python, Perl,
and R.

The final goal is to demonstrate the application of our framework
through a series of experiments on a variety of data sets, including
well-known, classified data sets,  and synthetically generated data
sets.

Section by section, we have organized the paper as follows: In Section
\ref{sec:change}, we build a foundation for later sections by defining
necessary terms and explaining the motivation for analyzing
multi-dimensional series and the need to develop appropriate tools.
In Section \ref{sec:multidimensional-methods}, we present details of
the statistical methods that can be applied to the two-sample problem
in multi-dimensional space and presents the theoretical background of
those methods. This includes the following known methods: conformal
prediction in Section \ref{sec:conformal}, Kolmogorov's information in
Section \ref{sec:kolmogorov-information}, kernels methods in Section
\ref{sec:kernels}, as well as our proposed non-parametric method in
Section \ref{sec:order}. In Section \ref{sec:distance}, we introduce
non-parametric statistics that are suitable for single- and
multi-dimensional series. Finally, in Section \ref{sec:experimental},
we provide an overview of the experimental results. In Section
\ref{sec:mult-experiments}, we start with the results obtained by the
application of methods for multi-dimensional series. Finally, in
Section \ref{sec:uniexperimental}, we follow with the performance of
our non-parametric statistics, which are based upon cumulative
distribution functions (CDF), and we compare them against the
classical probability--distribution-function (PDF) statistics.

\section{Change Detection in Series}
\label{sec:change}

This section addresses two topics: we present our notation to describe
a series, and we present examples of change in series. The notation
is used throughout the work; in the experimental results, we are going
to present the analysis for most of the examples introduced in this
section.

\subsection{Terminology} 

A {\bf series} $S$ is composed of elements $s_i=(x_i,y_i)$ where $i$
is a strictly increasing, non-negative integer, called an epoch, which
represents time. The epoch helps ensure a global ordering of the
elements of $S$, $S \in (\N^+,\R^d)^*$. We identify the most recent or
the last element of $S$ by $s_t$.

The term $x_i\in \N^+$, the natural positive number set, is a time
stamp where the epoch $i$ is always in order and increasing. Note: the
time stamp of a sample does not necessarily have this constraint. The
distinction between epoch and time stamp is made because there are
instances where the order of the time stamp does not coincide with the
order of the epoch. For example, the epoch could describe when a
process finishes and data was collected; in contrast, the time stamp
is when the process started. In such cases, we want to reorder the
sequence accordingly, because the processing time is not under
consideration.  The term $y_i \in \R^d$ with $d\geq 1$ is the sample
vector (e.g., a single value when $d=1$).

We define the {\bf reference window} $R$ and {\bf test window} $W$ as
the ordered set of $N$ successive elements of $S$. We shall present
methods for which the {\em length} of the two windows need not be
equal. When reduced to vectors, these windows are represented as
$\Vc{r}$ and $\Vc{w}$. In practice, these windows typically do not
overlap, and either one can be made more recent or closer to ``now''
depending on the need for defining a new reference. In this work, we
will use the time stamp $x_i$ to build intervals that are ordered sets
of points in time; thus we do not use the epoch. We explicitly use the
epoch only when we need to handle the last sample point to compare it
against a set of points that were collected in the past.

To simplify the presentation, we will overlook the difference between
epoch and time stamp. In fact, we can interchangeably use $s_i$, in
$\N^+\times \R^d$, and $y_i$ in $\R^d$ without loss of information,
because the original order should not matter once the intervals have
been created.

A {\bf change} occurs any time when $W$ is {\bf different} from
$R$. 
In the following section by using examples, we present an intuitive
explanation of change.

\subsection{Examples of Series and Change}

% Here we give a few and representative examples of series and change.

\mypar{Hardware Clusters} One can measure the read--write latency
of the four disks on each machine in a 1000-node cluster during a
stress test. Each disk is considered independent of the others;
therefore, each disk latency measure is an independent data point.
Over time, a multi-variate series is generated: 1,000 independent series
each with four dimensions, for a total of 4,000 series.
Alternatively this could be viewed as one series with 4,000 dimensions.

Any variation in the average latency could indicate a mechanical defect
and thus a possible rejection of the disk lot. Any variation in the
variance of the latency, an increasing deviation in the latency, could
indicate the possibility of a pending failure or data inconsistency.

\mypar{Sites and URLs} In the process of generating a web graph, sites,
hosts, and URLs are annotated with hundreds of features. Each feature
can be considered orthogonal and independent. Consecutive builds of a
web graph will generate billions of series composed of hundreds of
dimensions. 

A sudden change in the measure space may occur as a result of an
internal or external error. Similarly, a change in variance could hint
at potential issues.

\mypar{Hardware Counters} With the increased complexity of modern
hardware components, it is becoming more common to embed a variety of
hardware counters to monitor performance statistics. These counters
can be used to monitor the performance of software running on the
underlying hardware. Multi-dimensional series can be generated by
polling these counters over time. 

Quantifying the stochastic distance between series generated by
different software allows the grouping of software with similar
hardware processing profiles. We could match applications with
specific hardware accordingly. This example uses the magnitude of the
difference between series to identify similarity.

\mypar{Histograms in Time} Selection rank algorithms, such as PageRank,
attempt to order URLs by ranking them. These ranking methods are useful
in generating histograms, which are valuable inputs for machine learning
and self-adjusting ranking tools. A histogram can be thought of as a
vector or a multi-dimensional point. 

The histograms evolution over time results in a series which can be
used to monitor for internal or external failures in the collection
system.

\mypar{Stock Index} A stock index is a set of stocks which act as
an indicator for a specific sector of the market. Again, a stock
index can be thought of as a multi-dimensional series; although
it should be noted that the individual stocks may not necessarily
be independent. Also note that the historical data about a specific
stock can also be thought of as a multi-dimensional series composed
of such attributes as average price, volume, opening price, closing
price, high, and low. 

The benefits of quickly identifying changes should be readily
apparent.

\mypar{Multi-Dimensional Series} Instead of analyzing a multitude of single-
dimension series, it is sometimes easier and more natural to join sets of
single-dimension series into one multi-dimensional series. This is
important to consider, because we will show that certain feature changes
are easier to detect when looking at a multi-dimensional series, as opposed
to considering each dimension individually.

In general, given a sample of points from the reference interval $R$
and a sample from the moving interval $W$, we are able to quantify the
degree to which the two intervals originate from the same stochastic
process, and hence are indistinguishable and independent. Note that we
have not mentioned the correlation between dimensions or the
independence of samples. These topics raise such questions as
determining the important dimensions of a series, and whether certain
dimensions can be ignored without loss of information. We will touch
on these issues in the context of this paper.

In summary, given a series and a change, we would like to have a
quantitative and automatic method to detect the change. What follows
is a description of tests that are available in the literature, for
which we will provide experimental results. Finally, our original
contribution will be presented in Section \ref{sec:order}

% ##########################################################

\section{Methods for Multidimensional Series} 
\label{sec:multidimensional-methods}

The literature is rich and spans many disciplines. There is a wealth
of statistical tests and methods for organizing data in sets, and
numerous approaches for identifying relations across different
features or dimensions.

A recent work by Sriperumbudur et al. \cite{SriperumbudurGFLS2009},
clearly attempts to classify and understand the power of different
families of measures. For example, the authors draws a connection
between $\phi$-divergence based methods, such as those of
Kullback-Leibler \cite{KullbackL1951} and Jensen-Shannon
\cite{Jensen1906,Shannon1948} (Section \ref{sec:contribution}), and
integral probability based measures, such as reproducing kernel
Hilbert space methods \cite{BorgwardtGRKSS06} and Kolmogorov-Smirnov's
\cite{Kolmogorov1933e} method. The authors find only one metric, the
variation distance \cite{Pinsker1960,AliS1966}, which is common to
both types of methods. Others have found further commonality, such as
Jensen-Shannon's distance being embedded into a Hilbert space
\cite{FugledeTopsoe}, resulting in the application of
$\phi$-divergence into a space where integral probabilities are more
common.  Interestingly, the authors in \cite{SriperumbudurGFLS2009}
suggest that $\phi$-divergence methods are {\em difficult} to estimate
in high dimensions; either being too expensive computationally or not
powerful enough.  We will show that this is not the case.

Borgwardt et al. and Gretton et
al. \cite{BorgwardtGRKSS06,GrettonBRSS06,GrettonBRSS08} provide the
first extensive comparison of the same family of measures that we use
in this work \footnote{We introduce compression methods.}.  Their test
results show that kernel methods and conformal prediction are
insensitive to the dimensionality of the series, while previous tests
based on \cite{BiauG2005,FriedmanR1979} show a loss of discriminative
power as the number of dimensions increase.

What distinguishes our work from others is the focus on the
computational aspects of implementing each method in the context of a
set of statistical tools. A clear understanding of the computational
requirements of a method lead to insights about the method itself. As
we are building a set of tools, our target audience are those
researchers who may be familiar with one or two methods and want to
explore the effectiveness of other methods \footnote{Novice users may
  find that this work is lacking explanatory examples, while advanced
  users may find this work too verbose.}.  We feel that we have taken
the works of Bickel \cite{Bickel1969} and Friedman--Rafsky
\cite{FriedmanR1979} and succeeded in extracting the common features,
combining the data sorting algorithms, and deploying
non-parametric statistical tests that are independent of the data
ordering.

With respect to our original contribution, we show that the
$\phi$-divergence measures can be generalized, extended, and applied
to multi-dimensional series (i.e., $\R^d$ with $1\leq d \leq 10^3$) in
a manner that makes them as discriminative as other measures.  We show
that the complexity of $\phi$-divergence methods is $O(d(n+m)^2)$
where $n=|R|$ and $m=|W|$ by using spanning trees in conjunction with
all-to-all distance computations. The complexity becomes
$kd(m+n)\log_2 (m+n)$ by using sorting poset algorithms, where $d$ is
the dimensionality of the series and $k$ is the number of parallel
points where comparison is undefined. If $k\sim m+n$, then the
complexity becomes $O(dk^2)$, which means that the sample is not large
enough to represent the probability sufficiently. Note: this
complexity is comparable to the other methods (e.g., kernels methods).

In Section \ref{sec:conformal}, we present conformal prediction
methods along with the implementation details. This is followed by a
discussion about the similarity measure based on Kolmogorov's
information complexity measure in Section
\ref{sec:kolmogorov-information}. We describe the minimum mean
discrepancy measure as computed from reproducing kernels in a Hilbert
space in Section \ref{sec:kernels}. In Section \ref{sec:order} we
introduce our extension of the distribution comparison using a
poset-based and a minimum--spanning-tree topological ordering. We
complete the extension of our method to multi-dimensional series with
a look at methods of applying single-dimensional distribution function
comparison measures to the topological ordering in Section
\ref{sec:distance}.

\subsection{Conformal Prediction}
\label{sec:conformal}

We present the following methods under the assumption that the series
is composed of independent samples. Consider the series ${\bf s}_i$,
where $i$ is an integer $0\leq i \leq N$, and ${\bf s}_i=(x_i,y_i)$
where $y_i\in \R^d$. The event ${\bf s}_i$ should be independent of
previous and successive events. The importance of the independence
condition lies with the ability to fully count the contribution of a
single event towards the description of the process that generated the
event. In absence of this condition, it is possible that the event is
redundant and could be ignore entirely.

In this section we explain what an independence test is. This is followed
by a discussion of how different change detection methods are designed to
capture both independence and change.

The hypothesis of independence states that ${\bf s}_i$ can be
described by a distribution function $P$ where $P[{\bf s}_0,{\bf
    s}_1,\dots,{\bf s}_N]$ = $\prod_{i=0}^NP[{\bf
    s}_i]$. Alternatively, this could be expressed as
$\prod_{i=0}^NP[{\bf y}_i]$. The hypothesis of exchangeability is
based on the idea that the sequence of ${\bf s}_i$ is generated with a
probability $Q$. Thus we could obtain a probability under $Q$, such
that, the permutation ${\bf s}_{\sigma(i)}$ is distributed as the
original ${\bf s}_i$; that is, $P[{\bf s}_0\dots{\bf s}_N] =P[{\bf
    s}_{\sigma(i)}\dots{\bf s}_{\sigma(N)}]$ for any permutation
$\sigma()$. Independence implies exchangeability, although the reverse
is not true.

We are interested in exploring on-line
independence--interchangeability tests. For each new data point, ${\bf
  s}_t$ where $t\geq N$, we determine whether or not ${\bf s}_t$
belongs to the series based on the information that has been seen so
far. If not, then a change has occurred. In other words, if the
probability of a data point occurring is similar to that of the points
already seen, and if independent of the sequence of events, then truly
there is no detectable change.

\subsubsection{Individual Strangeness Measure} 

Consider a particular interval of a series $R = \{ {\bf s}_i | m\leq i
< m+N \}$ where ${\bf s}_i \in S$, a multi-dimensional space, such
that $s_i=(x_i,y_i)$ and $y_i\in \R^d$.

Now consider, a family of measurable functions $\{ A_j | j\in\N \}$,
where $A_j:S^N\rightarrow \R^N$. More specifically, $A_j:\R^{N\times
  d}\rightarrow \R^N$ is an {\bf individual strangeness measure} when
for any $A_j$ of any permutation $\sigma$ of the time stamps in
$[m,m+N)$, and for any ${\bf s}_i \in \Z^N$ and $i\in [m,m+N)$ and any
    $\alpha_i \in \R$
\begin{equation}
  (\alpha_m,\dots,\alpha_{m+N-1}) = A_j({\bf s}_m,\dots,{\bf
    s}_{m+N-1})
\end{equation}
is equal to 
\begin{equation}
  (\alpha_{\sigma(m)},\dots,\alpha_{\sigma(m+N-1)}) = A_j({\bf
    s}_{\sigma(m)},\dots,{\bf s}_{\sigma(m+N-1)}).
\end{equation}

\mypar{Example} With the term $A_j$ we identify a distance function,
such that for every ${\bf s}_k\in R$ it returns a real number
$\alpha_k$, which quantifies the strangeness of the point ${\bf s}_k$
with respect to $R$ or $R \setminus {\bf s}_k$ (the interval without
the point in consideration).  This becomes the means by which we can
represent a multi-dimensional series, $\R^{d\times N}$, as a single
vector in $\R^N$, for which we can estimate a distribution
function. In the next step, we shall show how to reduce it to a single
real number.

\subsubsection{Transducers: $\Transducer{f}_A$}

A {\bf deterministic transducer} is a function $\Transducer{f}_A :
(S)^* \rightarrow [0,1]$, where $A$ is a strangeness measure. We
define the transducer as
\begin{equation} 
  \Transducer{f}({\bf s}_{m+N-1}|R\setminus {\bf s}_{m+N-1}) =
  \frac{|\{\alpha_i\geq \alpha_{m+N-1}\}|}{N}
\end{equation} 
where 
\begin{equation}
  (\alpha_m,\dots,\alpha_{m+N-1}) = A_j({\bf s}_m,\dots,{\bf
    s}_{m+N-1})
\end{equation}
The transducer takes an interval and a new data point, for which a
measure of strangeness is computed, and it returns the number of
points that are of equal or greater strangeness.

There is another type of transducer, the {\bf randomized transducer},
which introduces a randomized multiplicative term, uniformly generated
from the real interval $[0,1]$, to break ties (i.e., $\alpha_i=
\alpha_{m+N-1}$).  The randomized transducer is used to ensure that,
in the case of no change, the output of the transducer will be
uniformly distributed over $[0,1]$. 

This is an important concept and we should clarify the meaning and the
power of a randomized transducer. In other words, if we have a set of
samples that have the same strangeness and we use a deterministic
method, the transducer is going to produce a sequence of
ones. Equality of strangeness is a strong signal that the new points
belong to the set but the output value will be skew towards the value
one.  It will not be uniform. The randomization has no effect if
there is no ties, and the transducer's distribution will be
evident. The randomization has effect only with a lot of ties, and the
distribution of the output will be artificially uniform.

Very briefly, in case of no change, the output of the random
transducer will be uniformly distributed. In case of change, the
distribution should be skewed. The type of transducer will determine
the extent of skewness.

We deploy deterministic transducers only, because they are easy to
understand. Furthermore, we propose an alternative to coping with an
output that is non-uniformly distributed.

\mypar{Example} The transducer--strangeness pair is an attempt to
estimate the distribution function $P[X={\bf s}_{m+N-1}]$ of the
process that generates the series.  If we knew the
distribution of the process, it would provide a direct method for the
computation of the transducer--strangeness pair.

Although we implemented a few transducers, we shall turn our attention
to transducers based on a minimum distance measure (i.e., nearest
neighbor) or an average distance measure, as either can be computed in
linear time $O(N)$ for each new point. In fact, with $O(N^2)$ space to
store an adjacent matrix, we can compute the update of the minimum and
average distance between $N$ points with $N$ comparisons.

If we apply the transducer to the series, we compute a different
series $p_i = \Transducer{f}({\bf s}_i)$. This transforms the problem
from one using multi-dimensional series data to single-dimension
series data.

The question now becomes one of how we can use the output series of a
transducer in such a way to detect change in a series.  To this end,
we present two approaches: the Martingale method and a non-parametric
method (and we can use them separately or together).  The Martingale
method simulates gambling to exploit a consecutive sequence of lucky
or unlucky bets: skewed distribution to specific discrete points. The
non-parametric method measure the change of distribution in its
entirety: change of the type of distribution from uniform to
exponential. One approach does not subsume the other. Both methods are
aimed at detecting variations in the transducer's output. At the time
of writing this paper, we have started experimenting with kernel
methods as well.

\subsubsection{Martingale Methods}

We know that the transducer $\Transducer{f}$ provides an
estimate of the distribution function $P$. As such, the series $p_i$
is an approximation of the probability that the sample ${\bf s}_i$
belongs to the series as seen so far. The Martingale method is based on
the idea that successful bets result in exponential gains if a long
enough sequence of successes is found. An example of a sequence of
successes would be a sufficiently long sequence of $p_i \sim 0$,
which consists of points determined to be strange with respect to the
reference interval.

The Martingale ${\cal M}_n^{(\epsilon)}$ measure is defined as
\begin{equation} 
\label{eq:martingaleupperbound}
  {\cal M}_n^{(\epsilon)} = \prod_{i=0}^n\epsilon p_i^{\epsilon-1}=
  \frac{\epsilon}{p_n^{1-\epsilon}}*{\cal M}_{n-1}^{(\epsilon)}
\end{equation}
where $\epsilon \in [0,1]$ and $\int_0^1\epsilon p^{\epsilon -1}dp=1$.
The Martingale measure will increase exponentially if $p_k\sim 0.01$ for
a sufficiently long sequence of points. In the literature, we found
two tests related to the Martingale measure and its maximum increase.

\mypar{Property, Without Proof \cite{HoW2010}} Given the hypothesis
that there is no change, we can accept the hypothesis as long as
\[
0< {\cal M}_{n}^{\epsilon} < \lambda
\]
and reject the hypothesis when ${\cal M}_{n}^{(\epsilon)} \geq
\lambda$. In fact, for any $\lambda>0$ and bounded $n>0$ we have
\begin{equation}
  \lambda P\big(\max_{k\leq n}{\cal M}_k\geq \lambda)\leq E[{\cal
    M}_n] 
\end{equation}
If $E[{\cal M}_n] = E[{\cal M}_1]$, that is, if we take an interval of
time where the Martingale starts and finishes with a steady point, then

\[ P\big(\max_{k\leq n}{\cal M}_k\geq \lambda)\leq \frac{1}{\lambda} \].

\mypar{Property, Without Proof \cite{Ho2005}} We can use the
derivative of the Martingale measure to reject the test in the case
of ${\cal M}_{1}^\epsilon = 1$ when

\begin{equation}
  P\big( |{\cal M}_{n}^\epsilon-{\cal M}_{n-1}^\epsilon| \geq t\big) \leq 
  2e^{\frac{t^2}{2(\epsilon (1/n)^{\epsilon -1} -1)^2}}
\end{equation}

In fact, we have
\[ 
P(|Y_1| \geq t) \leq 2e^{-\frac{t^2}{2c_1}}
\]
where $Y_1$ is a difference Martingale and $c_1$ is a proper constant.

The parameter $\lambda$ and $t$ are set according to the recommendations
in the literature by Ho et al. (\cite{HoW2005,Ho2005,HoW2010}); hence
$\lambda = 20$ and $t=3$.

\mypar{Property} The Martingale test approximates the sequential
probability ratio test (\cite{Wald1947}), which can be used in
combination with $\lambda$ to form a more robust test.
This is not investigated further.

In addition to the values for $\lambda$ and $t$ mentioned above, we
set $\epsilon = 0.95$; however, all three of these parameters should
be tuned accordingly with the size of the reference interval $N$.

\subsubsection{Non-Parametric Distance Application}

In this section, we present our contribution to the Martingale method
by taking a fresh look at $p_i$ as a series.

In the original works on the Martingale method, the transducers
provides an estimate of a probability function. If this is truly a
distribution function, we expect that $p_i = \Transducer{f}({\bf
  s}_i)$ is uniformly distributed on $[0,1]$. Let us assume we know
the nature of the process that generates the sequence ${\bf s}_i$;
that is, we know the distribution function $P_X$ and therefore we also
know $P_X[X={\bf s}_i]$. All of the above properties hold true if we
use $p_i = \Transducer{f}({\bf s}_i) = P_X[X={\bf s}_i]$.

In practice, $\Transducer{f}() \sim P_X[X={\bf s}_i]$ and, in the
work by Vovk \cite{Vovk1993}, the author defines the power of
transducers and rigorously demonstrates how they can be used
{\em instead of} the distribution function.

We make only one assumption about the sequence $p_i$, that is, if
there is a {\em change} in the original series ${\bf s}_i$, there is a
corresponding change in the $p_i$ series, and vice versa.  Instead of
making assumptions about the distribution of $p_i$, such as $p_i$
being uniformly distributed on $[0,1]$, we create a reference sequence
$R_{p_i}$ by running the system on a series for which there is no
change. We also create a moving window $W_{p_i}$ consisting of $p_i$
as the system evolves in time. This creates a two--$N$-samples problem
for which we can apply all the stochastic distance measures in the
literature. In practice, we will apply our generalized measures as
described in Section \ref{sec:distance}.

\mypar{Remark} We will show that this test, which is built in parallel
with the Martingale method, will have more than a supporting role.
We will also show that it is orthogonal to the Martingale method,
capturing global variations of the series $p_i$, not just temporal
variations (i.e., lucky/unlucky bets). This test can also be used to
reset the Martingale measure to 1, resulting in a faster response to
changes. For example, at steady state where there are no changes, the
Martingale measure will tend to decrease (the result of consecutive
losing bets) to a value as small as $1/10^{20}$. A side effect of this
small value, is that the method requires a longer string of changes
before it can recognize a changes has occurred, because it takes more
effort to recover. A periodic check of the sequence $p_i$ will allow
us to safely restart the Martingale method from a value of 1, where
it will be more responsive to change.

\mypar{Remark} Resetting the value to 1 is beneficial only when the
Martingale value becomes extremely large or small, hence the
Martingale value is slow to return to a steady state. It is also very
important to carefully choose the moment of the reset. If there is a
temporal change in $p_i$ with no corresponding change in the
distribution, it may be disruptive to reset the value at this time as
a change is just being detected. We will return to this topic in the
experimental results section.

This section's references are
\cite{VovkNG2003,ShaferV2008,VovkGS2005,Ho2005,HoW2010,Vovk1993,Wald1947,EinmahlK2001,Kulldorf1997}

\subsection{Normalized Compression Distance}
\label{sec:kolmogorov-information}

The measure that we discuss in this section is known by several names
in the literature, including {\em the similarity metric} describing
the universal nature of the measure, the {\em algorithmic information
distance}, and the {\bf information distance}. We feel the term
normalized {\bf Kolmogorov's information} measure is more precise and
distinguishes it from the Kolmogorov-Smirnov measure and other
information-theoretic measures like Jensen-Shannon or Kullback-Leiber
measures.

To describe Kolmogorov's information measure we need to introduce
Kolmogorov's complexity. Consider a descriptive process $E$ as a set
of pairs $(x,y)$ where $x$ is the description of $y$ and both are
binary strings such that $y$ can be described by a chain of
descriptions $x$s. $E$ can be seen as an algorithm.

The complexity $K_E(y)$ of an object $y$ is the minimum length of the
description $x$ such that $(x,y)\in E$: 
\begin{equation}
K_E(y) = \min_{(x,y)\in E} |x|
\end{equation}

Let us fix the $Y$, and thus the string set we need to describe. Let us
also consider a family of processes ${\cal U}$ that describe $Y$ and are
associated with algorithms such that Kolmogorov's hypotheses apply as
follows: for all $E \in {\cal U}$ there exists an optimum $A$ such
that $K_A(y) \leq K_E(y)+c_E$. For the sake of brevity, we can drop the
specifier so the complexity of $y$ is denoted simply as $K(y)$. Thus, we
have found an optimal algorithm $A$ capable of using a shorter key $x$ to
retrieve the output $y$, while maintaining a simple mapping $(x,y)\in E$.

Armed with these concepts, Kolmogorov was able to provide the first,
widely-accepted, and formal definition of a random sequence: a
sequence ${\bf s}_0,\dots,{\bf s}_{n-1}$ is {\bf random}, if
\begin{equation}
  K({\bf s}_0,\dots,{\bf s}_{n-1}) \leq n-c
\end{equation}
where $c$ is a constant and is independent of $n$.

The details of the computability of $K()$ \cite{TerwijnTV2010} are
outside the scope of this paper. Instead, we will use the compression
algorithm from {\em zlib}, referred to as $C()$ for compression,
to approximate the Kolmogorov's measure. The compression
algorithm takes a binary string $y$ as input and
produces a shorter string $x$ as output. A loss-less compression
creates the mapping $(x,y)$. Hence, we can measure the complexity of
$y$ by the length of $x$ as generated by the compression algorithm.

Now the question arises as to how we compute the distance between
two intervals $R$ and $W$ in a series?

In practice, the intervals $R$ and $W$ are two arrays of double
precision numbers stored consecutively in memory. We can simply
represent the encoded data as $r$ and $w$.

The Normalized Compression Distance is defined as
\begin{equation}
NCD(r,w) = \frac{C(rw) - \min (C(r),C(w))}{\max (C(r),C(w))}
\end{equation}
where $rw$ is the concatenation of $r$ and $w$. We have $0\leq
NCD(r,w)\leq 1+ \epsilon$ where $\epsilon$ is a small term function of
the compression algorithm, which represents an artifact of the compression
algorithm. If $NCD(r,w) = 0$, then $r$ ($R$) is similar to $w$
($W$); conversely, if $NCD(r,w) = 1$, then $r$ is different from $w$.

Intuitively, $C(rw) - C(r)$ is an estimate of $K(w|r)$, or the
complexity of $w$ under the condition that $r$ has already been seen. We
interpret $C(rw) - \min (C(r),C(w))$ as the independent complexity of
$w$ with respect to $r$ (or $r$ with respect to $w$).

Assume that $R$ is generated by a stochastic normal process with a
${\cal N}(0,4)$ distribution, $W$ is generated by either
${\cal N}(0,4)$ or ${\cal N}(0,1)$, and they are of the same length
$|R| = |W|=n$. Because of the characteristics of the compression algorithm
and the nature of the input, the compression measure $NCD(r,w)$ will be
very close to 1, independent of the choice of $w$ (${\cal N}(0,4)$ or
${\cal N}(0,1)$). However, there will be a difference, regardless of how
small it is. To handle the cases we are interested in, we must provide a
confidence level, or p-value, and take advantage of this difference.

\subsubsection{Bootstrap} \label{sec:ncd-bootstrap} 
Consider the intervals $R$ and $W$, generated by the same process, as
a sequence ${\bf s}_j$ composed of $N$ points each. Applying a series
of swaps between the original series (i.e., $swap(r_0,w_0)\/ \dots
 \/swap(r_k,w_k)$), two new sequences $R'$ and $W'$ can be created to
generate $NCD(r',w')$. The distance values are sorted, so that a
distribution and a p-value can be determined. In computing $NCD(r,w)$,
the distance value can be used to obtain a significance level. Then,
$NCD(r,w)$ can be used as a minimum threshold, in combination with the
p-value, to provide a measure of the significance of the difference.

This bootstrapping process tunes the sensitivity of the NCD measure to
the training set. For example, if we are working with intervals which
are very similar, then the range of possible distance values will be
small, and the p-value will be sensitive to small variations; thus, we
have a measure for the process that is quick to reject the equality
hypothesis.  For most of the synthetic series in the experimental
results section, this sensitivity is a powerful discriminating
feature. However, if the measure becomes too sensitive, every interval
will be considered {\em different}, and the measure will fail to give
useful information.

The section's references are
\cite{Lof1969,KolmogorovU1987,BennetGLVZ1998,LiCLMV2004,CilibrasiV2005,TerwijnTV2010,KeoghLR2004}.

\subsection{Kernel Methods}
\label{sec:kernels}

The following distance measure is called {\bf integral probability
metric} (IPM) \cite{Muller1997},
\begin{equation}
\label{eq:IPM}
  \gamma_{\cal F}(P,Q) = \sup_{f \in {\cal F}}\big| \int_\Omega f dP - \int_\Omega fdQ\big|
\end{equation}
where ${\cal F}$ is the class of real-value bounded measureable
functions in $\Omega$, and $P$ and $Q$ are probability functions.

If $P{=}Q$, $dP{=}p(x)dx$, and $dQ{=}q(x)dx$, then $\gamma_{\cal
  F}(P,Q){=}0$ because 
{\small \begin{equation*}
\int_\Omega f dP - \int_\Omega fdQ =\int_\Omega f(x)(p(x)-q(x))dx \leq
M\int_\Omega (p(x)-q(x))dx =0.
\end{equation*}  }
Furthermore, if $\gamma_{\cal F}(P,Q){=}0$, then $P{=}Q$, that is, 
{\small \begin{equation*}
  0= \gamma_{\cal F}(P,Q) \geq \int_\Omega f(x)(p(x)-q(x))dx\geq
\int_\Omega(p(x)-q(x))dx;
\end{equation*}} 
thus, $p {=} q$ with probability 1.

For example, if we restrict the class ${\cal F}$ to the step function
${\bf 1}(t)$ (i.e., ${\bf 1}(t){=}1$ when $t{\leq} 0$, ${\bf
  1}(t){=}0$ otherwise), then 
{\small \begin{equation*}
\gamma_{ 1()}(P,Q) = \sup_{x\in \Omega}\ |F(x)-Q(x)| \equiv KS(P,Q),
\end{equation*}  }
that is the Kolmogorov-Smirnov test.

In the remainder of this section, we examine the class
${\cal F} = \{f : \|f\|_{\cal H}\leq 1\}$ of bounded continuous functions
where ${\cal H}$ represents a reproducing kernel Hilbert space with $K()$
as its reproducing kernel. This measure is called the
{\bf maximum mean discrepancy} (MMD), and is defined as:

\begin{equation}
\label{eq:MMD}
  MMD_{\cal F}(P,Q) = \sup_{\|f\|_{\cal H}\leq 1 \text{ s.t. } f \in
    {\cal F}}\big( E_P[ f(x)] - E_Q[f(x)] \big).
\end{equation}

We will explain the concepts of Hilbert space and kernels, and then
examine how to transform a multi-dimensional problem into a
covariance matrix computation, and then into a single-dimension
problem.

\subsubsection{MMD, Kernels and Notations}

In discussing a Hilbert space, let us consider ${\cal F}$ as a class
of real-valued functions forming a real vector space and restricting
multiplication to real constants only; that is, the addition of
functions is a function in ${\cal F}$, given a $f \in {\cal F}$ and
any $\alpha \in \R$, then $\alpha*f() \in {\cal F}$. Such a class of
functions ${\cal F}$ is called a {\bf real Hilbert space} if the
following two conditions are met: First, the norm $\|.\|$ in ${\cal
  F}$ is given by $\|f\| = \SPr{f}{f}= Q(f)$ where $\SPr{}{}$ is a
scalar product so that for any real $\epsilon_1, \epsilon_2$ and any
function $f_1, f_2 \in {\cal F}$:
{\small\begin{equation}
\label{eq:Q}
Q(\epsilon_1f_1+\epsilon_2f_2)= \epsilon_1^2Q(f_1)
+\epsilon_2^2Q(f_2)+2\epsilon_1\epsilon_2Q(f_1,f_2);
\end{equation}}  
second, ${\cal F}$ is complete. Complete means that any Cauchy
sequence $f_n$ such that $\lim_{n->\infty}f_n=f$, then every function
$f_n,f\in {\cal F}$.

Notice that the inner product $\SPr{}{}$ is a vector norm and it is
also a distance measure for a functions space. The linearity property
in Equation \ref{eq:Q} states that the domain is {\em well behaved};
the completeness property makes the domain closed for infinite series
and their linear combinations.

Now that we have a definition of a Hilbert space, let us introduce
what is a kernel. Assume that ${\cal F}$ is a Hilbert space defined in
$E$; that is, $f(x)$ with $x\in E$. The function $K(x,y)$ with $x$ and
$y$ in $E$ is called a {\bf reproducing kernel} if the following two
conditions hold: First, for every fixed $y=y_0$, then $K(x,y_0)\in
{\cal F}$; second, $\forall y\in E$ and $\forall f\in {\cal F}$, then
we have
\[ 
f(y) = \SPr{f(x)}{K(x,y)}
\]
In other words, $K(x,y_0)$ is a valid function, and, in combination
with the inner product, we can reproduce the original function. It is
important that $f\in {\cal F}$ be continuous, as this will ensure the
existence of a reproducing kernel $K(,)$ that will be unique. Any one
familiar with the Fourier transform will recognize the previous two
properties, thus the ability to reconstruct the original
signal/function. We have stated now the definition of kernels in a
Hilbert space.

There are occasions where finding the witness function, the function
that minimizes Equation \ref{eq:MMD}, is useful to shed a light to the
data. For our purposes, we do not really need the witness function and
in this work we do not pursue it any further. In practice, once the
reproducing kernel is set, the computation can be simplified by
directly finding the MMD bound, without the witness function. Indeed,
the kernels are a powerful tool set.

Now, we explore how to transform the problem from the
multi-dimensional space, in which the series is defined, to a
single-dimensional space, where ${\cal F}$ is defined by the scalar
product into the Hilbert space.  

The first problem is how to transform a multi-dimensional space into a
single-dimension space. As suggested in \cite{GrettonBRSS08}, the
authors in \cite{ScholkopfS2002} found the existence of a mapping
$\phi(x)$ from the original domain to a feature domain in $\R$ such
that $f(x) = \SPr{f}{\phi(x)}$ is in the Hilbert space. Using the
kernel $K(x,y) = \SPr{\phi(x)}{\phi(y)}$, where $x$ and $y$ are
defined in the original space, results in $\phi(x) =
\SPr{\phi(y)}{K(y,x)}$. This is a single dimensional space.

The existence of $\phi(x)$ assures the existence of
$K$. Unfortunately, this property does not really provide a
constructive description about the kernel $K()$ that can be used.

The second problem is how to simplify the computation such that it is
using kernels only. The authors in \cite{GrettonBRSS08} suggest using
expectations $\mu_P = E_P[\phi(x)]$ to rewrite the MMD as follows,
\begin{equation}
 MMD_{\cal F}^2(P,Q) = \|\mu_P- \mu_Q \|_{\cal H}^2. 
\end{equation}
We report the original proof in the following  as
\begin{equation*}\small
\begin{split}
  & MMD_{\cal F}^2(P,Q) \\ 
  & = \sup_{\|f\|_{\cal H}\leq 1 s.t. f \in{\cal F}}\big( E_P[ f(x)] - E_Q[f(x)] \big)^2 \\
  & = \sup_{\|f\|_{\cal H}\leq 1 s.t. f \in{\cal F}}\big( E_P[\SPr{\phi(x)}{f}] - E_Q[\SPr{\phi(x)}{f}] \big)^2 \\
  & = \sup_{\|f\|_{\cal H}\leq 1 s.t. f \in{\cal F}}\big( \SPr{E_P[\phi(x)]}{f} - \SPr{E_Q[\phi(x)]}{f} \big)^2 \\
  & = \sup_{\|f\|_{\cal H}\leq 1 s.t. f \in{\cal F}}\big( \SPr{E_P[\phi(x)]-E_Q[\phi(x)]}{f} \big)^2 \\
  & = \| E_P[\phi(x)]-E_Q[\phi(x)] \|^2 \\
\end{split}
\end{equation*}
In the Hilbert space, the right-hand norm can be computed in terms of
kernels only:
{\small \begin{equation*}
\begin{split}
 &  \| E_P[\phi(x)]-E_Q[\phi(x)] \|^2 \\
 =&   \SPr{E_P[\phi(x)]}{E_P[\phi(x)]}+\SPr{E_Q[\phi(x)]}{E_P[\phi(x)]} \\  
 &   - 2\SPr{E_P[\phi(x)]}{E_Q[\phi(x)]} \\
 =&   E_P[\SPr{\phi(x)}{\phi(x)}] + E_Q[\SPr{\phi(y)}{\phi(y)}] \\
 &   - 2E_{P,Q}[\SPr{\phi(x)}{\phi(y)}] \\ 
 =&   E_P[K(x,x)] + E_Q[K(y,y)] - 2E_{P,Q}[K(x,y)]  \\ 
\end{split}
\end{equation*}}
For finite samples $r_i$ and $w_i$ in $R$ and $W$ where $|R| = |W| = m$
this can be estimated by
{\small \begin{equation}
\label{eq:mmdu2}
\begin{split} 
 MMD_{U,\cal  F}^2(R,W) =  & { \frac{1}{m(m-1)}\sum_{i\neq j}^m K(r_i,r_j)+ K(w_i,w_j) - 2K(r_i,w_j)  } \\
\end{split}
\end{equation} }

Notice that, once the kernel $K(,)$ is known, it is not necessary to
compute: the witness function $f()$, the scalar product $\SPr{}{}$,
nor the mapping $\phi(x)$\footnote{In practice, how to choose the
  right kernel is often a art. We followed the suggestions of the
  original authors as we will explain in the following section.}.

\subsubsection{MMD in Practice}
\label{sec:kernel-linear}
Based upon the theoretical understanding of MMDs and the process of
transforming a multi-dimensional space into a single-dimensional
space, we will examine the details of the methods used in this paper.
This includes a discussion of what is computed in practice, how the
significance measure is determined, and which kernel $K()$ to use.

\mypar{MMD Computation} We shall consider two MMD measures. The
$MMD_u^2$ has already been described in Equation \ref{eq:mmdu2} and
has a computation complexity of $O(m^2)$; instead, $MMD_l^2$ is a
linear approximation of $MMD_u^2$.

Assuming that $|R|=|W|=m$ and $m$ is even, then $MMD_{l,\cal
  F}^2(R,W)$ is defined as follows:
\begin{equation}
\label{eq:mmdl2}
\begin{split}
  \frac{2}{m}\sum_{i=1}^{m/2} &  K(r_{2i-1},r_{2i})+ K(w_{2i-1},w_{2i}) \\
  & - K(r_{2i-1},w_{2i}) - K(w_{2i-1},r_{2i})  \\
  = \frac{2}{m}\sum_{i=1}^{m/2} & h( r_{2i-1},r_{2i},w_{2i-1},w_{2i})
\end{split}
\end{equation}

Consider $MMD_{U,\cal F}^2(R,W)$ to be the product 
\[
{\bf x}'{\bf \Sigma}{\bf x}
\]
where the matrix ${\bf \Sigma}$ is a semi-definite covariance matrix,
that is ${\bf v}'{\bf \Sigma}{\bf v}\geq0$ for any ${\bf v}$, which
is a proper difference measure. Therefore, the linear approximation
considers the contribution of the one upper diagonal only, which
should also be the dominant one. Now, consider ${\bf \Sigma}$ to be a
covariance matrix consisting of only those consecutive points in the
series at a distance of 1 from each other. Notice that the comparison
reduces to a localized pair-wise comparison of $r_i$ and $w_i$ and their
direct neighbors $r_{i-1}$ and $w_{i-1}$, without requiring an
all-to-all comparison.

\mypar{Significance Level} Having examined the computation of the MMDs,
we consider the question of determining whether the two samples are
similar or different.

For $MMD_{U,\cal F}^2(P,Q)$ we followed the practical approach
outlined in \cite{BorgwardtGRKSS06} Algorithm 1 \footnote{In
	\cite{GrettonBRSS08}, the authors present a better way of computing
	the significance level for $MMD_u^2$ than in \cite{BorgwardtGRKSS06}.}.
Particular care must be taken in the computation of $4\sigma^2/(m(m-1))^2$,
because for even moderate values of $m$ the denominator can grow so large
that it cancels the overall contribution. This results in an incorrect
estimate of the variance.

For $MMD_{l,\cal F}^2(R,W)$ we use the method described in
\cite{GrettonBRSS08} Corollary 22, and present the results here as
well. The variance is computed at the same time as the distance for
both methods. This shows that the variance has a normal distribution
with 0 mean and a parametric variance of $\sigma$.  Knowledge of the
distribution of the variance (${\cal N}(0,\sigma^2)$) along with the
actual variance permits the generation of a confidence level.

With the mild conditions presented in \cite{GrettonBRSS08},
$E[h^2]<\infty$
\begin{equation} 
\begin{split}
& \sqrt{m}( MMD_{l,\cal F}^2 - MMD_{\cal F}^2) \rightarrow {\cal N}(0,\sigma^2)\\
& \sigma^2  = 2(E[h^2]-E^2[h])
\end{split}
\end{equation}
 
We show that, for a stochastic process composed of independent
variables with an obvious simplicity and speed, the linear method for
computing $MMD_{l,\cal F}^2(P,Q)$ provides a very good approximation
for $MMD_{U,\cal F}^2(P,Q)$. Among the methods presented, the linear
method is the fastest, having a complexity of $O(kN)$ where $N$ is
the number of points and $k$ is the number of dimensions of the series.

\mypar{Kernels} In this paper, we use the Gaussian kernel as
per \cite{BorgwardtGRKSS06,GrettonBRSS08}
\begin{equation}
K(x,y) = e^{-\frac{\|x-y \|^2}{2\sigma^2}}.
\end{equation}

Note that this kernel function is parametric, where $\sigma$ must be
estimated from the data before the kernel can be used to compute the
MMDs. Here we describe the process of computing $MMD_l$. For all
$r_{2i-1},r_{2i},w_{2i-1},w_{2i}$ we first compute $k^{xx}_i = \|
r_{2i-1} - r_{2i}\|^2$, $k^{yy}_i = \| w_{2i-1} - w_{2i}\|^2$,
$k^{xy}_i = \| r_{2i-1} - w_{2i}\|^2$, $k^{yx}_i = \| w_{2i-1} -
r_{2i}\|^2$. We compute the median for $k_i^*$ that will be the
$\sigma^2$. Then, we compute the sum of the terms such as
$e^{\frac{k_i}{\sigma}}$ and thus the kernel value.

Having chosen a Gaussian method means that the kernel methods are very
similar to Fourier methods and the Parseval--Plancharel theorem as
described in \cite{MeintanisI2008}.

The Section's references are
\cite{Aronszajn1950,GrettonFTSSS2007,BorgwardtGRKSS06,ZhangSGS08,BiauG2005,GrettonBRSS06,GrettonBRSS08,MeintanisI2008}.

\subsection{Sorting and Distribution Functions in $\R^d$}
\label{sec:order}

This section introduces our original contribution to the field. It is
based on the idea of creating a topological ordering of the data,
generating an empirical cumulative distribution function, and then
applying our statistical test. Given two arbitrary, empirical CDFs
$F_R$ and $F_W$, the test will generate a distance and a confidence
level for said distance. These pairs allow us to quantify how similar
$F_R \sim F_W$ or different $F_R \nsim F_W$ the two distributions are.
If $F_R \sim F_W$ is true, the two intervals have the same
distributions, and that there is a high probability that they are
instances of the same stochastic process. Section \ref{sec:distance}
contains a formal introduction.

Let us pose the questions of what $F_R \sim F_W$ means, and how a test
can measure such a difference.

% Let us pose two questions which we shall address. What does $F_R \sim
% F_W$ mean, in practice? How can a test measure such a difference?

Recall that an empirical cumulative distribution function (CDF) for an interval
$R$ means that for a given point ${\bf x} \in R$,
$F_R({\bf x}) = \frac{|\{ {\bf y} : {\bf y} \leq {\bf x} \text{ and } {\bf y} \in R\}|}{|R|}$,
where the ordering $\leq$ is satisfied for all components of the
vectors {\bf y} and {\bf x}. Here, it is not necessary that ${\bf x}$
belongs to interval $R$ or $W$. Notice that $F_R \sim F_W$ means that
for any ${\bf x}$ we have the same number of points from each $R$ and $W$,
which is proportional to the size of each interval when creating the
larger interval $R+W$. This is the same idea proposed by the
Mann--Whitney U statistics for testing whether one random variable is
stochastically larger than another random variable \cite{MannW1947}.
Our goal is to extend this idea to a multi-dimensional series.

In principle, if we can compute $F_R$ and $F_W$, we can also apply the
Kolmogorov-Smirnov test as is (see Section \ref{sec:df} and
Equation \ref{eq:ks}), such that

\begin{equation}\small
\begin{split}
  KS(F_R,F_W) &= \sup_{y}|F_R(y) -F_W(y)| \\ &\geq
  \max_{y=s_y\in R\cup W}|F_R(y) -F_W(y)| \\
\end{split}
\end{equation}

This is possible because the Kolmogorov-Smirnov test is based on the
image of the distance function, or the maximum difference of the
distribution functions, and is independent of the domain dimensions.
The weakness of this direct approach is intrinsic in the density of
the domains and our ability to estimate the distribution functions.
Simply, the larger the space is, the more points are required to
estimate the real distributions to ensure that the test can discriminate
properly between different distributions. In other words, if $R$ and $W$
do not have enough points to provide a good sample of the distribution
function domains, then the test results are often inconclusive and the
intervals are stochastically indistinguishable. This is often referred
as the {\em curse of dimensionality}.

In the literature, there are new tests being developed to handle the
case of multi-dimensional data. For these tests, we use the term
{\em statistical solutions} to differentiate them from the methods we
propose here. Instead, we will use the term {\em algorithmic solutions}
to refer to our methods. We make a clear distinction of these methods
in the following sections. In the following sections, we shall describe
two methods that are designed to circumvent the problem of sparse
samples by building on our understanding of single-dimension series
anomaly detection without requiring the introduction of new tests. The
first method is completely original, and based on the poset-sorting
algorithm (Section \ref{sec:poset-algo}), while the second method is based
on the minimum-spanning-tree (Section \ref{sec:mst-algo}). From our
observations, we have noticed that CDF measures used for single-dimension
series are easily applied, and build upon well established statistical
and computational grounds.

\subsubsection{Partial Ordered Topological Ordering}
\label{sec:poset-algo}
In the previous sections, we introduced the definition of distribution
functions and showed methods for comparing two empirical
distributions. Recall, that the definition of a distribution function
is based on the concept of order among the points in the interval. In
other words, ${\bf y} \leq {\bf x}$, where this condition is true for
all components when a point ${\bf y}$ is actually a vector ${\bf y}\in
\R^d$ with $d\in \N^+$.

For $d=1$, the condition ${\bf y} \leq {\bf x}$ is always defined as
either true or false. For $d>1$, cases may exist whereby the
inequalities ${\bf y} \leq {\bf x}$ and ${\bf x} \leq {\bf y}$ are
both not defined. In this case, the two points are parallel and
denoted as ${\bf x} || {\bf y}$. This is a partial ordered set or {\bf
  poset}.

The computation of the empirical distribution function turns out to be
exactly the same computation as the length of all the paths in the
poset (without repetition). It should be clear that building the poset
from $R$ and $W$ is based on a poset sorting algorithm which creates
links between those points for which the relation '$\leq$' is
defined. Once the poset is generated, given a point, we can compute
the number of points satisfying the relation '$\leq$' (i.e., the
distribution function). We should emphasize that the parallel points,
those for which '$\leq$' is {\bf not} defined, are {\bf not} used in
the distribution comparison.

Given two intervals $R$ and $W$ with $N$ points of dimension $d$, a
poset-based sorting algorithm can be used to build a poset directed
acyclic graph (DAG). We implemented a variant of the sorting algorithm
in \cite{FaigleT1988} as described by \cite{DaskalakisKMRV2009}, which
has a complexity of $O(Nwd\log N)$ where $w$ is the maximum number of
parallel points. In practice, we take advantage of the lexicographical
order of the points to further reduce the complexity by a constant.
We did not implement or test the faster version suggested in
\cite{DaskalakisKMRV2009}.

Given the set of points in $R$ and $W$, a source point $\dashv$ always
exists from which all other points in the DAG can be reached; also, a
sink point $\vdash$ always exists that can be reached from all other
points. In fact, each dimension of a vector can be defined as
$\vdash_i = \min x_i$ such that $\vdash\leq {\bf x}$, and each
dimension of a vector can be defined as $\dashv_i = \max x_i$ such
that ${\bf x} \leq \dashv$. Such points can always be added to the
poset above.

\mypar{Remarks} Once the poset DAG is generated, the computation of
the empirical distribution is trivial, as the points that satisfy the
relation '$\leq$' can be easily computed for each point in the
DAG. Unfortunately, the use of these distribution are not practical,
because for points close to $\dashv$ a large variation may occur as a
result of the ordering method chosen. Let us explain this problem:
from any point in the DAG we can always reach $\dashv$, this implies
that the distribution value of $F(\dashv)=1$; if we reach $\dashv$
from above the cluster of points, the distribution will likely
increase in small steps because we are counting also parallel node,
but if we reach from below the cluster, the increase will be much
faster because parallel nodes will counted only very close to
$\dashv$; it implies that the distribution may have large variation as
a function how we approach $\dashv$ in its close neighborhood.  This
would result in the method falsely identifying all intervals as
different, even for processes with small dimensions. This problem can
be resolved by taking into account the neighboring parallel points.

The question becomes one of how we can derive a strong order from a
partial order. This would permit the use all points and thus the
parallel points for the comparison as well.

We create a topological ordering by using a breadth-first search algorithm
to traverse the points from $\vdash$ to $\dashv$. The topological ordering
is an ordered and unique partition of the graph $R+W$ (the union of both
intervals with a possible intersection):
\[
{\cal P} = \{\vdash\},X_1, X_2, \dots X_{s-2} ,\{\dashv\}
\]

The topological ordering infers a strong order between points in $X_i$
and $X_j$ where $i<j$. Furthermore, a strong order within $X_i$ can be
inferred by using a lexicographical ordering because points in $X_i$
are parallel points, and thus the strong order extension is arbitrary
and harmless as long as $|X_i|$ is small, having few
points\footnote{If $|X_i|$ is large, any predefined order will skew
  the comparison}. Let us associate the color red with the points from
$R$ and the color white with the points from $W$.  Using the
topological ordering of the graph, we can perform another
breadth-first search to compute the empirical distributions for $R$
and $W$ such that $F_R(\vdash) =F_W(\vdash) =0$ and $F_R(\dashv)
=F_W(\dashv) =1$.  Then $F_R({\bf x})$ is the number of red points
that appear before ${\bf x}$ in the breadth-first search divided by
the total number of red points.  $F_W({\bf x})$ is similarly defined
for the white points.

To perform the breadth-first search, we visit each edge of the poset
DAG, which theoretically takes $O(N^2)$, but on average is closer to
$O(wN)$. Notice that this approach to ordering can be applied, without
modification, for an arbitrary number of intervals.

By using a topological ordering, the construction of a density function
is circumvented; thus our data partition is implicit within the ordering.
As the sample space is not explicitly partitioned (as the authors recommend
in \cite{WangKV2005}), bins may contain just a single point. This is not
a problem, as we are interested only in their cumulative contribution.
Now, we can apply any single-dimension non-parametric approach, including
those describe in Section \ref{sec:distance}.

\mypar{Remarks} Given a poset derived from $R$ and $W$, the topological
ordering is also known, and the partition ${\cal P}$ is unique as will
be explained shortly. We can identify the sets $X_i$ as bins, and can
think of the partition ${\cal P}$ as a histogram induced by the topological
ordering of the data. Each bin contains parallel points, where the maximum
number of parallel points, or the height of the bins, is referred to as
the parallelism of the poset. We do not differentiate between the points
in $X_i$ based on the ordering when considering the bins. Hence, we
consider this partition representative of, and unique across all partitions
that can be obtained by any point permutation of the bin $X_i$.

In other words, if $R$ and $W$ are obtained from the same stochastic
process, then the set $X_i$ in the topological ordering should have
an even mix of red and white points. For this partition to be
effective, the parallelism of the poset should be small and the number
of bins large. Otherwise, each bin will not have an even mix of points
and we will end up with a few bins with high parallelism. This result
is undesirable, because it means that the ordering within each bin
will be based on the lexicographical order, which is arbitrary and
provides no real information.  This discussion is not restricted to
the lexicographical ordering, but extends to any fixed ordering of the
points performed without any prior knowledge of the data.

The main limitation of this method is the inverse relationship between
the number of points and the number of dimensions. For example, if we
study 20-point intervals from a series with 500 dimensions, then the
probability of building a meaningful poset is very slim indeed, with a
high probability that all 40 points are parallel. As a result of the
high degree of parallelism, it is not possible to infer any '$\leq$'
relations amount the points. We will show that this poset approach is
ideal for series with less than 10 dimensions, if the intervals consist
of about 200 points each.

\mypar{Remark} A natural extension of this method would use the
dominant direction of the data instead of the fixed one used by the
poset sorting algorithm, and then define the '$\leq$' ordering
accordingly. For example, one could use the direction of the dominant
eigenvector to describe a new ordering, or to perform a domain
rotation, which would facilitate the use of the regular sorting method.
Where possible, the easiest solution is to increase the number of
samples accordingly.

\mypar{Remark} A multi-dimension problem is reduced to a
single-dimension problem by using comparisons and counting.  The
capability of this method can be extended by using the quantitative
contribution of each dimension distance instead of a bare
comparison. In the next section, we will introduce an approach that
extends the poset-based method to work on series with thousands of
dimensions by quantifying the distance between points based upon the
contribution of each vector component.

\mypar{Remark} The appeal of the poset-based topological ordering is that
poset ordering reduces to the usual ordering when applied to a single
dimension series.

%\section{Dominant eigenvector  sorting}

\subsubsection{Minimum-Spanning-Tree Based Topological Order}
\label{sec:mst-algo}

In this section, we examine the use of a well known approach used in
the determination of clusters: the minimum-spanning tree (MST) as
proposed by Friedman and Rafsky \cite{FriedmanR1979}.

Given two intervals, $R$ and $W$, in a series with dimension $d$, an MST
can be built using Dijkstra's algorithm. This method involves computing
the distances between each of the points in $R$ and $W$ (i.e., $N(N-1)/2$
distances or edges for $N$ points), sorting the edges in increasing order
according to their distance, and then attempting to insert each edge into
the tree provided it does not create a loop. If the introduction of an edge
would result in a loop, it is ignored. This process is continued until all
points have been added to the tree, which produces the MST.

The most expensive part of the computation is the computation of the
distances, which takes $O(dN^2)$.

In general, an MST is not necessarily unique, but it is the
minimum-cost tree connecting all points. The leaves of the MST are
easily recognizable because these are all the points with only a
single edge. By conducting a breadth-first search from the leaves it
is possible to build a topological ordering and the partition ${\cal
  P}$:

\[
{\cal P} = \{ leaves \},X_1, X_2, \dots X_{s-2} ,\{ roots\}
\]

Because the search is started from the leaves, it is possible to end
up with either one or two roots. Notice that the partition can be
determined in $N$ steps, as the tree is a DAG with $N-1$ edges.

The partition that is generated from the MST derives a strong ordering
among the points, but without a fixed orientation, which makes it more
resilient to the dimensionality of the space than the poset
topological ordering. In our implementation points in a bin $X_i$ are
parallel and ordered lexicographically, instead of in decreasing
order of the distance from the root \cite{FriedmanR1979}. This choice
simplifies the computation at the expense of possibly biasing the
comparison.

With a partition, and thus a strong order, the empirical distribution
functions can be inferred. The distributions allow us to perform the
statistic tests that we will explain in the following section.

\mypar{Remark} The MST-based topological ordering derives its order
from the data points and adapts to them; ordering the points from
the outskirts of the cluster towards the center. As noted, the
poset-ordering method is positively correlated to the number of
points and negatively correlated to the number of dimensions. In
other words, it is more sensitive to series with a small number
of dimensions and a large number of points. In contrast, the
MST-based ordering method is less sensitive to the number of points
and more sensitive to the number of dimensions. This difference
arises from the fact that an increase in dimensions contributes
to the distance between points; thus, the ability to differentiate
between points becomes more sensitive.

In \cite{FriedmanR1979}, the authors deployed the Wald--Wolfowitz test
and the Smirnov test on MST data. Briefly, the Wald--Wolfowitz use a
coin based test: head, we visit a node in $R$ and tail, a node in $W$;
a path in the DAG should have the distribution of a sequence of coin
tosses. The Smirnov test is basically the comparison of distribution
using the Kolmogorov-Smirnov's test. In this section we will describe
the {\em Radial Smirnov} test, for which the original authors showed
evidence of its power in finding changes in the variance, as opposed
to tests which can identify changes in the average. We have chosen to
postpone the implementation of the Wald--Wolfowitz test because of
contradicting evidence of its effectiveness. Early literature
\cite{Siegel1959} suggests that the Kolmogorov--Smirnov method is
preferable to the Wald--Wolfowitz method; while more recent works
\cite{MagelW1997} paint a more complex picture (even for very small
samples of up to $20$).  Finally, the \cite{GrettonBRSS08} manuscript
suggests that both methods tend to break down for large dimensions. We
are aware that the original implementation using MST and the
Wald--Wolfowitz test may provide better discriminative power due to
its sensitivity of changes of the average (see Section
\ref{sec:synthetic}).

\subsubsection{Single Dimension, Poset, and MST}

In the experimental results Section \ref{sec:mult-experiments}, we
will compare the discriminative powers of all these methods. We will 
show that the poset-based method performs better than any other
method on data with 10 or fewer dimensions, excelling in speed and
discriminative power with respect to changes in both the average
and variance. The MST-based method performs better on data with 10
or more dimensions and when detecting changes in variance. We will
show that for changes in the average, we are better off using other
methods.

This section's references are
\cite{FriedmanR1979,HallT2002,KimF1987,Bickel1969}.

This Section's reference are
\cite{Muller1997,SongWCR2007,DasuKVY06,SriperumbudurGFLS2009,KulldorffMDYKP2007,Bickel1969,BickelRS2006}.

\subsection{Single-Dimensional Series: Statistical Test, Unified Measure, and Notations }
\label{sec:distance} 

We start by introducing the terminology and definitions used from
hereafter.

Given two arbitrary empirical CDFs $F_R$ and $F_W$, a {\bf
  non-parametric statistical test} is composed of the following three
components:
\begin{enumerate}
\item The null hypothesis $H_0$: $F_R \sim F_W$ the assumption that
  the distributions are the same.
\item The measure $\D(F_R,F_W)$: it quantifies the distance between
  the distributions.  Here, the term ``measure'' is used after the
  same fashion as \cite{AliS1966}.
\item The statistical significance: it is the probability of judging
  the CDFs as different but they are similar; typically, we will
  consider significance levels at 0.05 but we can make it closer to
  zero in order to make the comparison stronger.
\end{enumerate}

In the following, we shall work with measures that are suitable to
capture similarity, thus designed to confirm the null hypothesis. In
the literature there are available measures that are more suitable to
capture difference and thus to confirm the complementary null
hypothesis ($H_1$). These {\em difference} measures can be always
added but for now they are beyond the purpose of this work. We shall
show that similarity measures are more suitable for the framework we
are going to build.

\mypar{Unified Measure} We present a generalized measure $\D$ that
unifies all the measures in this paper. It also helps us to abstract
the components of a measure and provides a notation for them.
\begin{equation}
\begin{gathered}[t]
  \D : \R^{N}\times \R^{N}\rightarrow \R \\
  \begin{split}
    \D : \D_{p,\f,\g_s,\h_k}(\Vc{r},\Vc{w})= &\f(N)*\g_s\big(\| \h_k(\Vc{r},\Vc{w}) \|_{p}\big) \\
    & p\geq 0 \\
    &\f : \N \rightarrow \R                 \\
    &\g_s : \R \rightarrow \R  \\
    &\h_k : {\R}^{N}\times {\R}^{N}\rightarrow {\R}^{N} 
   \end{split}
   \label{eq:distance-general}
\end{gathered}
\end{equation}

In Equation \ref{eq:distance-general}, the measure $\D$ takes in the
$N$-element vectors $\Vc{r}$ and $\Vc{w}$, which represent the input
distributions $F_R$ and $F_W$ respectively. It produces the final
output by performing these four steps:
\begin{enumerate}
\item Compare the elements of the input vectors on a one-to-one basis
	using the function $\h_k$.
\item Aggregate the results using a vector $p$-norm $\|.\|_p$.
\item Scale the result using the function $\g_s$.
\item Normalize the result using the function $\f$ so that the final
	result will be independent of $N$ for large $N$.
\end{enumerate}

Each of the functions in the definition of $\D$ takes different
forms for different measures, as shown in Table~\ref{table:summary}.
The only exception is the vector $p$-norm, which is defined as follows:

\begin{equation*}
\begin{split}
\|{\bf x}\|_p &= \Big(\sum_i |x_i|^p\Big)^{\frac{1}{p}}, \\ 
\|{\bf x}\|_\infty &= \max_{i} |x_i|, \mbox{and}\\ 
\|{\bf x}\|_0 &= \sum_i x_i. \\
\end{split}
\end{equation*}

Notice that the definition of $\|{\bf x}\|_0$ is not standardized in
the literature. For example, in \cite{GolubVL1996} $\|{\bf x}\|_0$ is
$\sum sign(x_i)$; while some other sources refer to $\|{\bf x}\|_0$
as the number of non-zero elements of $x$. Our vector norm helps simplify
the definition of $\D$.

We compute the distance between the two windows $R$ and $W$ using
$\D(R,W)$, as defined in Equation~\ref{eq:distance-general}. To
compute the distance, we can use the input vectors for these windows
to represent an empirical PDF or an empirical CDF.

% We compute the distance between the two windows $R$ and $W$ using
% $\D(R,W)$, as defined in Equation~\ref{eq:distance-general}. To
% compute the distance, we can use the input vectors for these windows
% to represent a PDF or CDF, both empirical.

\begin{table*}%{5in}
\tbl{The measure $\D_{p,\f,\g_s,\h_k}(F_R=\Vc{x},F_W=\Vc{y})$. Note ${\bf \iota}$ = identity function.\label{table:summary}}{%
\centering
\footnotesize
\begin{tabular}{ll|l|llll}
\hline\hline 
{\bf Name} & {\bf Eq.} &{\bf Measure} & $\bf p$ & $\f(N)$ & $\g_s(x)$ & $\h_k(x,y)$  \\
\hline\hline 
Bhattacharyya  &\ref{eq:Bhattacharyya}    &$ \sum_{i}\sqrt{x_iy_i}$                                              &$0$& $NA$& $ \iota$& $ \sqrt{xy}$      \\      
\hline
Camberra  &\ref{eq:camberra}         &$ \sum_{i}\frac{|x_i-y_i|}{x_i+y_i}$                                         &$0$& $\frac{1}{\sqrt{N}}$&  $ \iota$& $ \frac{|x-y|}{x+y}$ \\ 
\hline 
$\chi^2$  &\ref{eq:chi}         &$ \sum_{i}\frac{(x_i-y_i)^2}{x_i}$                                &$0$& $1$&  $ \iota$& $ \frac{(x-y)^2}{y}$      \\  
\hline
Cramer--von Mises  &\ref{eq:cramermises} &$\sum_{i}\big(x_i-y_i\big)^2$                                              & $2$& $1$& $x^2$& $ x-y$ \\ 
\hline
Euclidean     &     &$ \sqrt{(\sum_{i}(x_i-y_i)^2)}$                                              & $2$& $1$& $\iota$& $ x-y$ \\ 
\hline
Hellinger   & \ref{eq:hellinger}       &$\frac{1}{2}\sum_{i}\big(\sqrt{x_i}-\sqrt{y_i}\big)^2$                       &$0$& $1$&  $ \iota$& $ (\sqrt{x}-\sqrt{y})^2$      \\  
\hline
Jin-K       &       &$ KLI\bigl(x,\frac{x + y}{2}\bigr)$                                      &$0$& $\frac{1}{\sqrt{N}}$& $ \iota$& $ x\log_2(\frac{2x}{x+y})$                                   \\ 
\hline
Jin-L        & \ref{eq:jin-l}      &$ KLI\bigl(x,\frac{x + y}{2}\bigr) + KLI\bigl(y,\frac{x + y}{2}\bigr)$   &$0$& $1$& $ \iota$& $ x\log_2(\frac{2x}{x+y})+y\log_2(\frac{2y}{x+y})$          \\ 
\hline
Jensen--Shannon  & \ref{eq:js}   &$ \frac{1}{2}\bigl(KLI\bigl(x,\frac{x + y}{2}\bigr) + KLI\bigl(y,\frac{x + y}{2}\bigr)\bigr)$&$0$& $1$& $ \iota$& $ 0.5(x\log_2(\frac{2x}{x+y})+y\log_2(\frac{2y}{x+y}))$   \\ 
\hline
Kolmogorov--Smirnov & \ref{eq:ks}&$ \max_{i}|x_i -y_i|$                                                  & $\infty$& $\sqrt{N}$& $\iota$& $ x-y$\\ 
\hline
Kullback--Leibler-I & \ref{eq:kli}&$\sum_{i}x_i\log_2\frac{x_i}{y_i}$                                     & $0$& $\frac{1}{\sqrt{N}}$& $ \iota$&  $ x\log_2(\frac{x}{y})$                                 \\ 
\hline
Kullback--Leibler-J & \ref{eq:klj} &$\sum_{i}(x_i-y_i)\log_2\frac{x_i}{y_i}$                               & $0$& $1$& $ \iota$& $ (x-y)\log_2(\frac{x}{y})$                                \\ 
\hline
$K_r$    & \ref{eq:gkr}          &$\frac{1}{(r-1)} \log_2\big(\sum_{i}x_i^ry_i^{1-r}\big)$         & $0$& $NA$& $log_2(x)$& $x^ry^{1-r}$\\       
\hline
$K_s$    & \ref{eq:gks}          &$\frac{1}{s-1}\big(-1+\sum_{i}x_i^sy_i^{1-s}\big)$               & $0$& $NA$& $\frac{(x-1)}{s-1}$& $x^sy^{1-s}$\\      
\hline
$K_s^2$  & \ref{eq:gk2s}          &$\frac{1}{(s-1)s}\big(-1+\sum_{i}x_i^sy_i^{1-s}\big)$         & $0$& $NA$& $\frac{(x-1)}{s(s-1)}$& $x^sy^{1-s}$\\       
\hline
Minkowsky  & \ref{eq:minkowsky}       &$ \bigl(\sum_{i}|x_i-y_i|^r\bigr)^{\frac{1}{r}}$                        & $r=3$& $\log_2N$& $\iota$& $ x-y$ \\
\hline
$\phi$    & \ref{eq:phi}         &$ \max_{i}\frac{|x_i -y_i|}{\sqrt{\min\bigl(\frac{x_i+y_i}{2}, 1-\frac{x_i+y_i}{2}\bigr)}}$ & $\infty$& $\frac{\sqrt{N}}{\log_2N}$& $\iota$& $\frac{|x-y|}{\sqrt{\min(\frac{x+y}{2} 1-\frac{x+y}{2})}}$\\ 
\hline
Variational  & \ref{eq:variance}      &$ \sum_{i}|x_i-y_i|$                                                               &$1$& $\frac{1}{\sqrt{N}}$& $ \iota$& $|x-y|$      \\      
\hline
$\Xi$      & \ref{eq:ksi}        &$ \max_{i}\frac{|x_i -y_i|}{\sqrt{\frac{x_i+y_i}{2}*\bigl(1-\frac{x_i+y_i}{2}\bigr)}}$  &$\infty$& $\frac{\sqrt{N}}{\log_2N}$& $\iota$& $\frac{|x-y|}{\sqrt{\frac{x+y}{2}(1-\frac{x+y}{2})}}$\\ 
\hline
\hline
\end{tabular}}
\end{table*}

\section{Distance Measure Specification}
\label{sec:distance-measures}

For a finite number of samples, a measure is the quantitative
comparison of the distance of two vectors. For example, the Euclidean
distance of two $n$-dimension vectors is a norm and a metric; that is,
$E\geq0$, $E(\Vc{a},\Vc{b}) = E(\Vc{b},\Vc{a})$ and $E(\Vc{a},\Vc{b})
+E(\Vc{b},\Vc{c})\geq E(\Vc{a},\Vc{c})$. In this spirit, we can extend
the measures commonly used for vectors ---i.e., the Euclidean
distance--- or  for PDFS ---i.e., information-theoretic
measures--- and apply them to CDFs as inputs.

Let us consider the case when we want to compare two series in $\R$.
We can {\em always} define intervals using CDFs, which means we can
{\em always} compare two CDFs as vectors, or without any arbitrary
determination of buckets or reduction to discrete values. Moreover,
two series drawn from the same process will converge to the same CDF
(to the same vectors or PDFs), and all the measures will converge to
zero. 

However, the measure output for {\em different} CDFs can be literally
bounded only by the number of points in the comparison, and, certainly
larger than in the case of PDFs; for example, $[0,2]$ for the measure
in \cite{Jin1991} otherwise most PDFs measures are always smaller than
one. 

Symmetric measures, such as the Kullback--Leibler--J
Equation~\ref{eq:klj}, the Jensen--Shannon Equation~\ref{eq:js}, or
the Variational Equation \ref{eq:variance}, are {\em more} suitable
for our needs.  In fact, the symmetry property ensures that the
measure is not biased by the reference window and both windows can be
interchanged if necessary. Nonetheless, we can find applications for
positive asymmetric measures, such as $\chi^2$ in
Equation~\ref{eq:chi}, because these measures offer better
discriminative power when applied to empirical distributions,
especially when observations are few or when the reference is {\em
  more } trustworthy than the running window \cite{Lee1999}.

We show that 17 of the measures in Table~\ref{table:summary} generate
output CDFs that are independent of the input CDFs. For example, the
Kolmogorov--Smirnov measure has a limit distribution that is normal,
independent of the input stochastic processes.
In \SectionRef{sec:input-distribution-independence} we explore this
independence, explain the reason for using the Kolmogorov--Smirnov
measure and present experimental evidence. Unfortunately, we found that
the generalized functions $K_r$, $K_s$, and $K_s^2$, used by the PDFs
(Equation~\ref{eq:gks}, \ref{eq:gkr}, and \ref{eq:gk2s}), tend not to
work for the CDFs, because we cannot find a CDF for their output
measures (see \SectionRef{sec:window-independence}).

Finally, we must clarify that there are several measures which we
chose not to investigate, these include the geometric measure, the
relative frequency model, and the resistor distance. We did not use
the geometric measure $\cos(\Vc{a},\Vc{b})$ \cite{WangWY1992},
\cite{JonesF1987}, because the measure only compares the direction of
two vectors without considering their magnitude, which we consider
important. \cite{ShivakumarM95} proposed the relative frequency model
to overcome the drawbacks of the cosine measure when used with
histograms built from a set of discrete entities that are easily
classified into buckets, such as a bag of words.  The resistor
distance is described in \cite{JohnsonS2000} and is an alternative
symmetric version of the Kullback--Leibler measure.

\subsection{Information-Theoretic Measure Extensions}
\label{sec:contribution}

In this section, we present our contributions to the field of information-
theoretic measures. We will detail how we have extended these measures
and applied them to CDFs.

\mypar{Kullback--Leibler-I (KLI)~\cite{KullbackL1951}} KLI is an
asymmetric measure where $F_R,F_W \ne 0$, which assumes that undefined
values have no contributions. Notice that $KLI(F_R,F_W)=0$ {\em iff}
$F_R=F_W$; however if $F_R \ne F_W$, $KLI(F_R,F_W)$ can be arbitrarily
large.
\begin{equation}\small
\label{eq:kli}
KLI(F_R,F_W) = \sum_{y=s_y\in R\cup W}F_R(y)\log_2\biggl(\frac{F_R(y)}{F_W(y)}\biggr)
\end{equation}

\mypar{Kullback--Leibler-J (KLJ) \cite{KullbackL1951}, \cite{AliS1966}}
KLJ is a symmetric measure where $F_R,F_W \ne 0$. The KLJ measure is
similar to the KLI measure in that $KLJ(F_R,F_W)=0$ {\em iff} $F_R=F_W$,
and if $F_R \ne F_W$ then $KLJ(F_R,F_W)$ can be arbitrarily large.
\begin{equation}\small
\label{eq:klj}
KLJ(F_R,F_W) = KLI(F_R,F_W) + KLI(F_W,F_R)
\end{equation}
An alternative example of a symmetric adaptation of KLI is described in
\cite{JohnsonS2000}.

\mypar{Jin-L (JinL)~\cite{Jin1991}} JinL is a symmetric measure that
is always defined, assuming that $0=0log_2(0/0)$. $JinL(F_R,F_W)=0$
{\em iff} $F_R=F_W$; otherwise, $JinL(F_R,F_W)$ will not be larger
than $2N$ if $F_R \ne F_W$.
\begin{equation}\small
\label{eq:jin-l}
\begin{split} 
  JinL(F_R,F_W) = & KLI\biggl(F_R,\frac{F_R + F_W}{2}\biggr) + \\
                  & KLI\biggl(F_W,\frac{F_R + F_W}{2}\biggr)
\end{split}
\end{equation}

\mypar{Jensen--Shannon (JS)~\cite{Jensen1906,Shannon1948}} We describe
JS using the Kullback--Leibler measure; however, the JS has
historically been formulated using the {\em entropy} measure (i.e.,
$H(F_R)=-\sum_{i=0}^{k-1}F_R(i)\log_2F_R(i)$). We use
Kullback--Leibler as the generalization of this entropy.
\begin{equation}\small
\label{eq:js}
\begin{split}
JS(F_R,F_W) =& \frac{1}{2}\big[ KLI\biggl(F_R,\frac{F_R + F_W}{2}\biggr) \\
             &+ KLI\biggl(F_W,\frac{F_R + F_W}{2}\biggr)\big]
\end{split}
\end{equation}

\mypar{$\chi^2$ ($\chi^2$) \cite{Kagan1963,Vadja1972,Hope1968}}
$\chi^2$ is an asymmetric measure that is defined for $F_R \ne 0$;
again, the contribution is not considered when $F_R=0$. Notice that
$\chi^2(F_R,F_W)=0$ {\em iff} $F_R=F_W$; however, if $F_R \ne F_W$,
$\chi^2(F_R,F_W)$ can be arbitrarily large.
\begin{equation}\small
\label{eq:chi}
\chi^2(F_R,F_W) = \sum_{y=s_y\in R\cup W}\frac{\big(F_R(y)-F_W(y)\big)^2}{F_R(y)}
\end{equation}

\mypar{Hellinger (H)~\cite{Hahn1912,Vadja1972}, \cite{AliS1966}}
H is a symmetric measure that is always defined. The square root
operation {\em normalizes} the component values to ensure that the
component-wise comparison is less biased. The value of all components
are between 0 and 1, but components near the extremes (0 or 1) are
moved closer to $\frac{1}{2}$. Notice that $H(F_R,F_W)=0$ {\em iff}
$F_R=F_W$.
\begin{equation}\small
\label{eq:hellinger}
H(F_R,F_W) = \frac{1}{2}\sum_{y=s_y\in R\cup W}\big(\sqrt{F_R(y)}-\sqrt{F_W(y)}\big)^2
\end{equation}

\mypar{Bhattacharyya (B)~\cite{Bhattacharyya1943,Kailath1967}}
B is a symmetric measure that is always defined. Notice that
$B(F_R,F_W)<=N$ {\em iff} $F_R=F_W$, and will tend towards 0 if
$F_R \ne F_W$, $B(F_R,F_W)$. Also, if applied to the PDFs $\Vc{x}$
and $\Vc{y}$, Bhattacharyya and Hellinger measures are related in
following manner: $1-B(\Vc{x},\Vc{y})=H(\Vc{x},\Vc{y})$. However,
for CDFs, Hellinger is more suitable because we can determine a
significance measure. Bhattacharyya is presented for completeness.
\begin{equation} \small
\label{eq:Bhattacharyya}
B(F_R,F_W) = \sum_{y=s_y\in R\cup W}\sqrt{F_R(y)F_W(y)}
\end{equation}

\mypar{Variational Distance (V)~\cite{Pinsker1960,AliS1966}}
V is a symmetric measure that is always defined. Notice that $V(F_R,F_W)=0$
{\em iff} $F_R=F_W$; however, if $F_R \ne F_W$ then $V(F_R,F_W)$ will be no
larger than $2N$. This measure is also known as the Manhattan measure or
Kolmogorov's variance distance \cite{AliS1966}.
\begin{equation}
\label{eq:variance}\small
V(F_R,F_W) = \sum_{y=s_y\in R\cup W}|F_R(y)-F_W(y)|
\end{equation}

For the remaining measures, we use the following notation
${\bf \cal B}_x(F_R,F_W)$ \cite{Chernoff1952} to refer to the sum
$\sum_{y=s_y\in R\cup W}F_R(y)^xF_W(y)^{1-x}$.
\vspace{0.5cm}

\mypar{Generalized $K_r$ ($K_r$)~\cite{TanejaK2004}}
$K_r$ is a generalization of measures that are based on the
Kullback--Leibler methodology.
\begin{equation}\small
\label{eq:gkr}
K_r(F_R,F_W) = \begin{cases}
  \text{ {\bf if} } r=1 \hspace{0.2cm}KLI(F_R,F_W)   \\
  \text{ {\bf if} } r>0 \\
  \hspace{0.4cm} \frac{1}{(r-1)} \log_2\big({\cal B}_r(F_R,F_W)\big) 
  \end{cases}
\end{equation}

\mypar{Generalized $K_s$ ($K_s$)~\cite{TanejaK2004}}
For specific values of $s$ with PDFs, $K_s$ can generate Bhattacharyya,
Hellinger and Kullback--Leibler.
\begin{equation}\small
\label{eq:gks}
K_s(F_R,F_W) = \begin{cases}
  \text{ {\bf if} }s=1  \hspace{0.2cm} KLI(F_R,F_W)   \\
  \text{ {\bf if} } s>0    \\ \hspace{0.4cm}\frac{1}{s-1}\big(-1+{\cal B}_s(F_R,F_W)\big)      
  \end{cases}
\end{equation}

\mypar{Generalized $K_s^2$ ($K_s^2$)~\cite{TanejaK2004}}
For specific values of $s$ with PDFs, $K_s^2$ can generate Bhattacharyya,
Hellinger, Kullback--Leibler and $\chi^2$.
\begin{equation}
\label{eq:gk2s}\small
K_s^2(F_R,F_W) = \begin{cases}
  \text{ {\bf if} } s=1  \hspace{0.2cm}KLI(F_R,F_W)   \\
  \text{ {\bf if} } s=0 \hspace{0.2cm} KLI(F_W,F_R)   \\
  \text{ {\bf if} } s>0   \\
  \hspace{0.4cm}\frac{1}{(s-1)s}\big(-1+{\cal B}_s(F_R,F_W)\big)    \\
  \end{cases}
\end{equation}

Notice the equivalence relations $K^2_{1/2}(\Vc{x},\Vc{y}){=}
2K_{1/2}(\Vc{x},\Vc{y})=4(1{-}B(\Vc{x},\Vc{y})){=}4H(\Vc{x},\Vc{y})$
and $K^2_{2}(\Vc{x},\Vc{y}) {=} 2K_{2}(\Vc{x},\Vc{y}) =
\chi^2(\Vc{x},\Vc{y})$, where $\Vc{x}$ and $\Vc{y}$ are PDFs. Although
we present $K_s$, $K_s^2$, and $K_r$ for completeness, we could not
find a significance measure for most methods, excepting $\chi^2$ and a
few others.

\subsection{Classic Distribution-Function Measures}
\label{sec:df} 
Here we present the set of measures $\D(F_R,F_W)$ from the literature
that already use CDFs.

\mypar{Kolmogorov--Smirnov
  (KS)~\cite{Kolmogorov1933e,Kendall1991,FellerII}} KS is a symmetric
measure that is always defined. Notice that $KS(F_R,F_W)=0$ {\em iff}
$F_R=F_W$; and if $F_R \ne F_W$ then $KS(F_R,F_W)$ is no larger than
1.
\begin{equation}\small
\label{eq:ks}
\begin{split}
  KS(F_R,F_W) &= \sup_{y\in \R}|F_R(y) -F_W(y)| \\
              &\geq \max_{y=s_y\in R\cup W}|F_R(y) -F_W(y)| \\ 
\end{split}
\end{equation}

\mypar{$\phi$ ($\phi$) \cite{KiferBG2004}} $\phi$ is a symmetric
measure that is always defined. Notice that $\phi(F_R,F_W)=0$ {\em
  iff} $F_R=F_W$; and if $F_R \ne F_W$ then $\phi(F_R,F_W)$ is no
larger than 2.
\begin{equation}
\label{eq:phi} \small
\begin{split}
  & \phi(F_R,F_W) \\
  &= \sup_{y\in \R}\frac{|F_R(y)
    -F_W(y)|}{\sqrt{\min\bigl(\frac{F_R(y)+F_W(y)}{2},
      1-\frac{F_R(y)+F_W(y)}{2}\bigr)}} \\ 
  & \geq
  \max_{y=s_y\in R\cup W}\frac{|F_R(y)
    -F_W(y)|}{\sqrt{\min\bigl(\frac{F_R(y)+F_W(y)}{2},
      1-\frac{F_R(y)+F_W(y)}{2}\bigr)}}
\end{split}
\end{equation}

\mypar{$\Xi$ ($\Xi$) \cite{KiferBG2004}} The $\Xi$ is a symmetric
measure that is always defined. Notice that $\Xi(F_R,F_W)=0$ {\em iff}
$F_R=F_W$; and if $F_R \ne F_W$ then $\Xi(F_R,F_W)$ is no larger than
2.
\begin{equation}
\label{eq:ksi} \small
\begin{split}
  & \Xi(F_R,F_W) \\
  &= \sup_{y\in \R}\frac{|F_R(y)
    -F_W(y)|}{\sqrt{\frac{F_R(y)+F_W(y)}{2}*\bigl(1-\frac{F_R(y)+F_W(y)}{2}\bigr)}}\\
  & \geq \max_{y=s_y\in R\cup W}\frac{|F_R(y)
    -F_W(y)|}{\sqrt{\frac{F_R(y)+F_W(y)}{2}*\bigl(1-\frac{F_R(y)+F_W(y)}{2}\bigr)}}
\end{split}
\end{equation}

\mypar{Cram\'er--von Mises ($W^2$)~\cite{Melucci2007}} $W^2$ is a
symmetric measure that represents the Euclidean distance of a
vector. Notice that $W^2(F_R,F_W)=0$ {\em iff} $F_R=F_W$; and if $F_R
\ne F_W$ then $W^2(F_R,F_W)$ will be no larger than $2N$ (the number
of samples in the window). Recently, a new definition has been
proposed \cite{Melucci2007} that does not follow the original
definition by Anderson\cite{Anderson1962} exactly.
\begin{equation}\small
\label{eq:cramermises}
\begin{split}
W^2(F_R,F_W) & = \int_{-\infty}^{\infty}\big(F_R(y)-F_W(y)\big)^2dy \\
             & = \sum_{y_i=s_y\in R\cup W}(y_{i+1}-y_i)\big(F_R(y_i)-F_W(y_i)\big)^2 \\
             & \sim  \sum_{y=s_y\in R\cup W}\big(F_R(y)-F_W(y)\big)^2
\end{split}
\end{equation}

\mypar{Minkowsky ($M_r$)~\cite{Batchelor1978,WilsonM1997}} $M_r$ is a
symmetric, parametrized measure that generalizes both the Euclidean
($r=2$) and Variational ($r=1$) distance of a vector.  Notice that
$M_r(F_R,F_W)=0$ {\em iff} $F_R=F_W$. In our experiments, we set
$r=3$.
\begin{equation}\small
\label{eq:minkowsky}
M_r(F_R,F_W)  = \big(\sum_{y=s_y\in R\cup W}|F_R(y)-F_W(y)|^r\big)^{\frac{1}{r}}
\end{equation}

\mypar{Camberra (C)~\cite{Diday1974,WilsonM1997}} C is symmetric, and
a relative measure of the Euclidean distance, in the same fashion that
$\phi$ is a relative measure of the Kolmogorov--Smirnov
distance. Notice that $C(F_R,F_W)=0$ {\em iff} $F_R=F_W$;
\begin{equation}\small
\label{eq:camberra}
C(F_R,F_W)  = \sum_{y=s_y\in R\cup W}\frac{|F_R(y)-F_W(y)|}{F_R(y)+F_W(y)}
\end{equation}

\subsection{Rank Function Measures}
\label{sec:rank} 

For the sake of completeness, we discuss a few other measures which
are based on rank measures for a single-dimensional series. These
methods are used for the comparison of experimental results to show
that our measures have comparable discriminative powers. That being
the case, it means that we could use them instead of, or in
combination with the following measures. Of course, these measures are
not clearly defined in the case of multi-dimensional series.

\mypar{Wilcoxon-Mann-Whitney (Wilcox)~\cite{Wilcoxon1945,MannW1947}}
Wilcox is a symmetric test that is based on the rank of the events
that occur in each series. This is a standard test that is available
in R. We also used the {\bf t-test}.

%%%%% PAOLO Sept 3 2011 10:12 pm

\section{Significance or $p$-value of a Measure}
\label{sec:pvalue}

\begin{comment}
\doublefigure{0.5}{KolmogorovSmirnov}{Hellinger}{ The
  Kolmogorov--Smirnov-measure CDFs (left) and Hellinger-measure CDFs
  (right)}{fig:distributions}
\end{comment}

For some measures $\D$, the distribution of the measure values is well
studied. Some examples include $\sqrt{N}KS(F_R,F_W)$ for CDFs
generated from windows with $N$ points, or $\chi^2(f_R,f_W)$ for PDFs
with $N$ buckets. For others, the distribution can be determined by
simulation, as in the case of $\phi(F_R,F_W)$ or $\Xi(F_R,F_W)$.  Our
goal is to pre-determine the measure distribution and thus the measure
significance through the use of tables or simulations. We could use
bootstrapping instead; bootstrapping is a powerful approach,
computable on the fly, and adaptable to any series; however, it
requires a training set and, therefore, an {\em a priori} knowledge of
the series. This extra knowledge makes bootstrapping inconvenient for
the final user of these statistical measures; the final user simply
wants the measures to describe the data.

We have found, through empirical testing, that simulations are
sufficient in determining a simplified distribution and thus the
significance for most of the measures $\D(F_R,F_W)$ used in this
paper. However, we were {\bf not} able to find a distribution function
for the following measures:
\begin{enumerate}
\item KLI, because it produces negative measures.
\item Bhattacharyya, generalized $K_s$, $K_s^2$, and $K_r$, because we
	could not find a normalizing function $\f(N)$ (see Table
	\ref{table:summary}), and thus a CDF.
\end{enumerate}

\subsection{Simulation, $\D$, and its CDF} 

\begin{table*}%
\tbl{Simulation of the expectation for the null hypothesis $H_0$ measure(i.e., $E[\D]$).\label{tab:average_value}}{%
\centering \footnotesize
\begin{tabular}{|l|r|r|r|r|r|r|}
  \hline\hline
  $N$         &100	       &1000      & 10000	&100000	      &1000000         &$1/\f(N)$ \\ \hline\hline
  E[$\phi$]   &   0.28269   & 0.09925  &    0.03391  &     0.01141 &      0.00374   & $\sim log(N)/\sqrt{N}$ \\ \hline
  E[$\Xi$]    &   0.30643   & 0.10466  &    0.03523  &     0.01169 &      0.00383   & $\sim log(N)/\sqrt{N}$ \\ \hline
  E[KS]       &   0.11867   & 0.03830  &    0.01227  &     0.00383 &      0.00124   & $\sim 1/\sqrt{N}$ \cite{FellerII}\\ \hline
  E[KLI]      &   1.95563   & 1.00179  &    1.76800  &    -0.10939 &     51.66521   & N/A \\ \hline
  E[KLJ]      &   2.87245   & 2.84168  &    2.94492  &     2.81301 &      3.06639   & constant \\ \hline
  E[JnK]      &   0.63525   & 0.14532  &    0.51653  &    -0.40636 &     25.44923   & N/A \\ \hline
  E[JinL]     &   0.74388   & 0.71258  &    0.73634  &     0.70326 &      0.76659   & constant \\ \hline
  E[JS]       &   0.37194   & 0.35629  &    0.36817  &     0.35163 &      0.38329   & constant \\ \hline    
  E[$\chi^2$] &   2.11921   & 2.00276  &    2.04105  &     1.95051 &      2.12593   & constant \\ \hline
  E[V] 	      &   8.88756   &28.09942  &   89.08333  &   272.74629 &    915.27286   & $\sim \sqrt{N}$\\ \hline
  E[H]        &   0.26507   & 0.24784  &    0.25529  &     0.24374 &      0.26568   & constant \\ \hline
  E[B]        & 100.23492   &1000.25215&10000.24470  &100000.25625 &1000000.23429   & $\sim N$ \\ \hline
  E[W]        &   0.67146   & 0.66599  &    0.67193  &     0.63212 &      0.71241   & constant \\ \hline
  E[E]        &   0.76166   & 0.75571  &    0.75984  &     0.73908 &      0.77774   & constant \\ \hline
  E[$M_{r=3}$] &  0.35486    &0.23930   &   0.16416   &    0.10910  &     0.07781    & $\sim 1/log_3(N)$ \\ \hline
  E[C]        &  16.74404   &54.79298  &  177.49281  &   557.46331 &   1809.98318   & $\sim \sqrt{N} \sim N^{\frac{540}{1000}}$ \\ \hline \hline
\end{tabular}}
\end{table*}

The simulation process can be described as follows. We select a measure
$\bar\D$, choose the number of samples $N$, and then randomly generate
$M$ pairs of $N$ samples each, as taken from the same stochastic process.
One example of a simulation run might use the following parameters:
$\bar\D=KS$, $N=1000$, $M=5000$.

We generate a CDF from the measure values $x$, which is denoted
as $F_N^1 (x)$. Repeating the simulation $k$ times results in
different CDFs $F_N^i$ for $i\in[1,k]$, which produces a {\em cloud}
of functions $\{F_N^i\}_{1\leq i\leq k}$. By extension, changing the number
of samples $N$, results in {\em clouds} of functions,
$\{F_N^i\}_{(1\leq i\leq k,N)}$. For any number of samples $N$, we
want to determine the normalizing function $\f(N)$ that makes it
possible to compare the measures with respect to other sample sizes,
$F_{N_0} \sim F_{N_1}$. In Figure~\ref{fig:distributions}, we show
the simulation results of $\sqrt{N}*KS()$, where $\f(N)=\sqrt{N}$,
for different sample sizes $N\in[1,20]*100$, and the resulting cloud
of distributions.

To rationalize the deterministic nature of $\f(N)$ with the stochastic
nature of the measure values, it is necessary to estimate $\f(N)$.  To
do so, we use $\frac{1}{\f(N)} \sim E[\D_{\f=\iota}]$. This is the
average distance for the different measures when no normalizing factor
is applied, see \TableRef{tab:average_value}. Then, we apply the
values to the measure $\D$. Before proceeding further, a bibliographic
note about how to estimate the average $E[\D]$ is in order. The
estimate boils down to the properties of a random walk and the area
beneath its path. Even though there is no clear and complete treatment
for all the measures, our experiments confirm the results in the
literature for the variational distance $E[V(R,W)] =
\frac{1}{4}\sqrt{\pi N}$, see \cite{Harel1993,Takas1991}.

The simulations generate CDF clouds as a function of $N$. We define a
{\bf representation of the behavior of the CDF of the measure} as a
stochastic function
\begin{equation}
  F_\D(x) \in {\cal  N}(\bar{\mu}(x),\bar{\sigma}(x)),
\end{equation}
where $\bar{\mu}(x)$ is an estimate of the representative CDF and
$\bar{\sigma}(x)$ is a function representing the confidence about the
representative function.

We assume that we have found a representative distribution when at
least 90\% of $F_N$ are included in the intervals
$\bar{\mu}(x)\pm2\bar{\sigma}(x)$, giving empirical justification of
the claim that the CDF functions $F_N(x)$ have a normal
distribution. Moreover, the empirical $\bar{\mu}(x)$ should be a
smooth function, not exhibiting anomaly accumulations or steps because
of the merging of $F_N(x)$ with different $N$. That being the case, we
may take $\bar{\mu}(x)$ as the representative distribution function of
the measure.

What follows is a presentation of our methods and findings. We start
by showing how to determine the functions $\bar{\mu}(x)$ and
$\bar{\sigma}(x)$, which is described in
\SectionRef{sec:window-independence}.  In
\SectionRef{sec:input-distribution-independence}, we show that
$\bar{\mu}(x)$ is a CDF that is independent of the input CDFs.

\subsection{Window-Size Independence}
\label{sec:window-independence}
\label{sec:simulaton-distribution} 

\singlefigure{0.5}{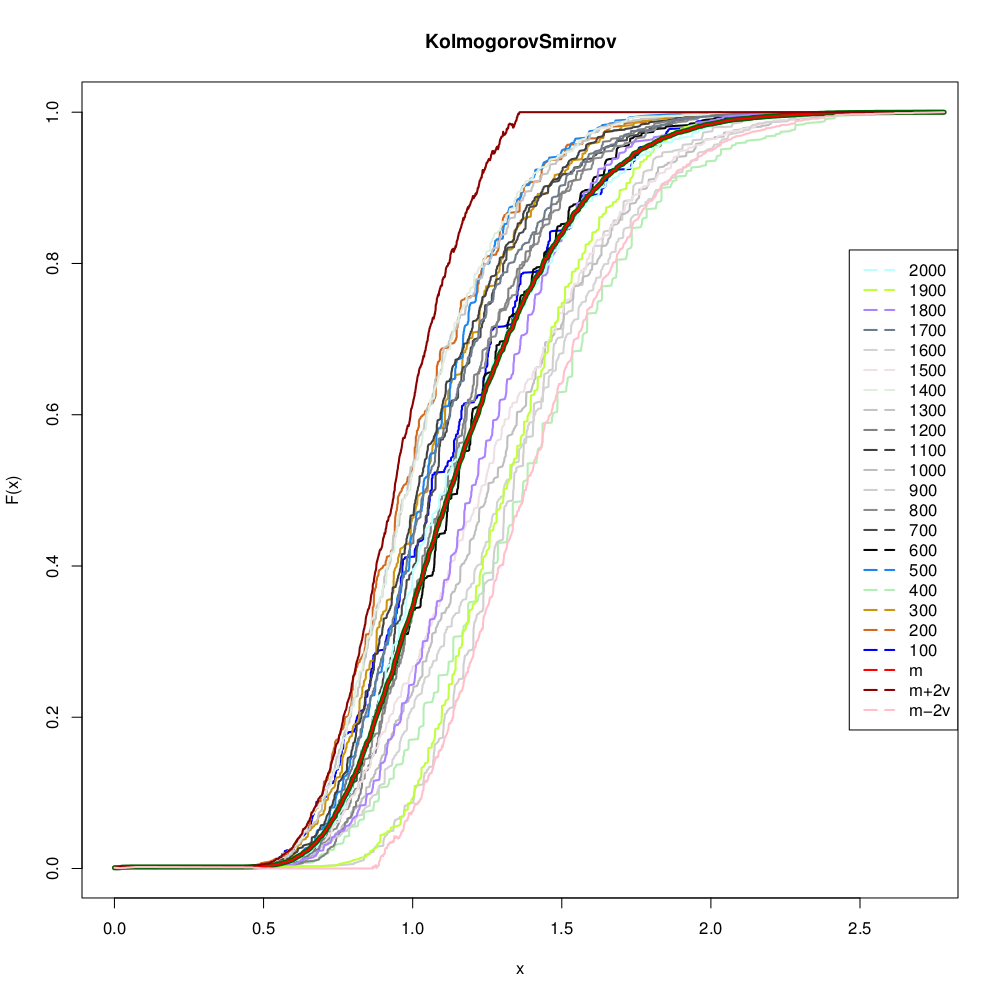}{The Kolmogorov--Smirnov-measure
  CDFs}{fig:distributions} 
\singlefigure{0.5}{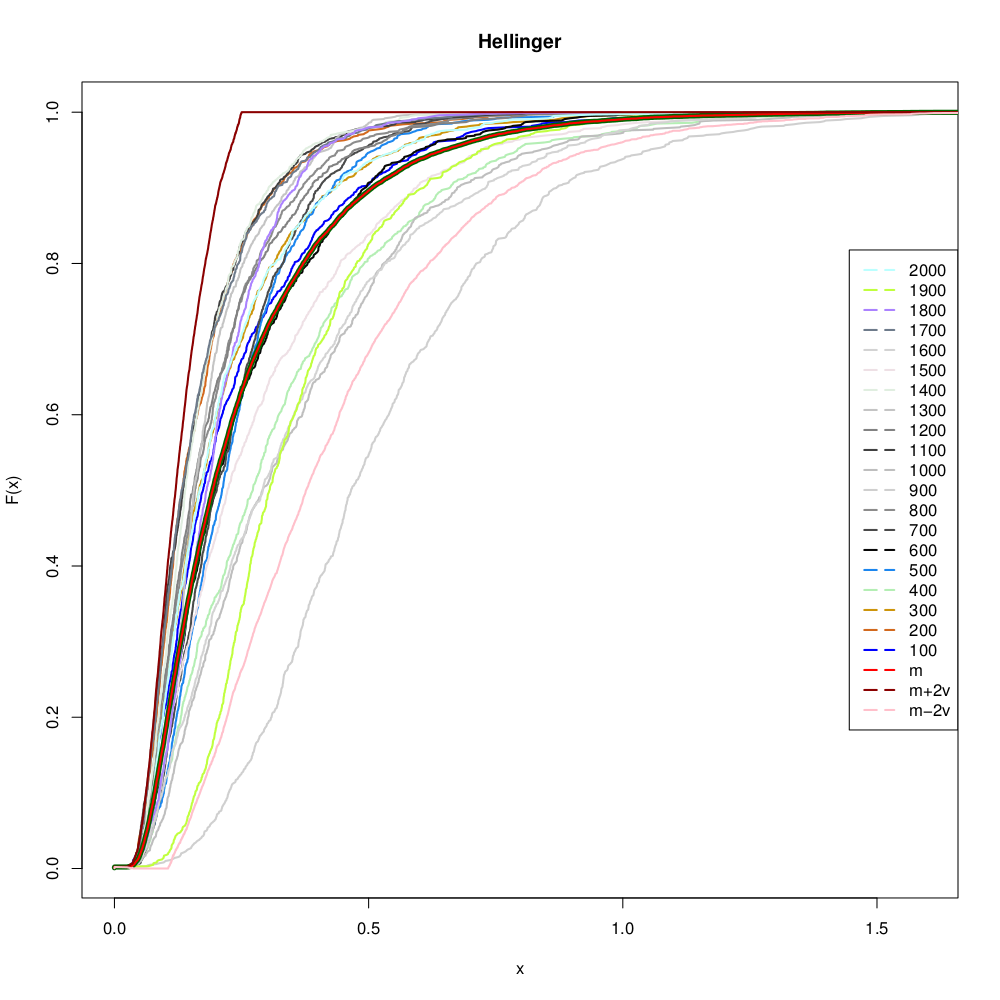}{The Hellinger-measure
  CDFs}{fig:distributions-h}

We present the results of the simulation of the Kolmogorov--Smirnov
($\sqrt{N}KS$) and Hellinger (H) measures in Figure \ref{fig:distributions}
and \ref{fig:distributions-h}, respectively. For window sizes between
100 and 2000, at increments of 100, we generated 1000 intervals drawn
from the same normal distribution ${\cal N}(0,1)$. Then we computed
the value of the measures to determine the CDF that supports the
similarity assumption $H_0$.

In Figure~\ref{fig:distributions} and \ref{fig:distributions-h}, for
window sizes $N_0{=}100$ (dark blue) and $N_1{=}2000$ (azure), both
measures have CDFs that are {\em similar}. This is made possible
because of $\f(N)$.

\mypar{Average $\bar{\mu}(x)$} For each window size, we have a different
CDF $F_N(x)$. We define the average of the distribution as,
\begin{equation} 
  F_{\bar{\mu}}(x) = \frac{1}{M}\sum_{N} F_N(x) 
\end{equation}

Notice that $F_{\bar{\mu}}$ is still a distribution and it could be
considered as representative of the family of distributions (e.g.,
even though $F_1(x) + F_2(x)$ is not a valid distribution,
$\frac{1}{2}(F_1(x) + F_2(x))$ is).  In
Figure~\ref{fig:distributions}, we draw the average $\mu$ in red. With
our assumption about the nature of the distribution function,
$F_{\bar{\mu}}(x)$ should tend to $\bar{\mu}(x)$.

\mypar{Variance $\bar{\sigma}(x)$} A natural definition of
distribution variance is
\begin{equation} 
  F_{\bar{\sigma}}(x) = \sqrt{\frac{1}{M}\sum_{N} (F_N(x) - F_{\bar{\mu}}(x))^2}
\end{equation}

In general, $F_{\bar{\sigma}}$ is not a distribution. Furthermore,
the use of subtraction and power prevents the result from being a valid
distribution, because the resulting $F_N(x) - F_{\bar{\mu}}(x)$ can be
negative for some $x$.  In Figure~\ref{fig:distributions}, we plot
$F_{\bar{\mu}}(x)+2F_{\bar{\sigma}}(x)$ in dark-red, and
$F_{\bar{\mu}}(x)-2F_{\bar{\sigma}}(x)$ in pink.

We assume that we have found a representative distribution when at
least 19 of the 20 CDFs, 90\% of them, are included in the intervals
$F_{\bar{\mu}}(x)\pm2F_{\bar{\sigma}}(x)$. This suggests that the CDF
functions $F_N(x)$ have a normal distribution
${\cal N}(F_{\bar{\mu}}(x),F_{\bar{\sigma}}(x))$. So, as $M$ gets
larger, this should converge to our assumption
${\cal N}(\bar{\mu}(x),\bar{\sigma}(x))$. Recall that empirically,
$F_{\bar{\mu}}(x)$ should be a smooth function that does not exhibit
any anomaly accumulations or steps because of the merging of $F_N(x)$ with
different $N$. Thus, we may consider $F_{\bar{\mu}}(x)$ to be the
representative distribution function of the measure. \footnote{Note
	that in practice, we could not find a CDF for the Bhattacharyya
	measure because we could not find a smooth CDF.}

\subsection{Input Distribution Independence}
\label{sec:input-distribution-independence}

Let us consider the output of a CDF to be a stochastic variable and
refer to $F_R(y)$ as $Y$. Now, assume that $G_R(Y)$ is the inverse
function of $F_R$, properly defined for a finite number of samples
$N$. Then the event $\{ F_R(y)\leq t \}$ is identical to the event $\{
y\leq G_R(t)\}$, which has the probability $F_R(G_R(t))=t$. This leads
to $P[Y\leq t] = t$ for $t\in[0,1]$ (see \cite{FellerII} Ch.1 Section
12). Thus, when we consider the input $Y= F_R(y)$, we actually obtain
a measure for which the distribution of the input should not affect
the distribution of the measure, because $F_R$ is uniformly
distributed independent of $R$.

\begin{figure*}
  \includegraphics[height=0.20\linewidth,width=0.25\linewidth]{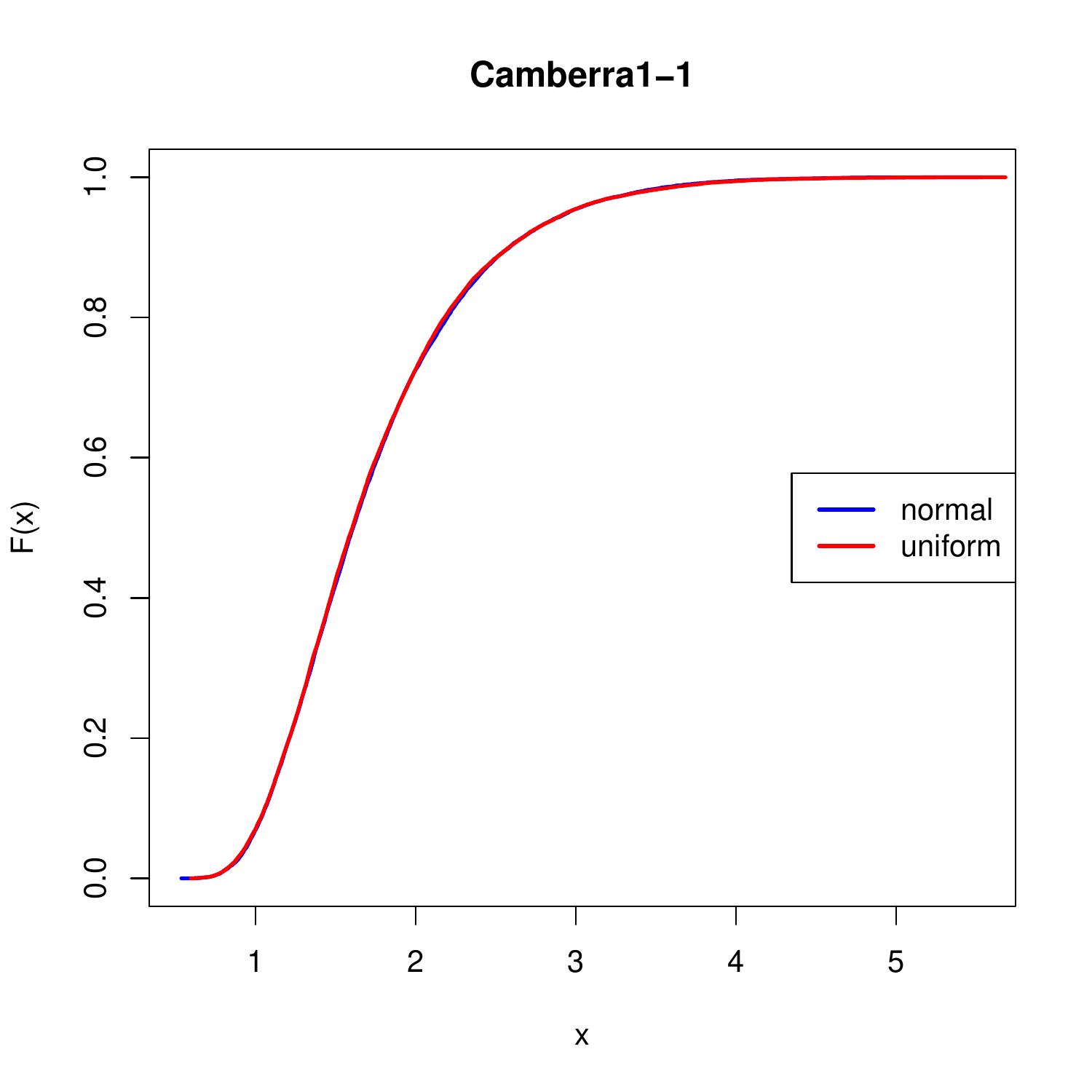}        \includegraphics[height=0.20\linewidth,width=0.25\linewidth]{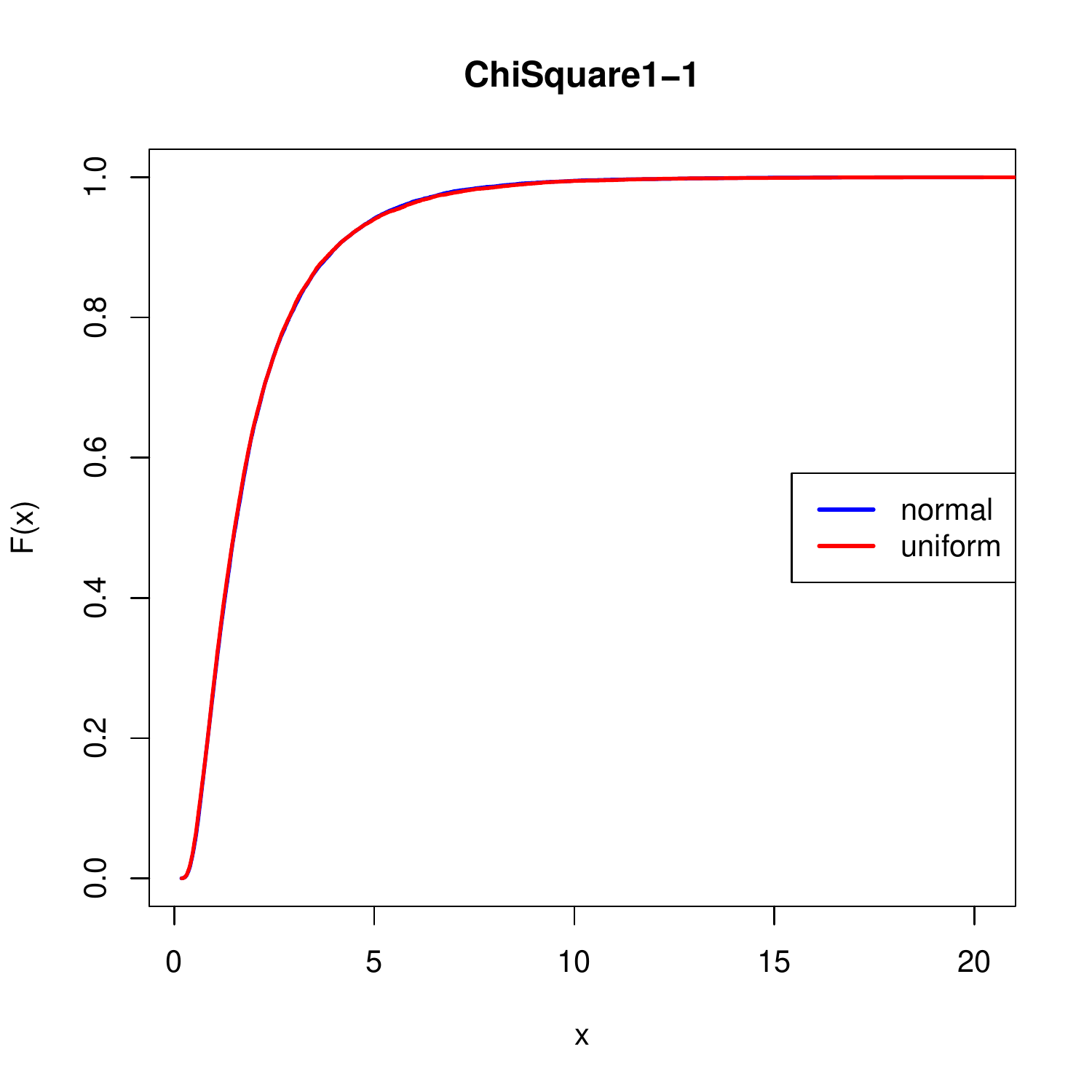}          \includegraphics[height=0.20\linewidth,width=0.25\linewidth]{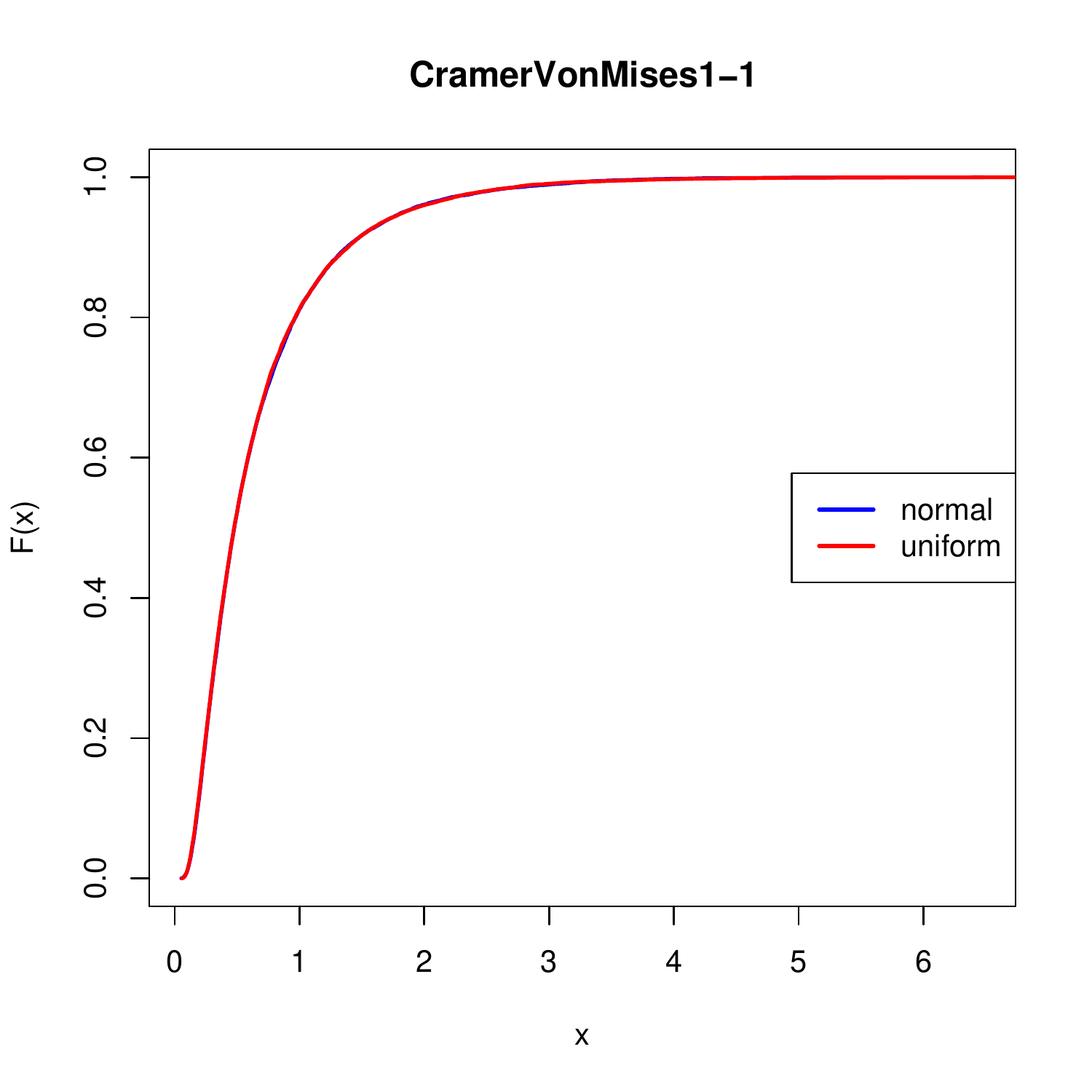} \\
  \includegraphics[height=0.20\linewidth,width=0.25\linewidth]{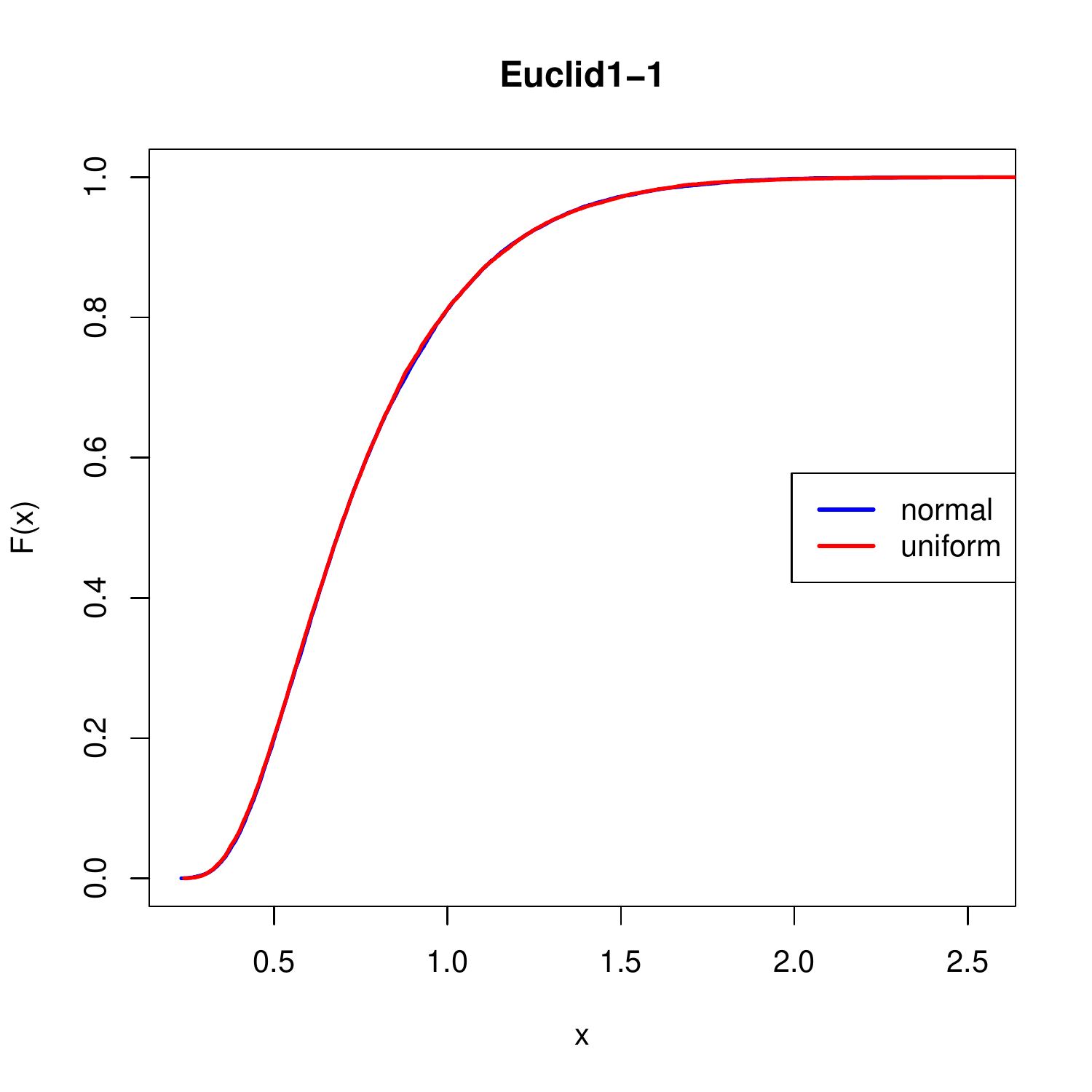}          \includegraphics[height=0.20\linewidth,width=0.25\linewidth]{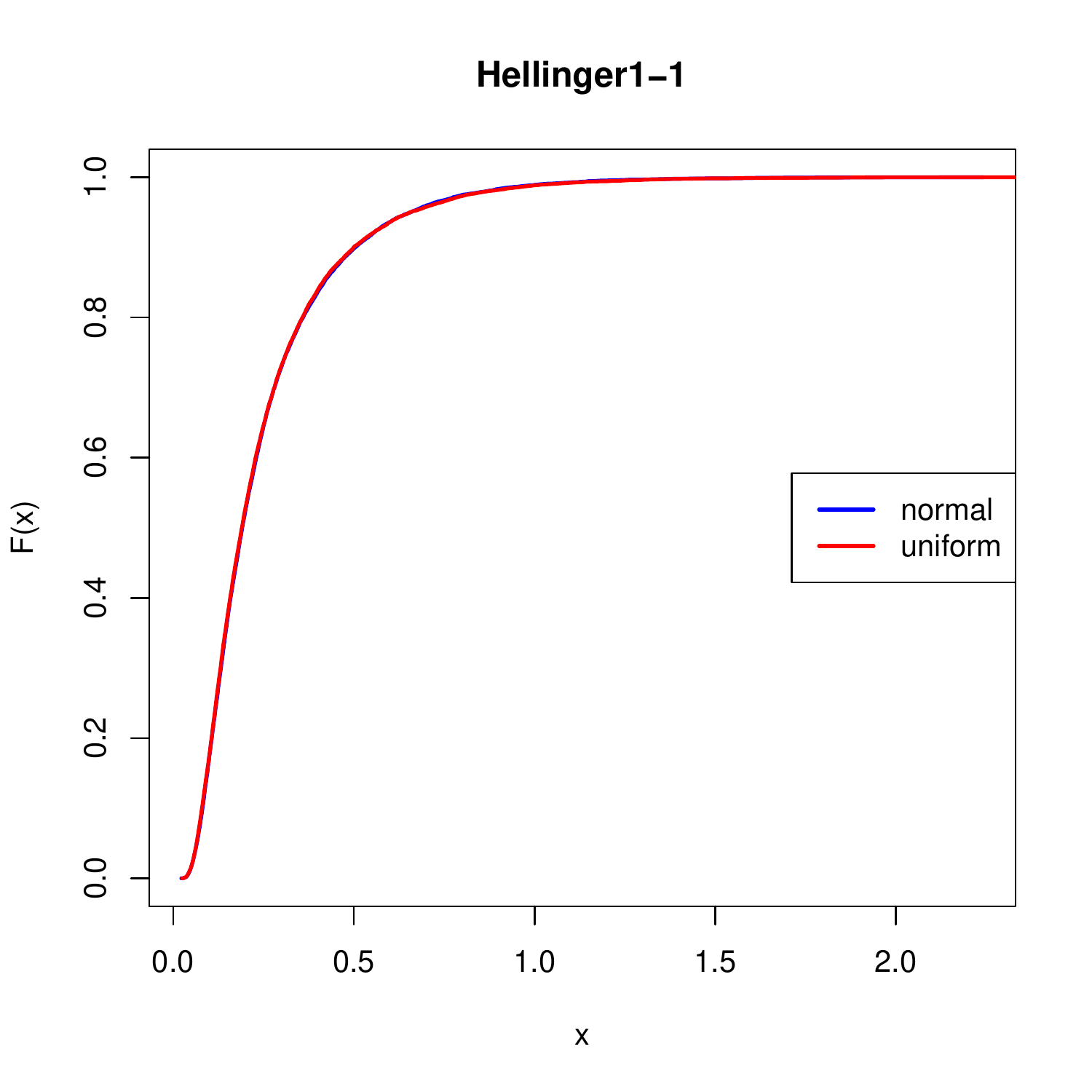}          \includegraphics[height=0.20\linewidth,width=0.25\linewidth]{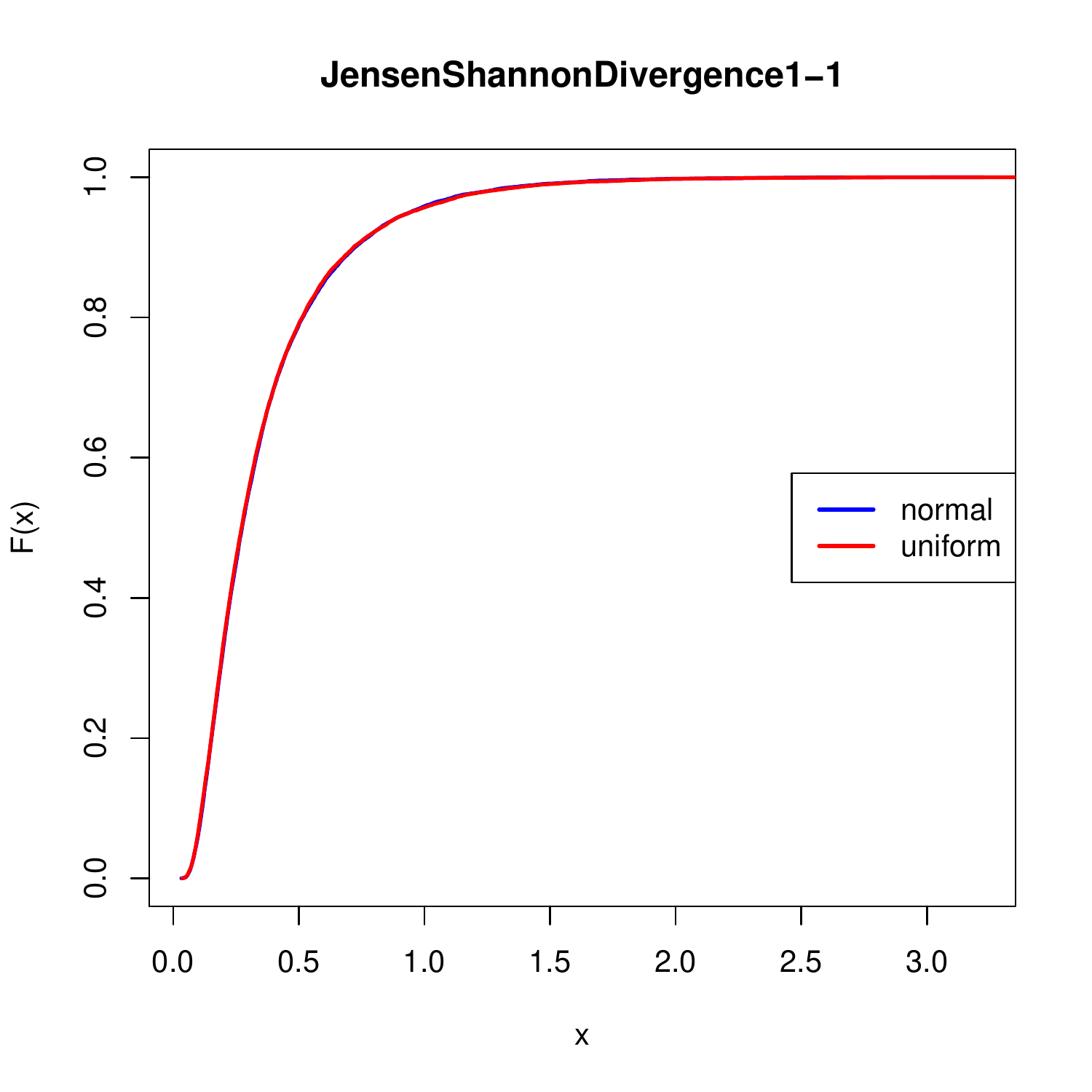}\\
  \includegraphics[height=0.20\linewidth,width=0.25\linewidth]{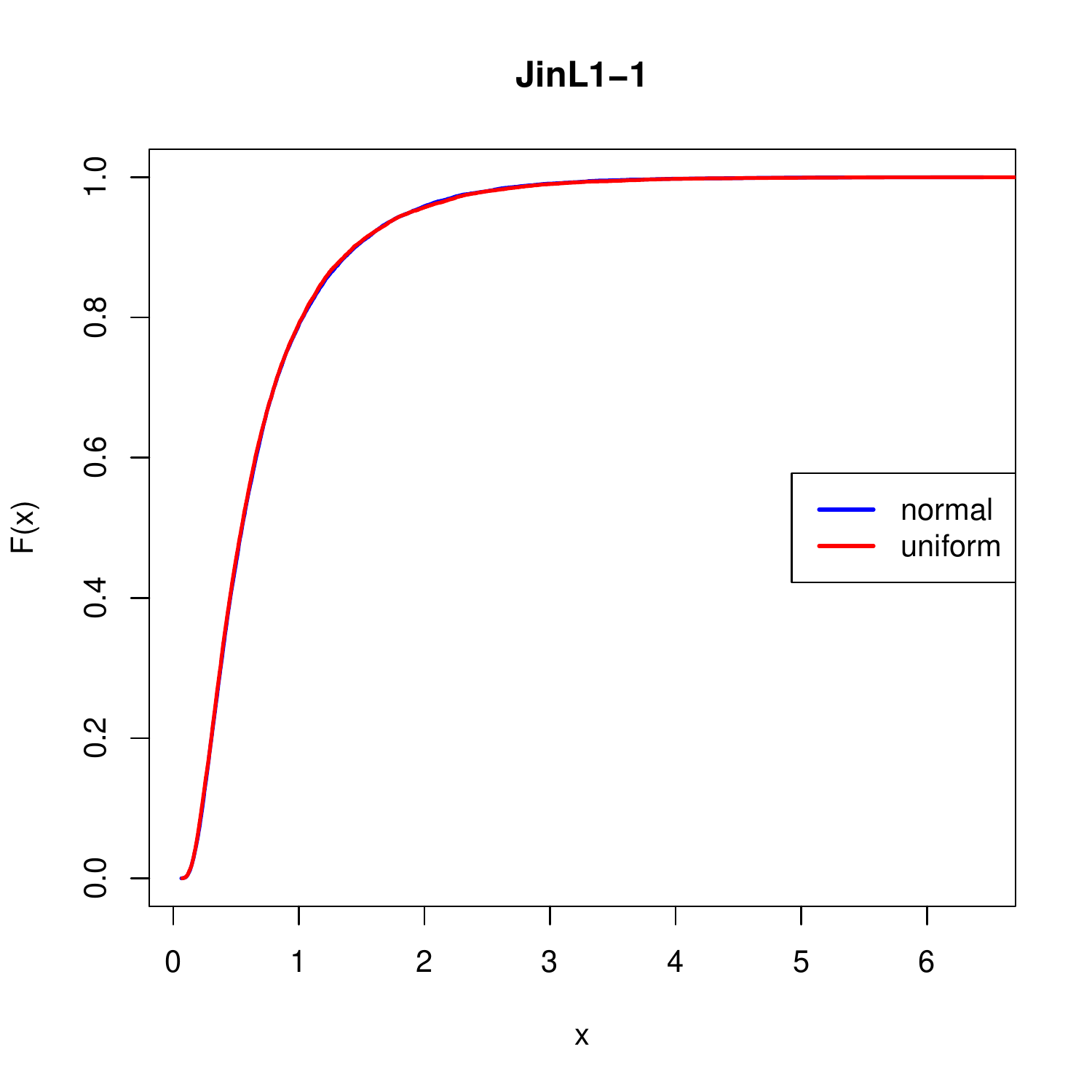}            \includegraphics[height=0.20\linewidth,width=0.25\linewidth]{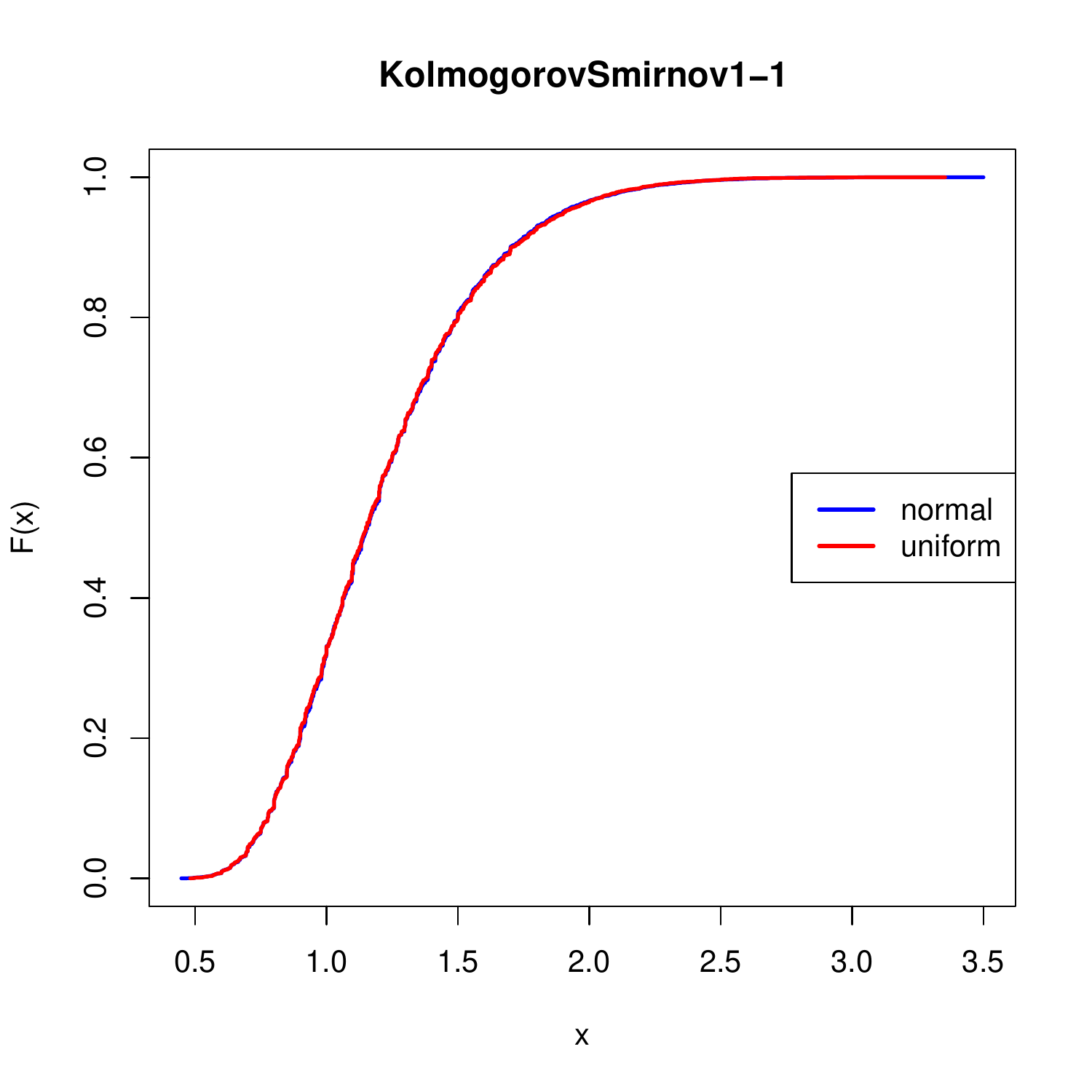}  \includegraphics[height=0.20\linewidth,width=0.25\linewidth]{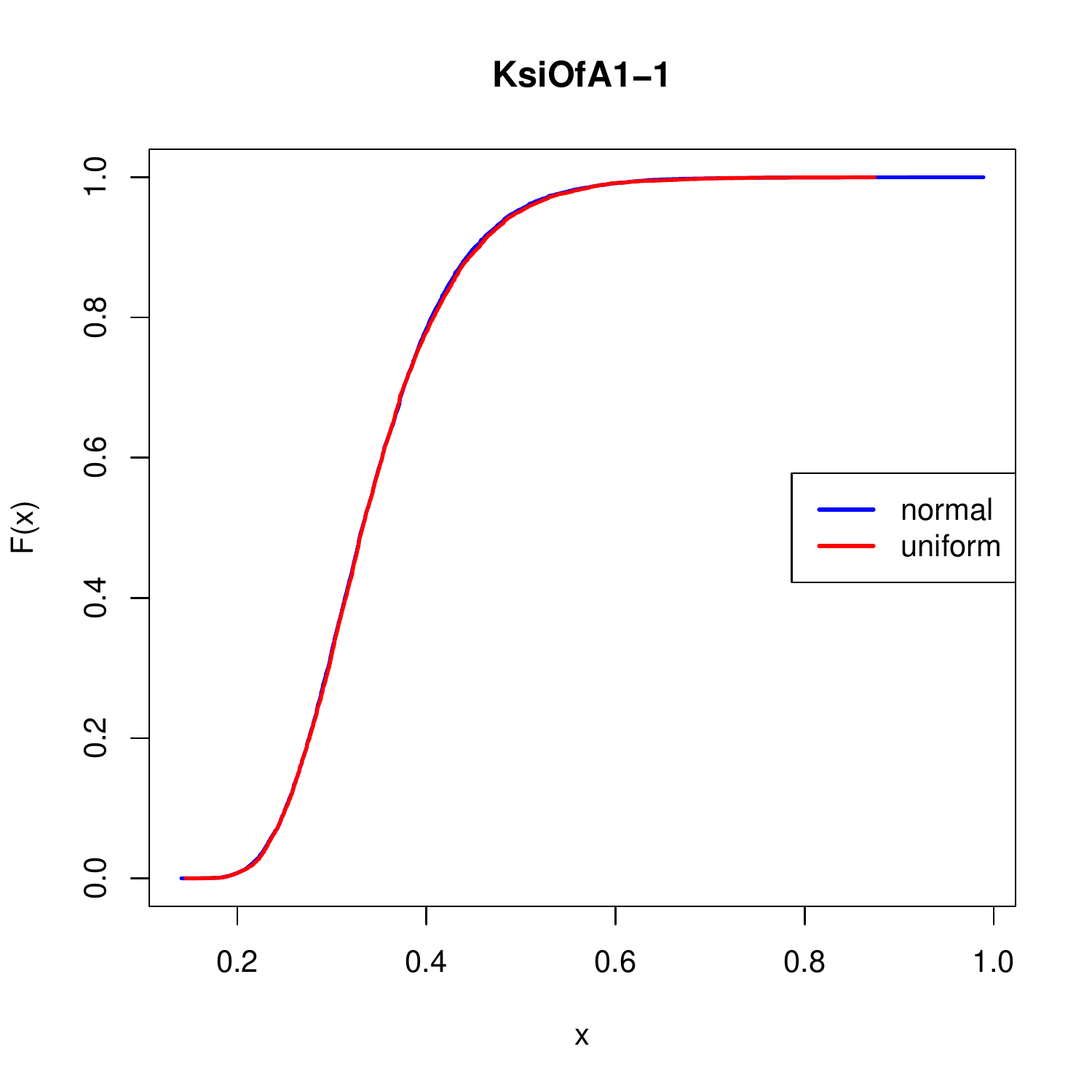}\\
  \includegraphics[height=0.20\linewidth,width=0.25\linewidth]{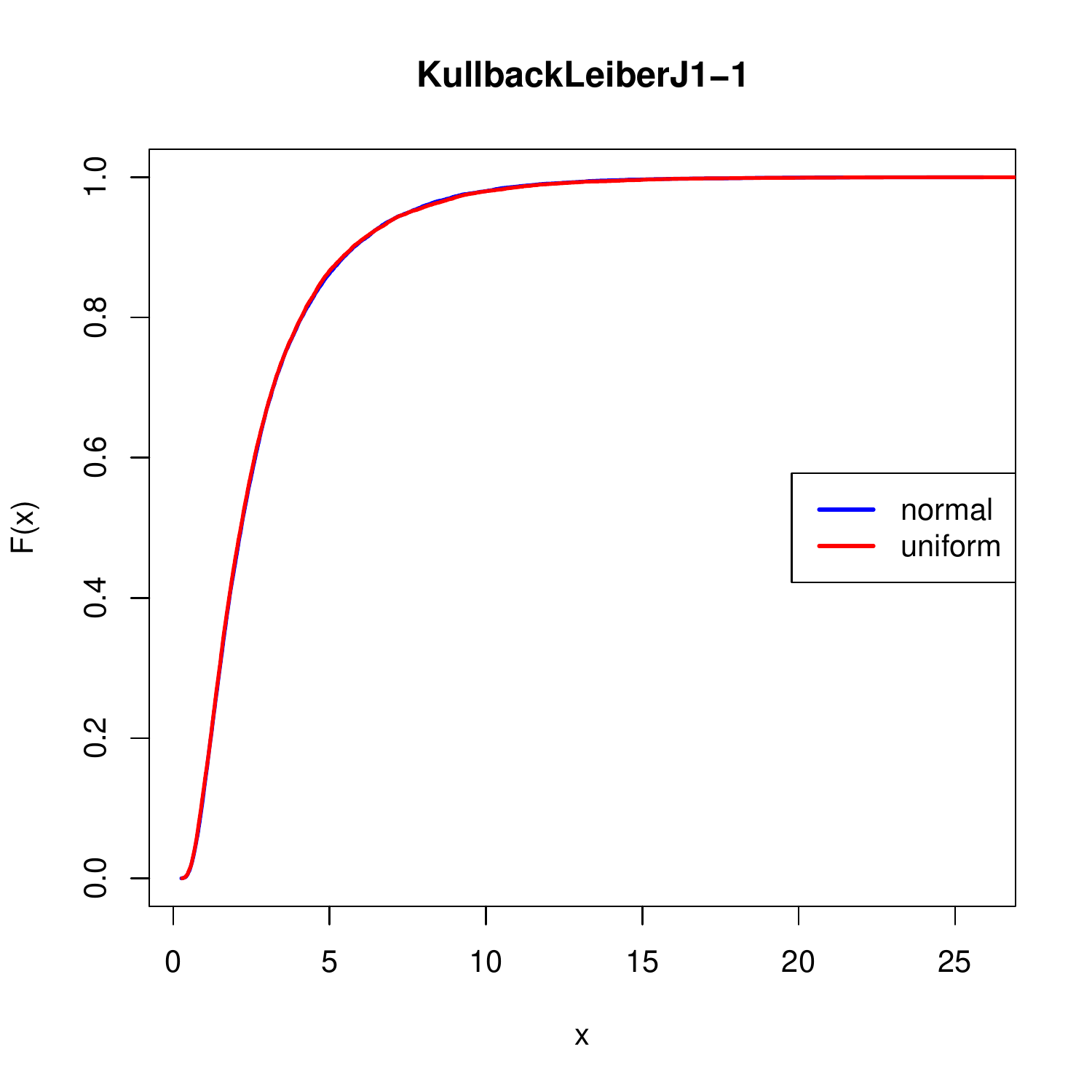} \includegraphics[height=0.20\linewidth,width=0.25\linewidth]{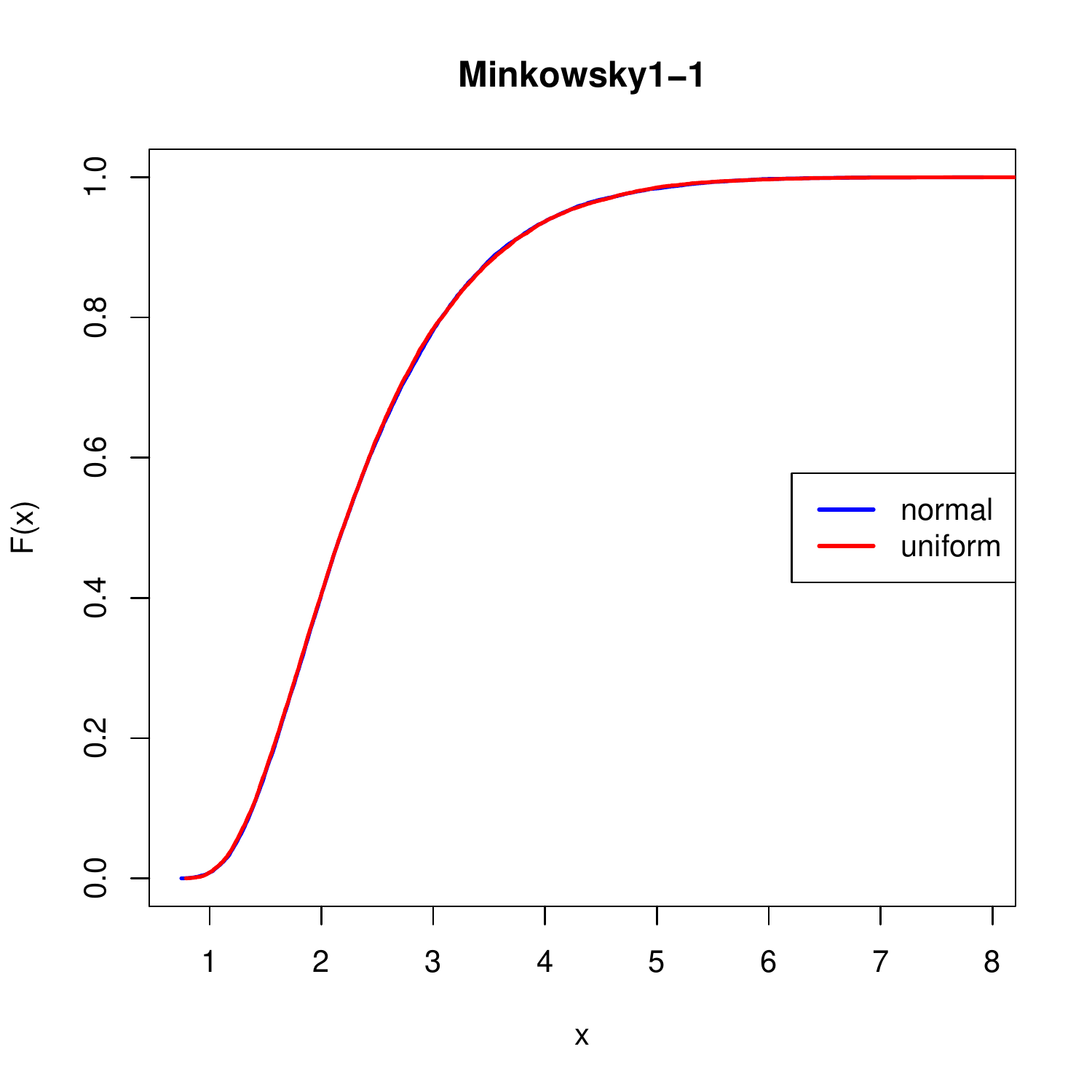}          \includegraphics[height=0.20\linewidth,width=0.25\linewidth]{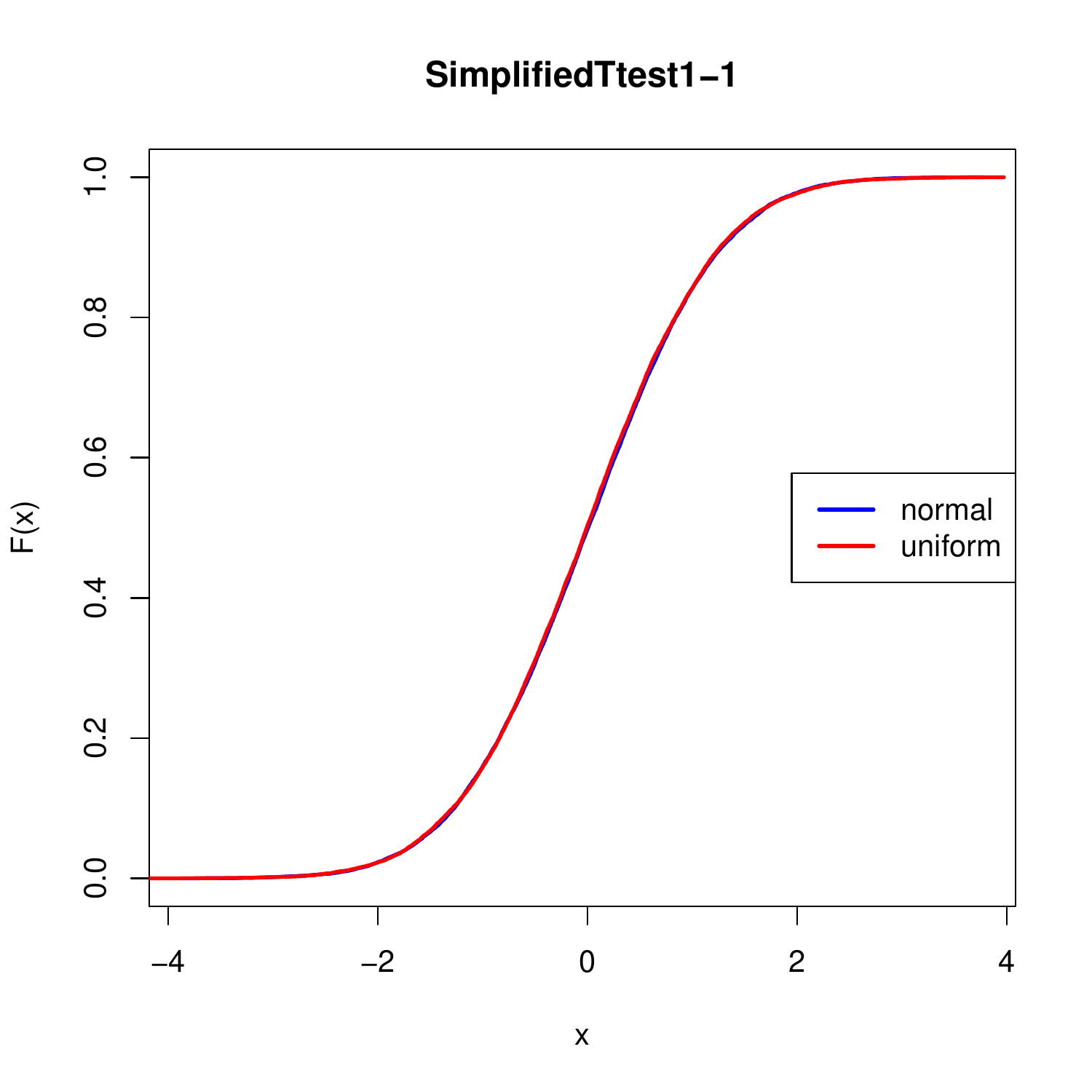}\\
  \includegraphics[height=0.20\linewidth,width=0.25\linewidth]{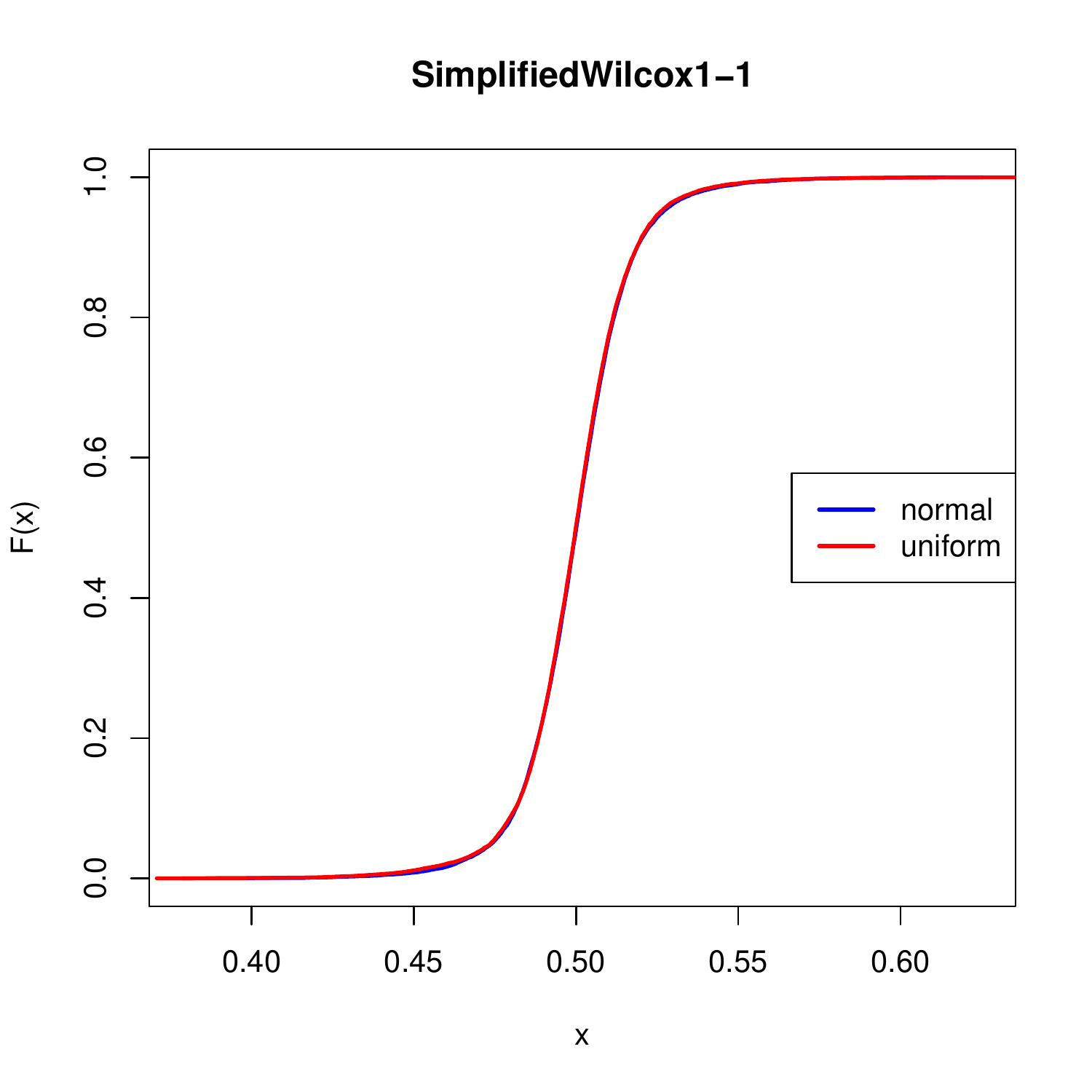}\includegraphics[height=0.20\linewidth,width=0.25\linewidth]{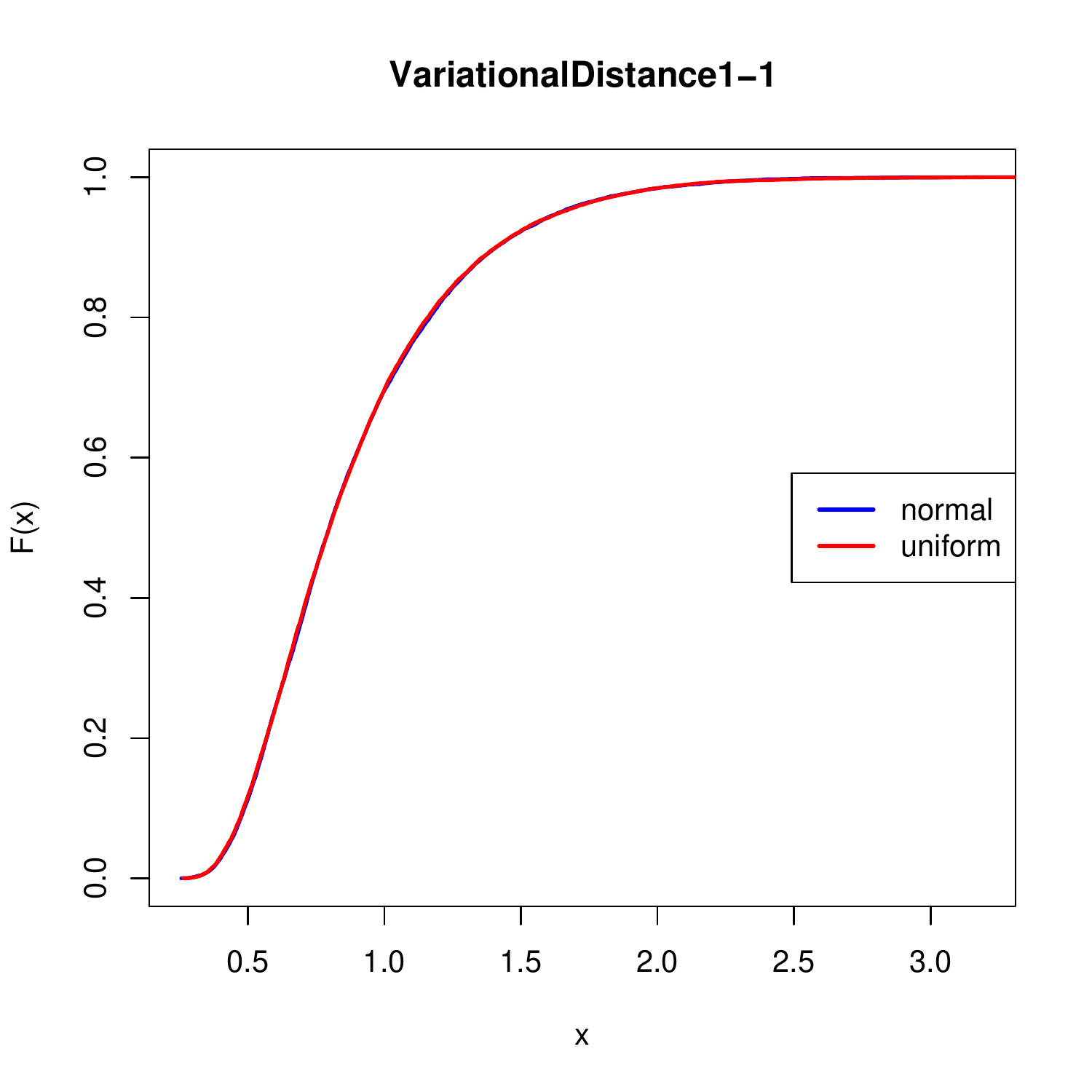}\includegraphics[height=0.20\linewidth,width=0.25\linewidth]{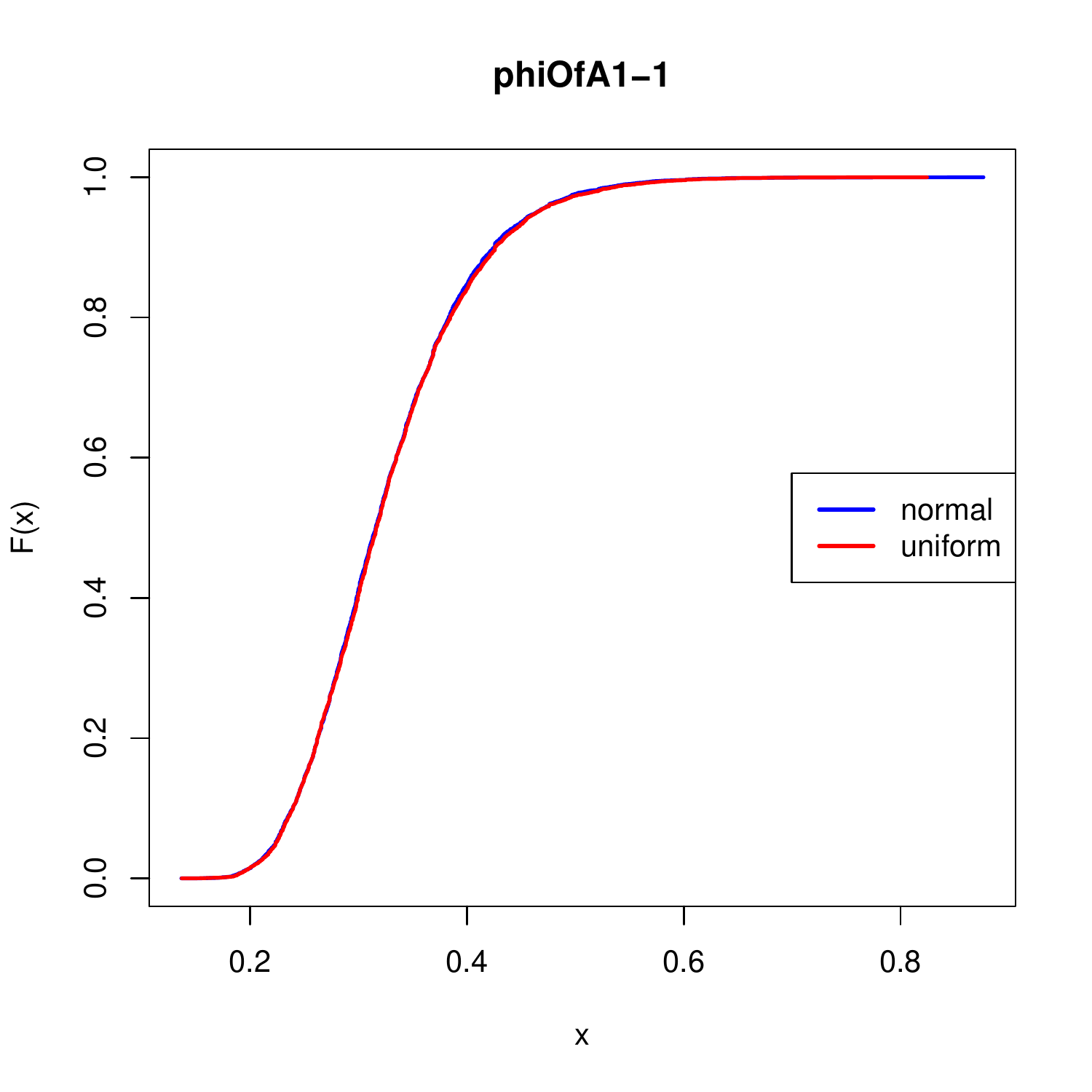}\\
  \caption{The 1000 samples for $R$ and $W$ drawn from the same
    normal ${\cal N}(0,1)$ and uniform ${\cal U}(0,1)$ stochastic
    processes with length 100, 200, ..., 2000. Note that we plot the CDFs
    obtained by the Wilcox and t-test, for which we know that the measure
    CDF is independent of the input.}
  \label{fig:distributions-all}
\end{figure*}

What we have found experimentally is that $F_{\bar{\mu}}(x)$ is
independent of the distribution of the inputs $R$ and $W$, as
we show in Figure~\ref{fig:distributions-all}. Moreover, the
distribution function $F_{\bar{\mu}}(x)$ can be used as a
representative distribution.

%\end{comment}

We repeat the simulation described in Section
\ref{sec:window-independence} for window sizes 100--2000, collecting
1000, 2000, 5000, and 10000 samples per window size, while using two
different stochastic processes, a normal distribution (${\cal
  N}(0,1)$) and a uniform distribution (${\cal U}(0,1)$). For each
stochastic process, we obtain four representative distribution
functions. We then compare the results in a $4 \times 4$ table (by
size and input distribution), and report the first tile only (1000
samples per window size) in Figure \ref{fig:distributions-all} due to
space restrictions.

By applying the standard measures, the same measure described in this
work, and by simple visual inspection, we can say that these new
stochastic distance measures have a distribution that is independent
of the window sizes and the distribution of the process we
compare. This gives us compelling evidence that our measures, such as
JS, have a measure distribution that need to be pre-computed by
simulation and only once.

We use the larger set (with 10,000 samples) to determine different
$p$-values, and then determine the measure thresholds for each
$p$-value. For example, we determine the threshold value having a
$p$-value of 95\% for each measure. In practice, this means that
if we generate $F_R$ and $F_W$ from two intervals $R$ and $W$, and
apply the measure $JS(F_R,F_W)$, then if that measure has a value
that is larger than the threshold, we know that only 5\% of intervals
drawn from the same stochastic process will have the same or larger
measure. However, we may still decide to reject the assumption that
$R$ and $W$ are similar because the probability is too small.

In the following experiments, we use the simulation distributions
to tabulate the $p$-values and the significance for each measure.

\subsubsection{Disagreement (with Multiple Measures)}
\label{sec:disagreement} 

A measure is designed to detect and to quantify the differences
between inputs. Different measures are sensitive to different
properties of the inputs, and therefore, they do not all perform
alike.

We investigate and quantify how the aggregation of different measures
can affect the sensitivity of a non-parametric measuring system. Consensus
is a simple approach by which we can use $M$ different measures and make
a decision only when a quorum of the measures agree. All the measures
presented in this paper are designed to perform {\em better} at verifying
that two distributions are statistically equivalent (the $H_0$ hypothesis
is true); otherwise, there is no equivalence.

We quantify the detection power of different measures and determine
the minimum quorum or rate for consensus (e.g., 10\% disagreement
means that only 1 measure in a set of 10 suggests that the two
distributions are different, 90\% agreement). In particular, we want
to show that our extensions of measures, as proposed in Section
\ref{sec:contribution}, are a good contribution.

In the following section, we discuss the experimental setup and
results (see Section \ref{sec:uniexperimental}).

This Section's references are
\cite{DAlbertoA2009,AliS1966,GolubVL1996,Jin1991,Lee1999,WangWY1992,JonesF1987,ShivakumarM95,JohnsonS2000,KullbackL1951,Jensen1906,Shannon1948,Kagan1963,Vadja1972,Hope1968,Hahn1912,Bhattacharyya1943,Kailath1967,Pinsker1960,Chernoff1952,TanejaK2004,Kolmogorov1933e,Kendall1991,FellerII,KiferBG2004,Melucci2007,Anderson1962,Batchelor1978,WilsonM1997,Diday1974,WilsonM1997,Wilcoxon1945,MannW1947,Harel1993,Takas1991,Feller1948,ChakrabartiSD1998}.

\section{Experimental Results}
\label{sec:experimental} 

We separate this section into two parts. We discuss the
multi-dimensional series (Section \ref{sec:mult-experiments}) before
providing a further investigation to the CDF-based measures (Section
\ref{sec:uniexperimental}).

%%% Paolo Sept 4 2011 11:25 PM

\section{Multi-dimensional Experimental Results}
\label{sec:mult-experiments}

In this section, we present the experimental results for
multi-dimensional series. For clarity, the discussions in this section
are further separated into the following topics: series generated from
synthetic data (Section \ref{sec:synthetic}), series generated from
classified data taken from UCI databases (Section
\ref{sec:classified-data}), series generated from hardware counters
for search engine properties (Section \ref{sec:hwcounters}), and
series generated from historical stock quotes (Section
\ref{sec:stockmarket}).

Note that for the first two data sets we have a full understanding of
the series data. We use this knowledge for the validation of the
methods and their discriminative capabilities. Specifically, we want
to apply the methods to detect the differences between benign and
malignant cancer.  For the last two data sets, we are more interested
in finding the similarity of series. Specifically, identifying
similarity in the hardware counter data helps to group applications
based upon run-time performance properties. For stock quotes, we show
how the methods can be applied as {\em scan statistics}
\cite{GlazNW2001}.

In the following, we describe the experimental set up for each method.

\mypar{MST/Poset Method Set Up} The MST and poset-based methods compare the
empirical CDFs using a quorum of the following ten measures: $\phi$,
$\Xi$, KS (Kolmogorov--Smirnov), KLJ (Kullback--Leiber--Jeffrey),
JS (Jensen--Shannon), $\chi^2$, H (Hellinger), $W^2$ (Cram\'er--von--Mises),
E (Euclid), and C (Camberra). We set the ratio for rejection by the
quorum to 20\%, meaning that at least two measures must reject the
equality hypothesis for the quorum to reject the hypothesis (see Section
\ref{sec:distance} or \cite{DAlbertoA2009}).

\mypar{Compression Method Set Up} We create a $p$-value range by
bootstrapping using 100 runs. The training set and bootstrapping
process are described in Section \ref{sec:ncd-bootstrap}.
Bootstrapping is performed for every series that we present.

\mypar{Martingale Method Set Up} We set the following parameters:
$\lambda = 20$, $\epsilon = 0.95$, and $t=3$ for the Martingale method.
Recall that $\lambda$ is the maximum value of the Martingale value
before a change is declared, $\epsilon$ is the confidence margin used
in rejecting the equality hypothesis even though the hypothesis could
be true (in this case, a maximum of 5\% error is tolerated), and $t$
is the maximum increase of the Martingale value in a single step. In
practice, $\lambda$, $\epsilon$, and $t$ are tunable parameters in
conjunction with the size of the reference interval $N$.

\mypar{Kernel Method Set Up} Having decided upon the Gaussian kernel,
we need to determine the value of the variance parameter $\sigma$.
The value of this parameter is estimated from the data of all the
series. As for other methods, the significance $p$-value of 5\% is used.

\subsection{Synthetic Variation}
\label{sec:synthetic}

\Doublefigure{0.80}{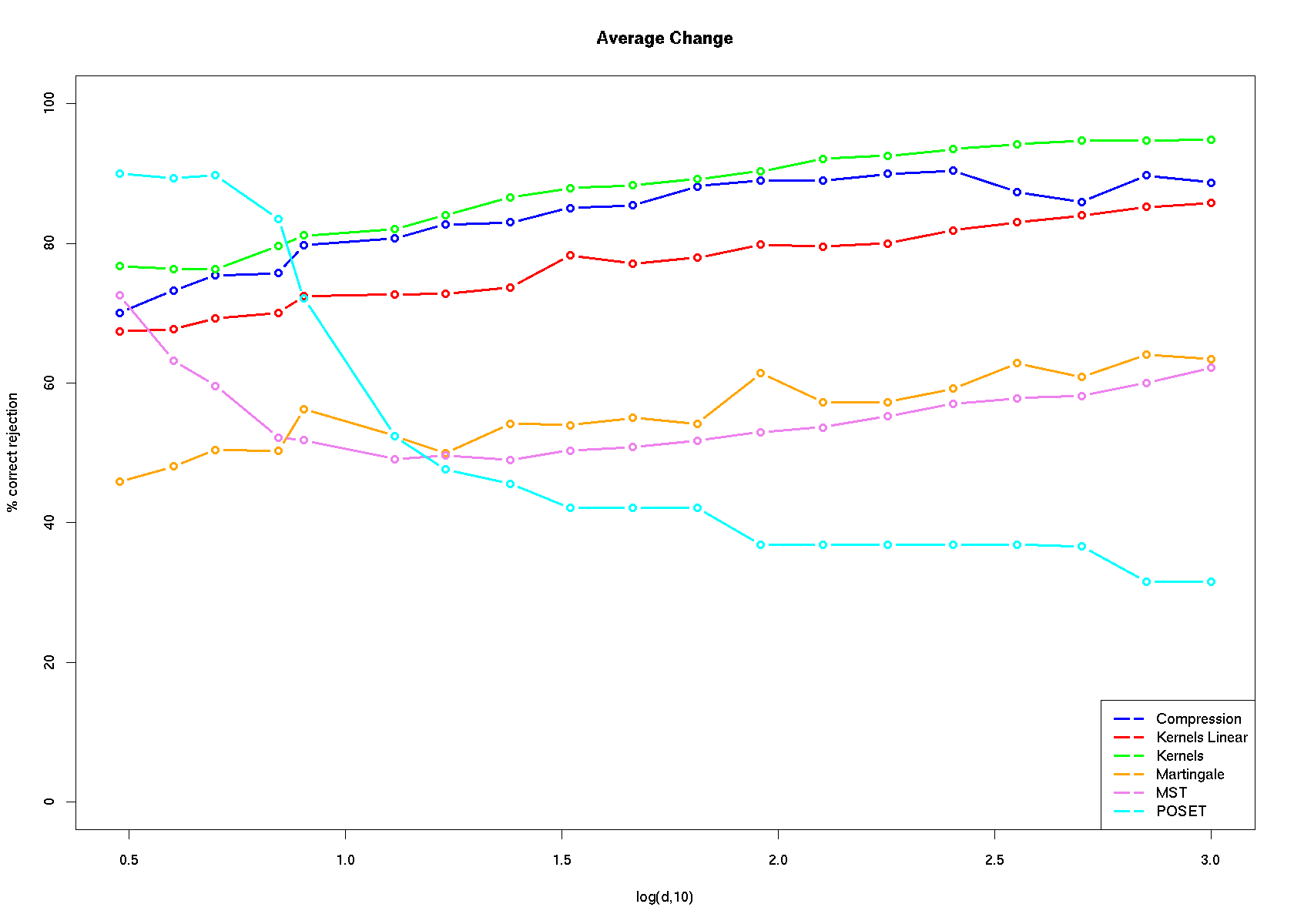}{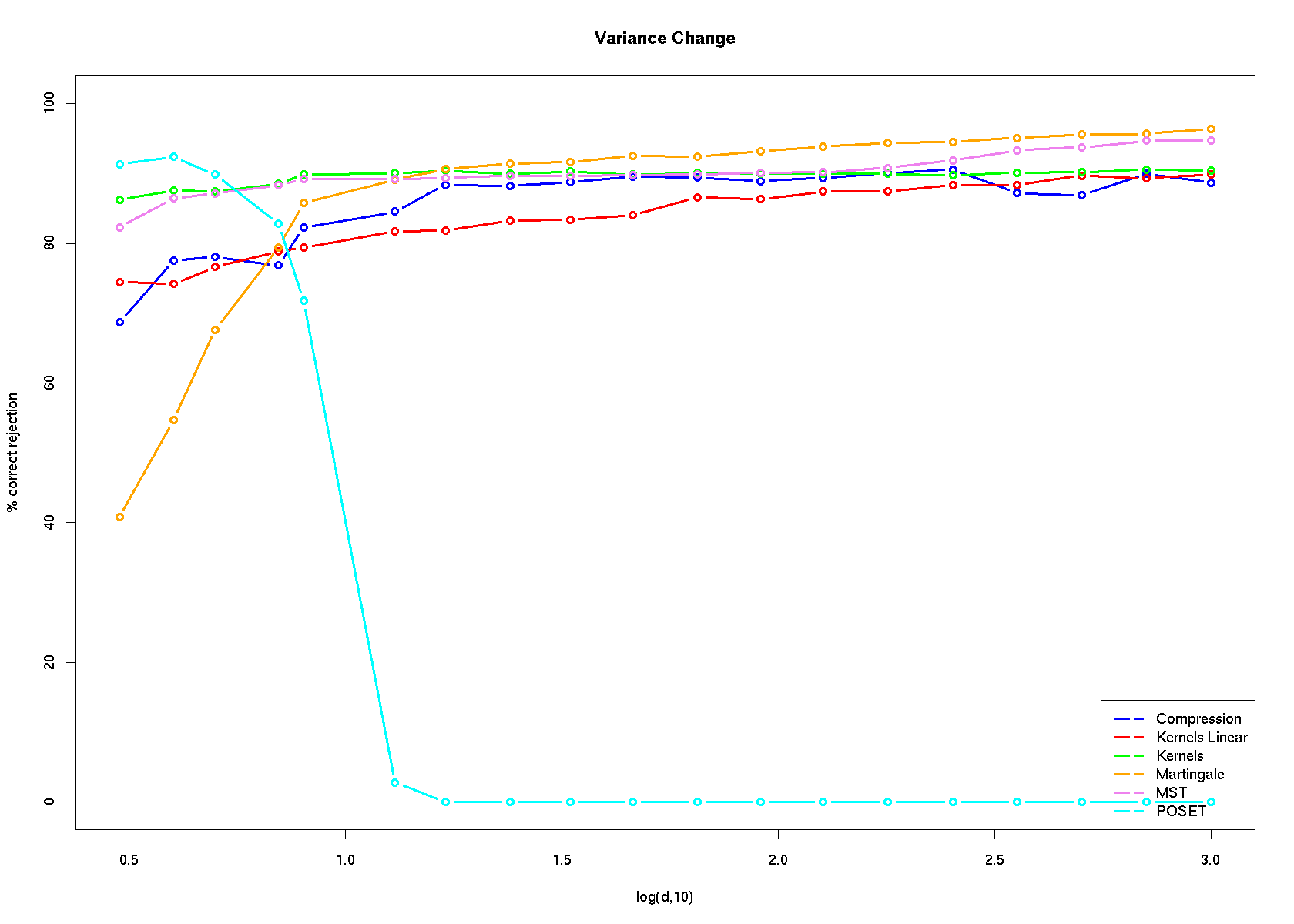}{Early
correct rejections: (top) average increasing exponentially, (bottom)
variance increasing exponentially}{fig:exponential-test}

We created three tests, similar to the experiments conducted by other
researchers. The tests are a change in average only, a change in
variance only, and a change in distribution from normal to uniform.
We took a series and divided it into 21 intervals, each consisting of
$N{=}250$ samples points. We chose to use 250 sample points per
interval to keep the tests compatible with experiments presented in
the previous work. The first two intervals are drawn from the same
stochastic process, which has the normal distribution ${\cal N}(0,1)$.
These two intervals can be used to bootstrap the compression method or
to tune the Martingale method. Nonetheless, it is a pre-requisite that
all methods recognize these two intervals as equivalent.

We constructed the series such that we increased the average, the
variance, or the distribution mix as the series progressed. We also
investigated the relationship between the dimensionality and the
sensitivity of the method to identify the variation as early as possible.
That is, for each method we measured the average shortest time or
the least number of data points necessary to identify the artificial
change.

For each series, the first interval is the reference interval $R$, and
does not move; while the moving interval $W$ slides through the series.
For all methods, except the Martingale method, the moving interval $W$
shifts by a full interval window size. The Martingale method scans the
series one point at a time.

The summary results displayed in Figure \ref{fig:exponential-test}
were generated in the following manner. For each method and each
of the three types of series change, we ran 100 simulations. Two
such results are presented in Figure \ref{fig:average-exponential-test}
and Figure \ref{fig:variance-exponential-test}. We recorded the earliest
time stamp when a method first identified a variation and rejects
the similarity hypothesis. More detail will be provided in the
following sections.

The rejection rate is computed as the ratio of the earliest time to
the overall length of the series. We define the ratio as:
\[ 
\text{rejection ratio} = \frac{1 - (\text{earliest point} - 2N)}{ \text{series length} - 2N}, 
\]
where $N$ is the number of samples in an interval, $N$=$250$; recall
that the first two intervals are drawn from the same distribution and
thus we must remove $2N$ points from the overall series.

Figure \ref{fig:exponential-test} presents the effect on the average
early rejection rate as the dimensionality of the series increases (on
a logarithmic scale).

\Doublefigure{0.85}{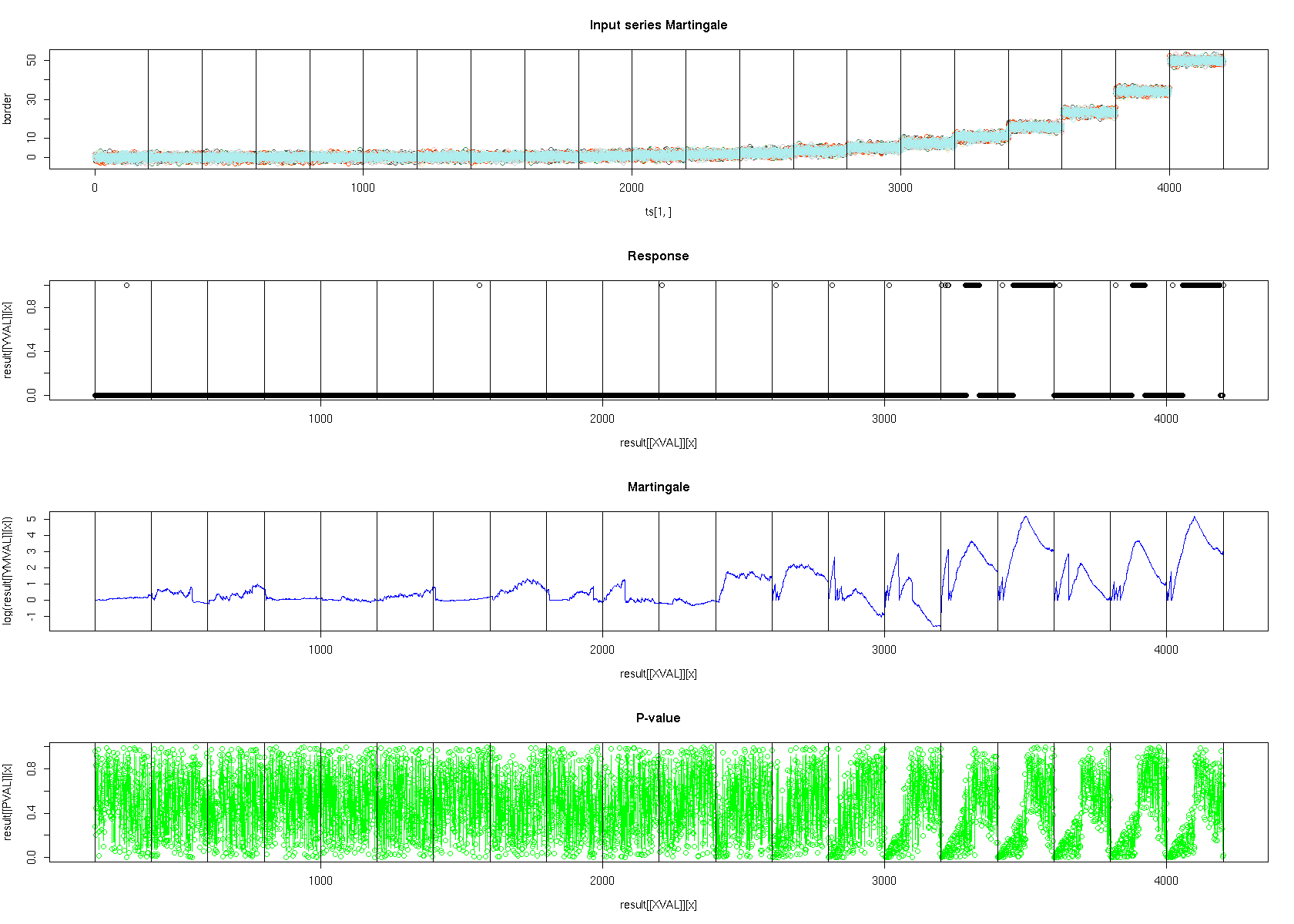}{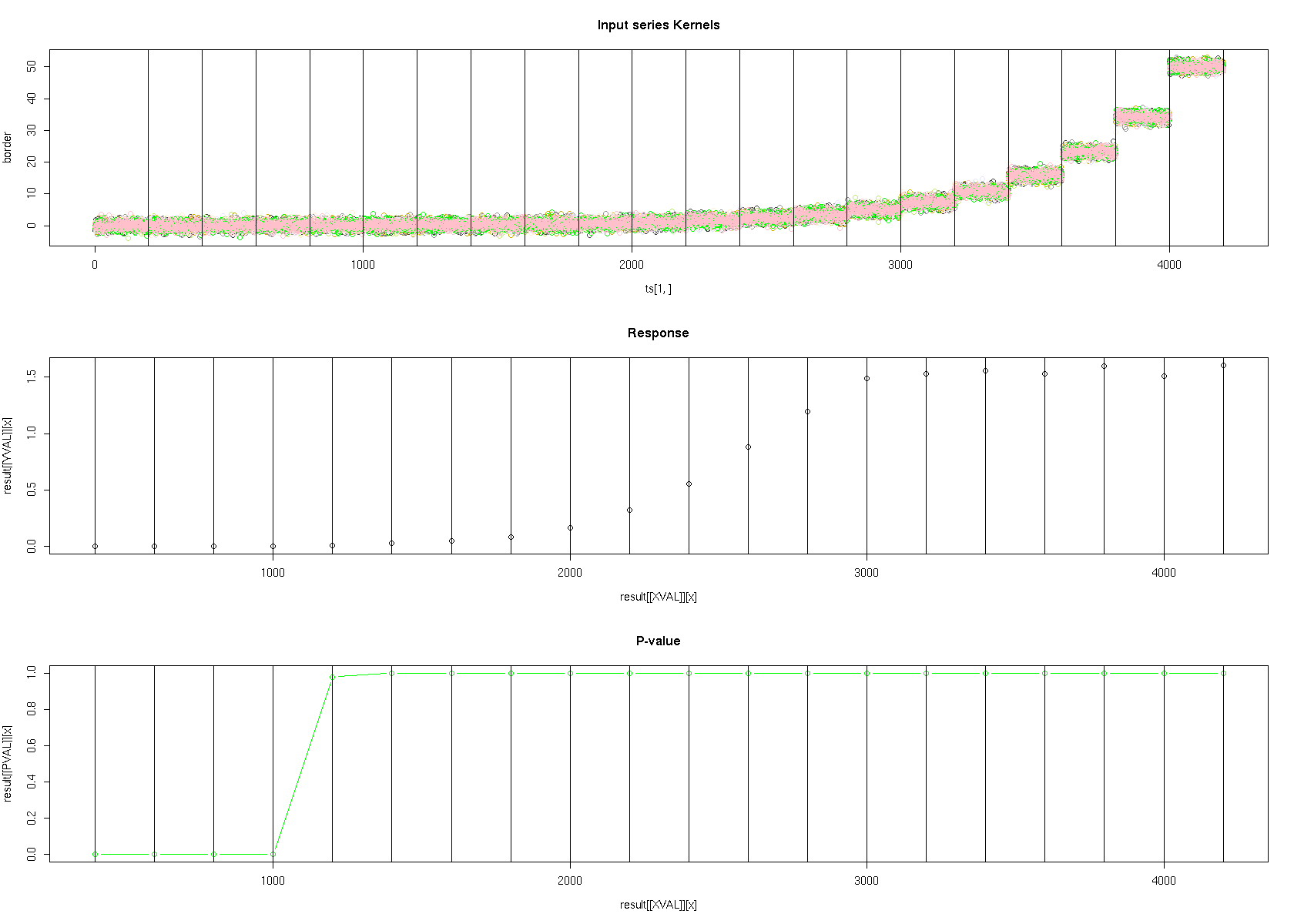}{Series with
  intervals marked and average increasing exponentially over time.
  Above, the Martingale method falsely rejects after 200 samples, and
  then successfully after 1600 samples. Below, the Kernel method
  successfully rejects the hypothesis after 1200 samples.
}{fig:average-exponential-test}

% Paolo Sept 7 2011  11:33PM

\subsubsection{Results for Changes in Average} In Figure
\ref{fig:average-exponential-test}, we show an example of a series and
an excerpt of the responses generated by the Martingale method.  The
Martingale method response is composed of the following four series
from the top to the bottom: The first series is the input; the second
is the response, which indicates a difference when the value is 1,
otherwise 0; the third is the Martingale value; finally, the fourth
series is the $p$-value from the transducer, which is the estimated
probability distribution.

We also show an excerpt from the Kernel
method, similarly displaying the input, the Kernel value, and the
$p$-value. A $p$-value larger than $0.95$ indicates a change with
5\% confidence. 

As mentioned above, we created 21 intervals from a normal distribution
with average values logarithmically spaced over the range of 0.05
to 50. Clearly, the earlier a method discovers a change in the
average, the more discriminating the method is. The overall performance
is shown in Figure \ref{fig:exponential-test}. The details are
discussed further in the summary section.

\Doublefigure{0.80}{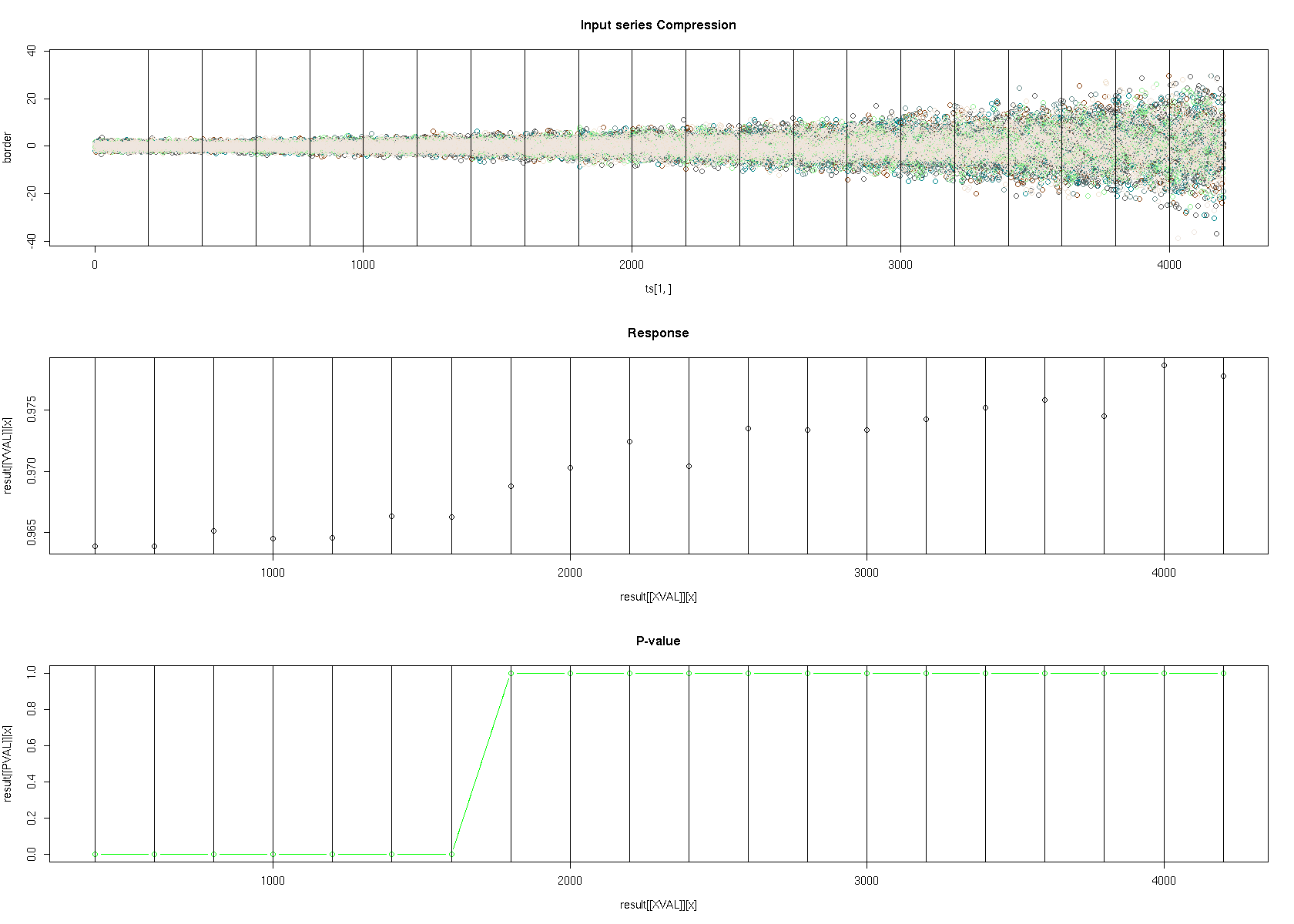}{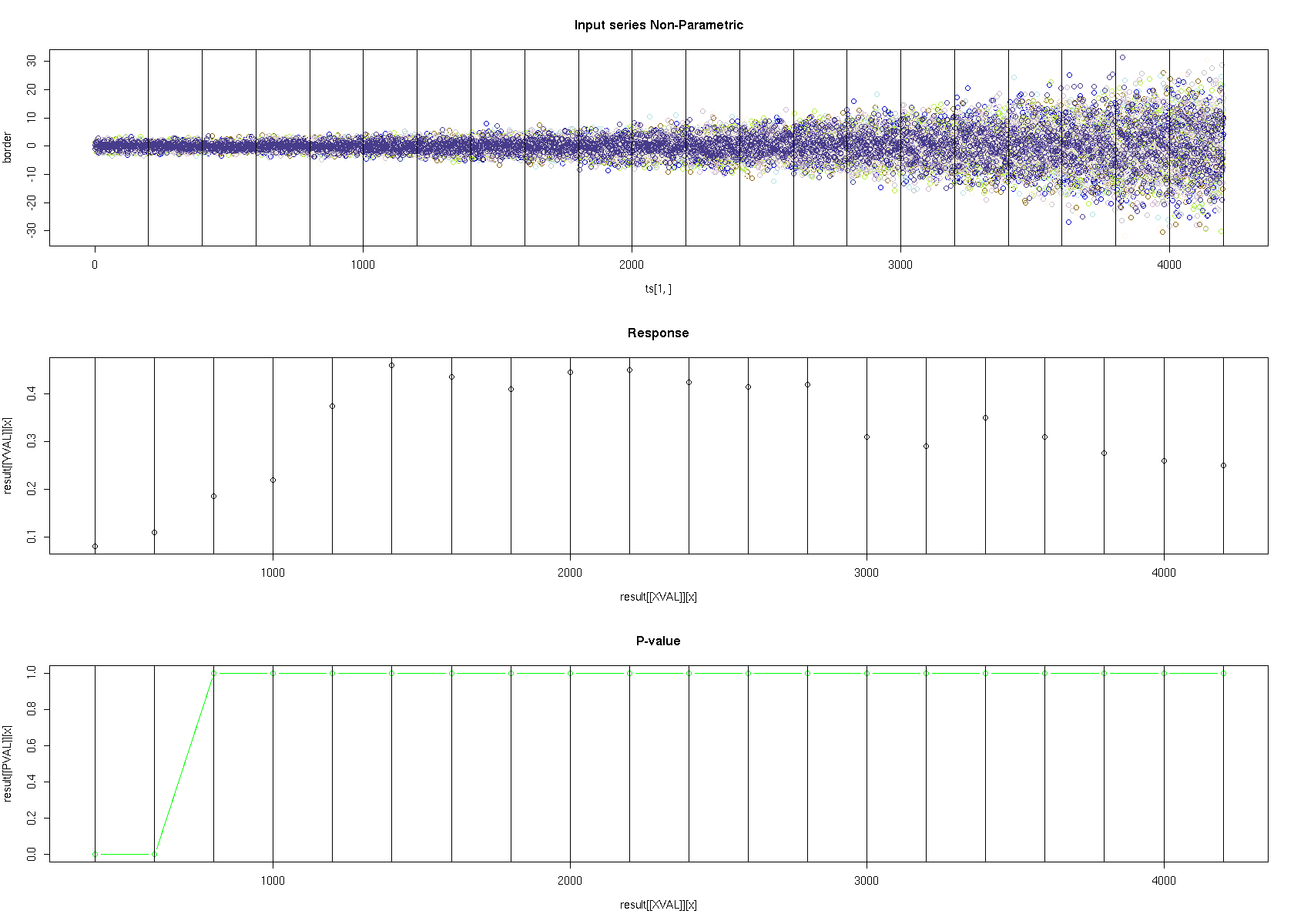}{Series with
	intervals marked and variance increasing exponentially over time.
	Above, the Compression method successfully rejects the hypothesis
	after 1800 samples. Below, the MST method successfully rejects
	after 800 samples}{fig:variance-exponential-test}

\subsubsection{Results for Changes in Variance} In Figure
\ref{fig:variance-exponential-test}, we show an example of a series
and an excerpt of the responses generated by the Compression method.
The Compression method response is composed of the following series:
the input, the NCD value, and the $p$-value as computed during
bootstrapping with 100 samples. We also show an excerpt from the MST
method, displaying the input, the minimum distance across 10 built-in
tests, and the 20\%-quorum based $p$-value, generated from the
$p$-values of the quorum measures.

Similarly as in the previous tests, we created 21 intervals from a
normal distribution with a null average and variance values
logarithmically spaced over the range of $10^{0.01}$ to 10. Clearly,
the earlier a method discovers a change in the variance, the more
discriminating the method is. The overall performance is shown in
Figure \ref{fig:exponential-test}.  The details are discussed further
in the summary section.

\subsubsection{Results for Changes in Distribution} This is the last of the
tests on the synthetically generated data sets. The purpose is to
quantify the ability of each method to identify when the distribution
changes (average and variance do not change significantly). We start
with an interval generated by a normal process with distribution
${\cal N}(0,1)$.  With successive intervals we mix points generated
from the uniform distribution ${\cal U}[-2.4,2.4]$, in such a
proportion that all the points in last interval are generated by the
uniform distribution.

\Doublefigure{0.85}{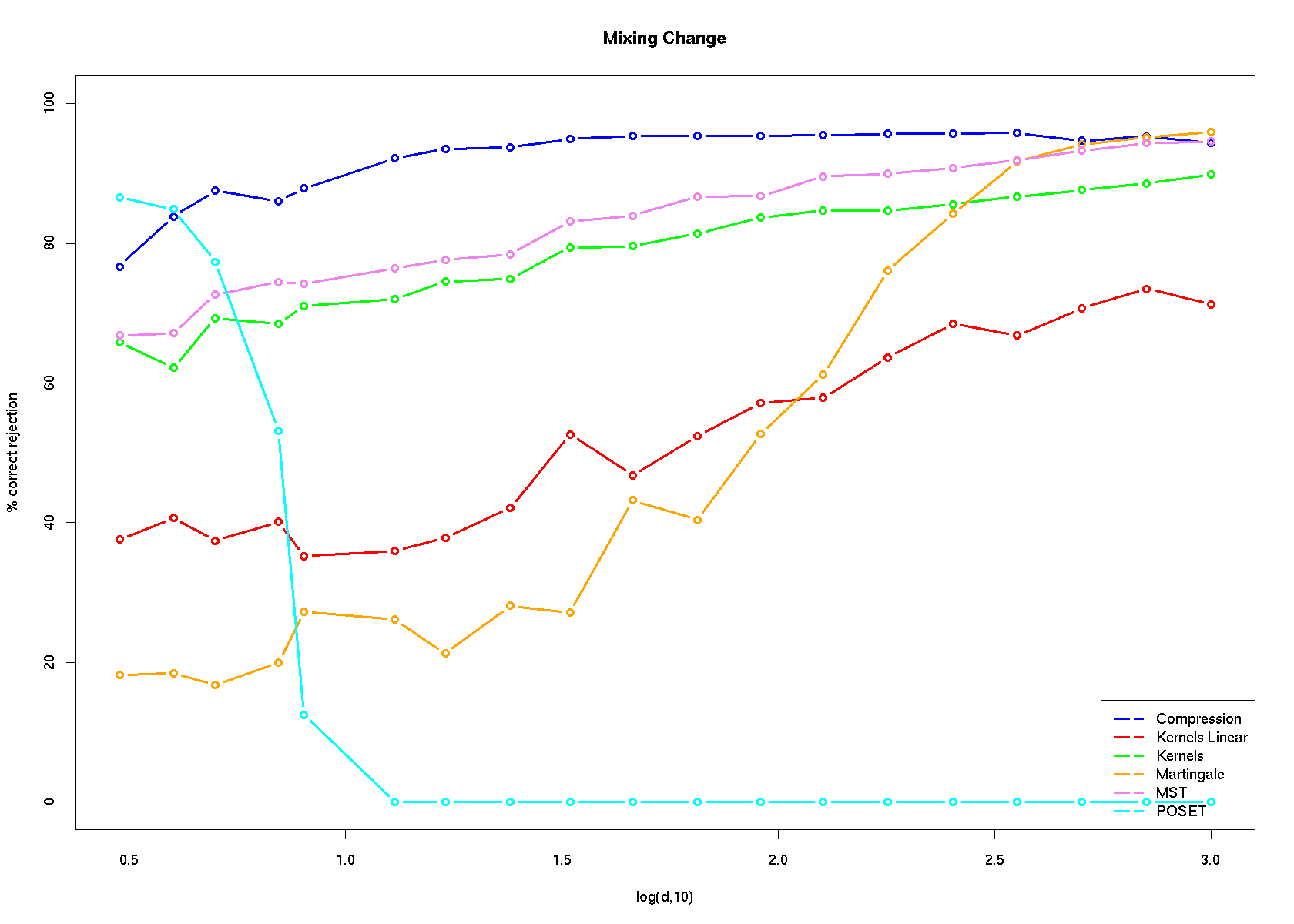}{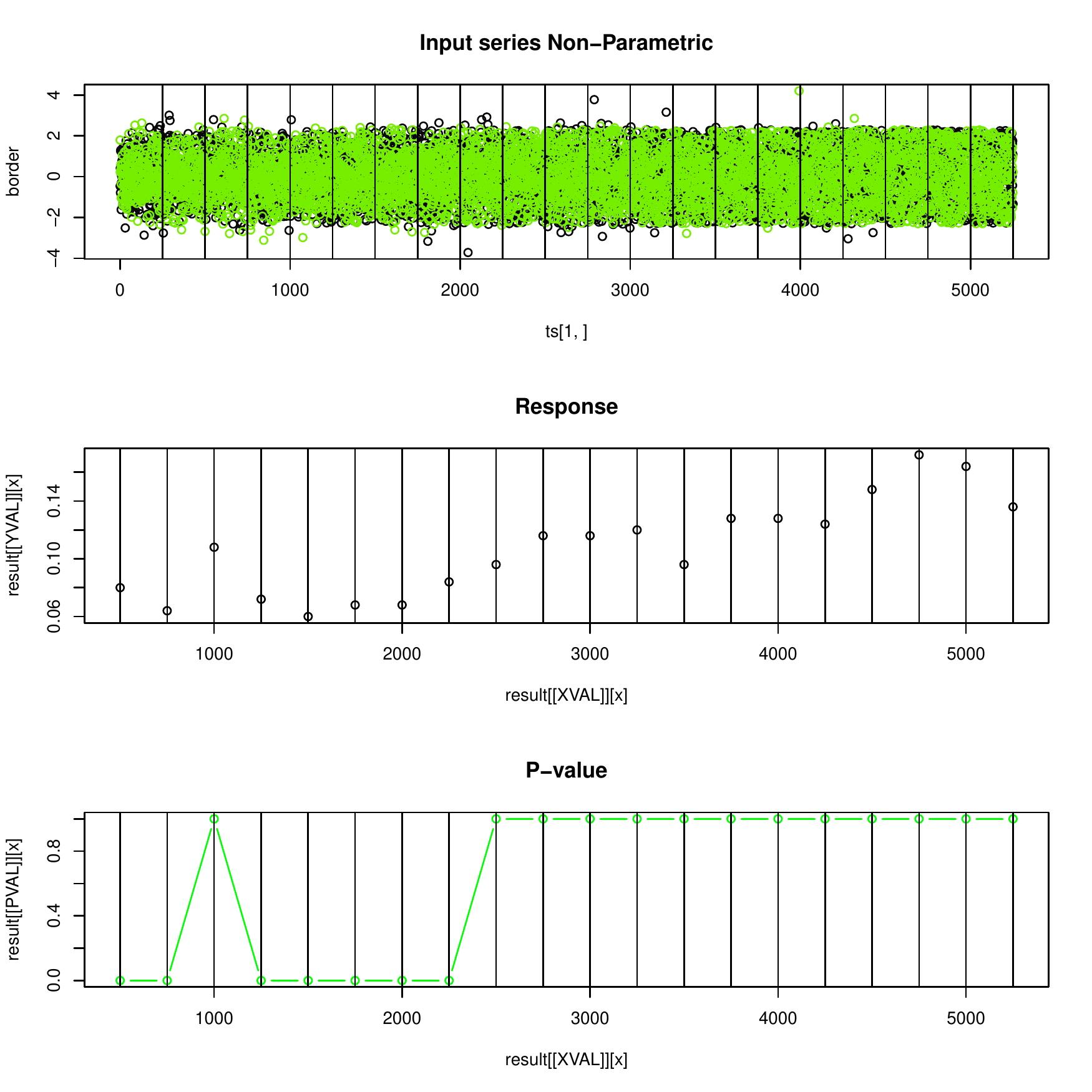}{Early correct
  rejections. Above, transition from a normal distribution ${\cal
    N}(0,1)$ to a uniform distribution ${\cal U}[-2.4,2.4]$. Below,
  poset analysis for a 2-dimensional series ---i.e., two colors are
  used}{fig:mixing-test}

As per the above tests, the first two intervals are drawn from the
same process. We gradually decrease the number of points from the
normal distribution by a factor of 1/20, while increasing the points
from the uniform distribution by the same factor. This results in a
gradual transition from a normal distribution to a uniform
distribution.  We repeated this process 100 times. Then we increased
the number of dimensions of the series so to test the effects of
dimensionality.  In Figure \ref{fig:mixing-test}, we show the average
early detection of distribution change for all methods (above), and an
example of the responses of the poset method on a two dimensional data
set (below).  The response consists of the input, the minimum distance
across 10 built-in tests, and the 20\%-quorum based $p$-value,
generated from the $p$-values of the quorum measures.

\subsubsection{Summary of Synthetic Data Results} Here, we present our
conclusions from the average, variance, and distribution tests
described previously.

\begin{itemize}
\item The methods exhibit different sensitivity to changes in average
  and variance; they tend to be more sensitive to changes in variance.

\item Although the poset method has the best discriminative power for
  few dimensions, it does not scale well. When a series has more than
  10 dimensions, the method is not able to detect any changes in
  variance or distribution. The reason for this can be understood if
  one considers a selection of points in $\R^{10}$; the probability of
  finding them in any particular order is very small. In practice, the
  partition of the topological ordering is composed of a few sets
  (3--5), too few to draw any conclusions.

\item The MST method is not very consistent. Although it performs
  poorly for detecting changes in average, it is the second most
  powerful method for detecting changes in variance or
  distribution. As the dimensions increase, the performance of the MST
  method improves.  Because the MST method is based on the relative
  distance between points, the information gained from the distance
  between points increases with more dimensions. 

\item The Kernel method performs consistently well and outperforms
  other methods for detecting change in the average.

\item The Martingale method works best for changes in the variance,
  but not so well for changes in the average. Notice that the moving
  window is relatively small and the transducer output, the $p$-value,
  has relatively short changes. Consequently, the Martingale value
  cannot reach large values, there is minimal skew in the $p$-value
  distribution, and the change is difficult to detect.  However, for
  the variance test, the $p$-value tends to be small because the new
  point is often out of the range of the moving interval (in this
  scenario the Martingale increases fast and the $p$-value has a
  skewed distribution).

\item The Compression method works well because bootstrapping is done
  before every run. The bootstrapping tunes the method to the
  properties of a particular series.

\item For the distribution test, the linear Kernel method performs
  poorly.  The Martingale method performs poorly on small and medium
  dimensions because the $p$-value of the transducer does not change
  enough to prevent the Martingale value from being reset to 1 (this
  is a problem with our implementation more than a failing of the
  method).  However, the Martingale method does work well on large
  dimensions.  The full Kernel method generally works well. The best
  method for the early detection of change is the Compression
  method. The poset for very small dimensions, and the MST methods for
  all dimensions are the best non-parametric methods. This comes as no
  surprise, as these methods are designed to capture changes in
  distribution.
\end{itemize}

\mypar{Sample Size and Dimensionality}
\begin{table}%
\tbl{Rejection ratio median for series in $\R^{10}$,
  based on 100 runs, with change in average and an increasing number
  of points in each interval
  \label{table:sensitivity}}{%
\centering
{\small \begin{tabular}{lrrr}
\hline\hline 
Method        & N=250 & N=500 & N=750 \\ \hline \hline
Compression   & 0.77  & 0.83  & 0.84  \\ 
Kernels       & 0.81  & 0.85  & 0.87    \\ 
Kernels Lin.  & 0.73  & 0.729 & 0.75  \\ 
Martingale    & 0.50  & 0.66  & 0.73  \\ 
POSET         & 0.66  & 0.76  & 0.79  \\ 
MST           & 0.49  & 0.49  & 0.49  \\ \hline
MST (variance) & 0.89  & 0.90  & 0.89  \\ \hline
\end{tabular}}}
\end{table}

In the previous tests, we fixed the length of the interval and change
the dimensions. The length of the interval is the same used in
previous work. In this section, we present an introductory study about
the relation between the interval length and the number of dimensions
of each point in the interval, see Table \ref{table:sensitivity}.

As we expected, increasing the number of points means there is more
information about the process being measured, which results in the
methods being more discriminative. The exception to this rule is the
MST method, as explained in Section \ref{sec:mst-algo}.  Although the
poset and MST methods share the same statistical tests, they order
data differently. In the case of the MST method, having more points
does not change the inherent structure of the MST, and therefore does
not impact its discriminative ability.

\Doublefigure{0.49}{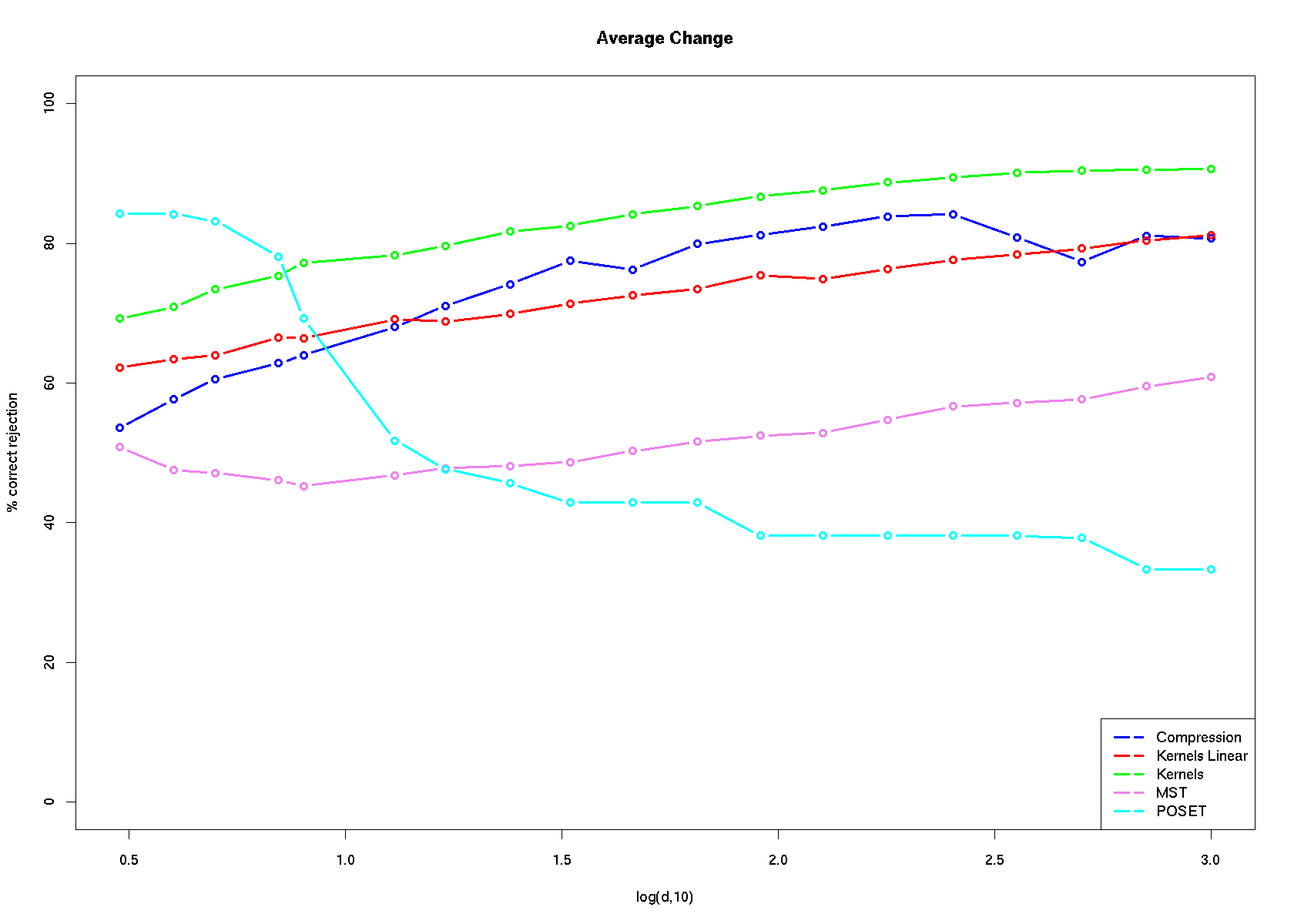}{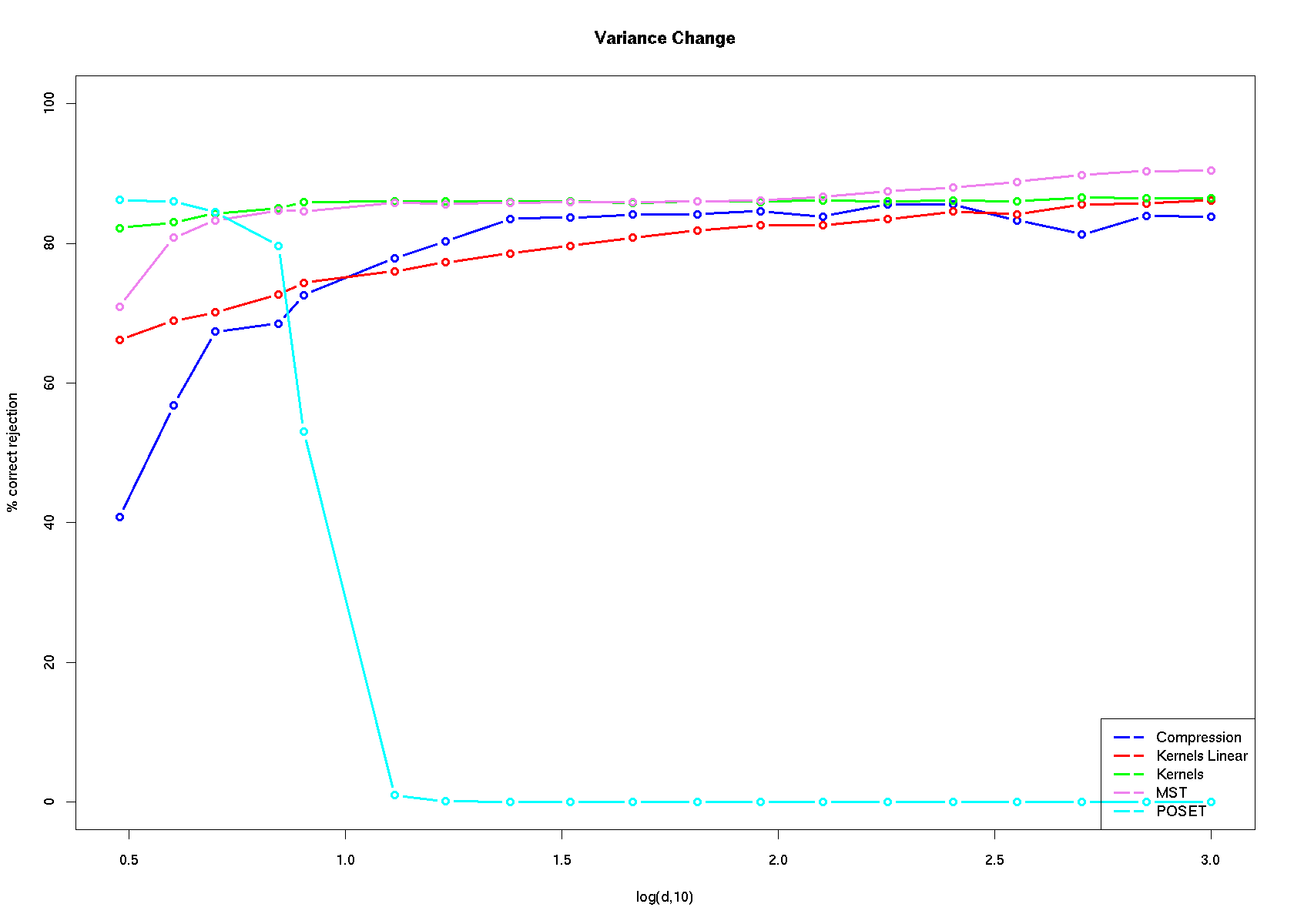}{Rate of correct
  rejections. Left, average increasing exponentially.  Right, variance
  increasing exponentially.}{fig:exponential-test-counts}

\mypar{Sensitivity: Early Rejection Versus Total Number of Rejections}

\singlefigure{0.9}{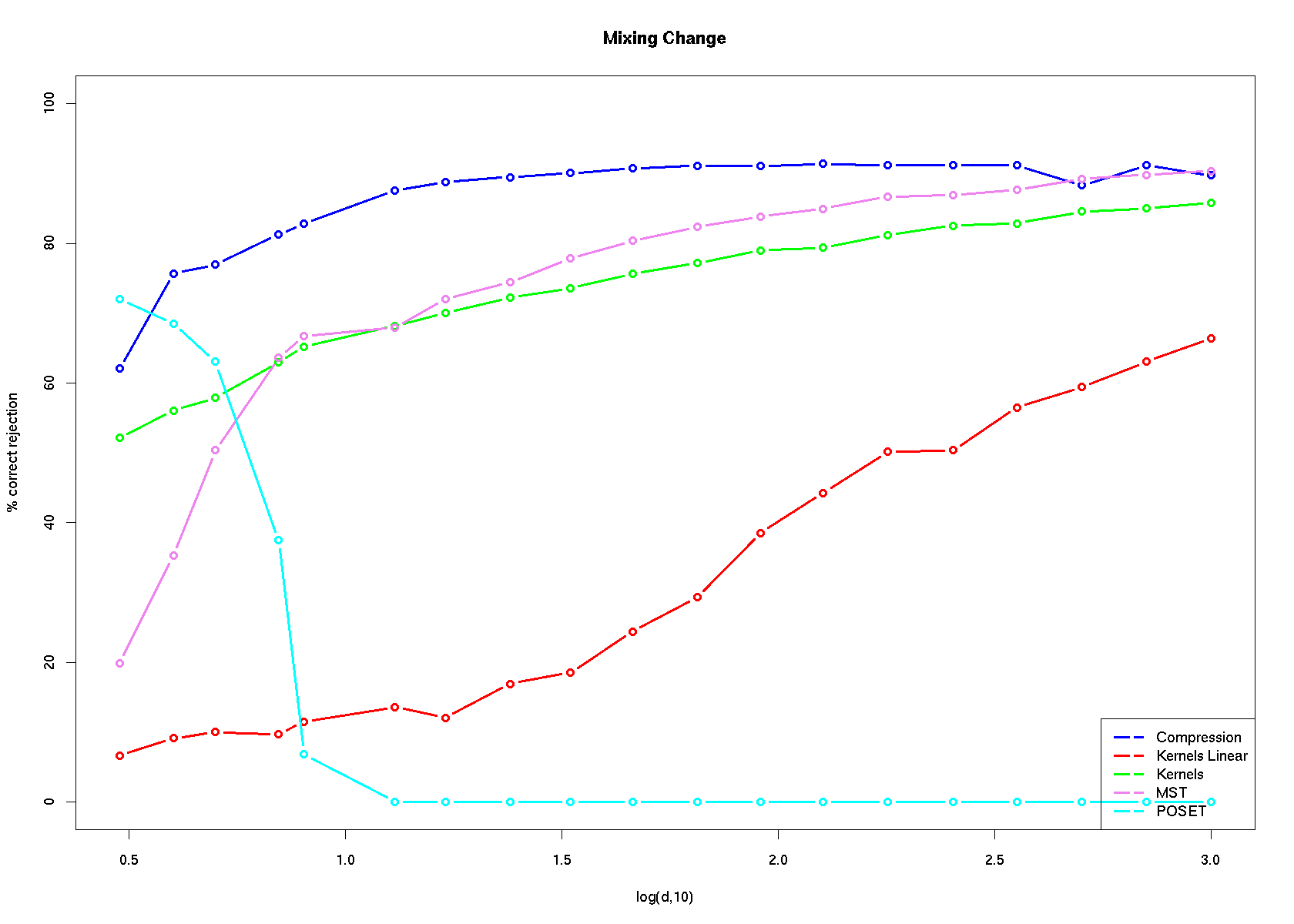}{Rate of correct rejections. Transition
	from a normal distribution ${\cal N}(0,1)$ to a uniform distribution
	${\cal U}[-2.4,2.4]$.}{fig:mixing-test-count}

Excluding the Martingale method, we can compute a sensitivity measure
for each method by counting the number of times a method identifies a
change (instead of how early a method detects a change). Recall that
21 intervals were constructed where the first two intervals are
intentionally similar (to the reference window) for the purpose of
tuning the method.  As the moving window covers a single interval, we
expect each method to identify 20 changes.

Figures \ref{fig:exponential-test-counts} and \ref{fig:mixing-test-count}
present the sensitivity measure. As before, the methods tend to become
more sensitive as the number of dimensions increases, excepting the poset
method for reasons mentioned previously. Again, the poset method is the
most sensitive for series with fewer than 10 dimensions.

As expected, the methods are consistent in the sense that an early
discrimination translates into a better overall discrimination.

\subsection{Pre-Classified Data}
\label{sec:classified-data}
In the previous section, we explained how we created synthetic data
sets with known properties, such that we could control the
introduction of differences caused by changes in average, variance,
and distribution.  In this section, we present a method for creating
data sets with well known properties based upon pre-classified
data. We use four freely available data sets from the UCI machine
learning repository \cite{FrankAsuncion:2010}. We used the following
data sets: yeasts with nine dimensions (also referred to as attributes
in \cite{HortonK1996}), abalone with eight dimensions
\cite{NashSTCF1994}, Parkinson's disease with 26 dimensions
\cite{TsanasLMR2010}, and breast cancer with 30 dimensions
\cite{StreetWM1993}.

We selected these data sets for several reasons, including the fact
that they have a managable number of dimensions, literal dimensions
can be easily translated into numerical values without any loss of
information, and we can safely assume that the attributes come from a
continuous distribution.

\mypar{From Data Set to Series} For each data set, we grouped the data
into intervals by using the class identifier as the partitioning key;
for the Parkinson's disease data set we include the person identifier
in the partitioning key. We concatenate these intervals in decreasing
length order. In practice, the first interval will contain the largest
number of points, while the last interval will contain the fewest. For
each interval we perform a random permutation of the points, which
helps to break down any previous bias within the interval. In the
experiments we use a moving window with a size that is often smaller
than the intervals.

\begin{figure*}
\includegraphics[width=0.5\linewidth]{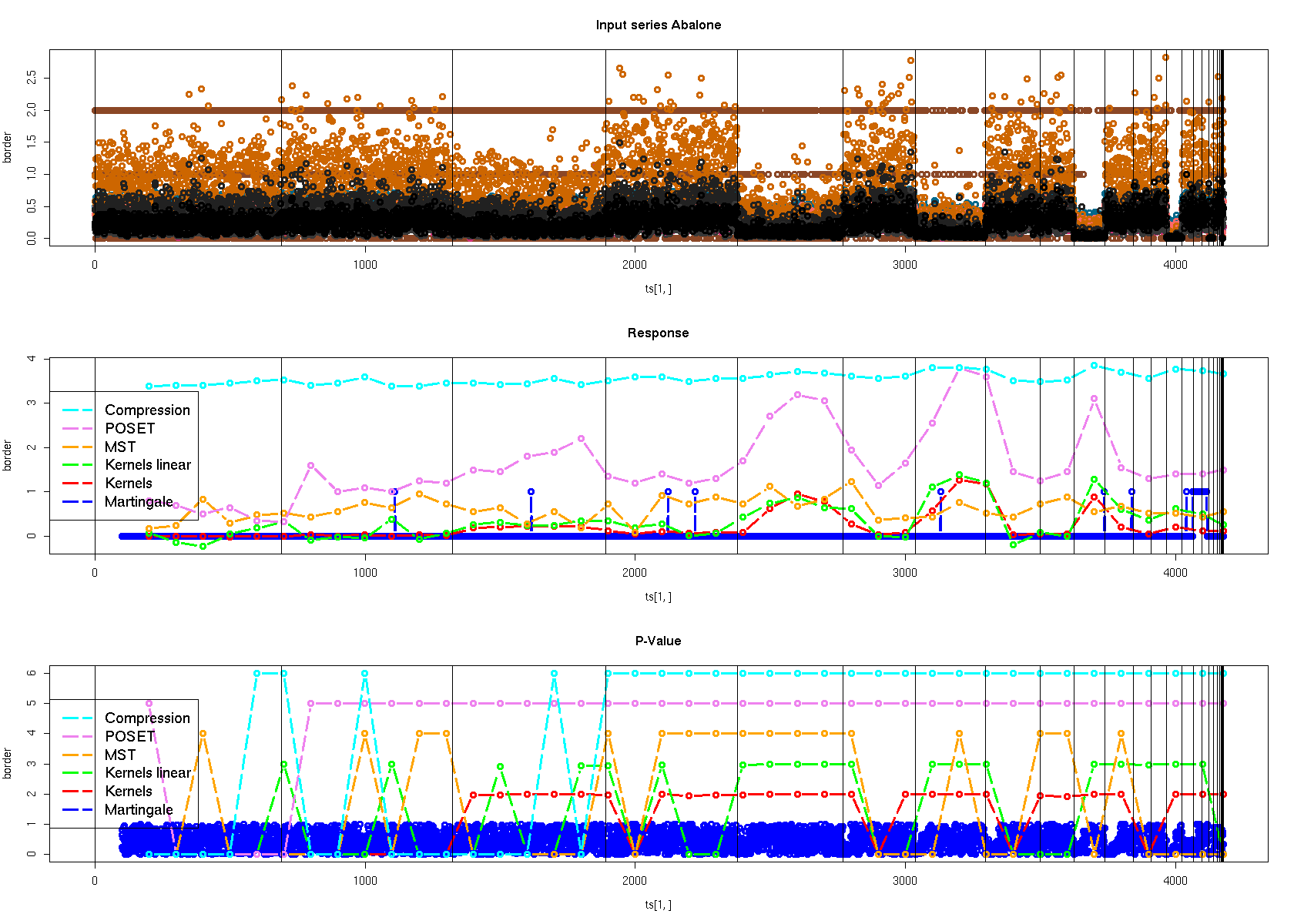} \includegraphics[width=0.5\linewidth]{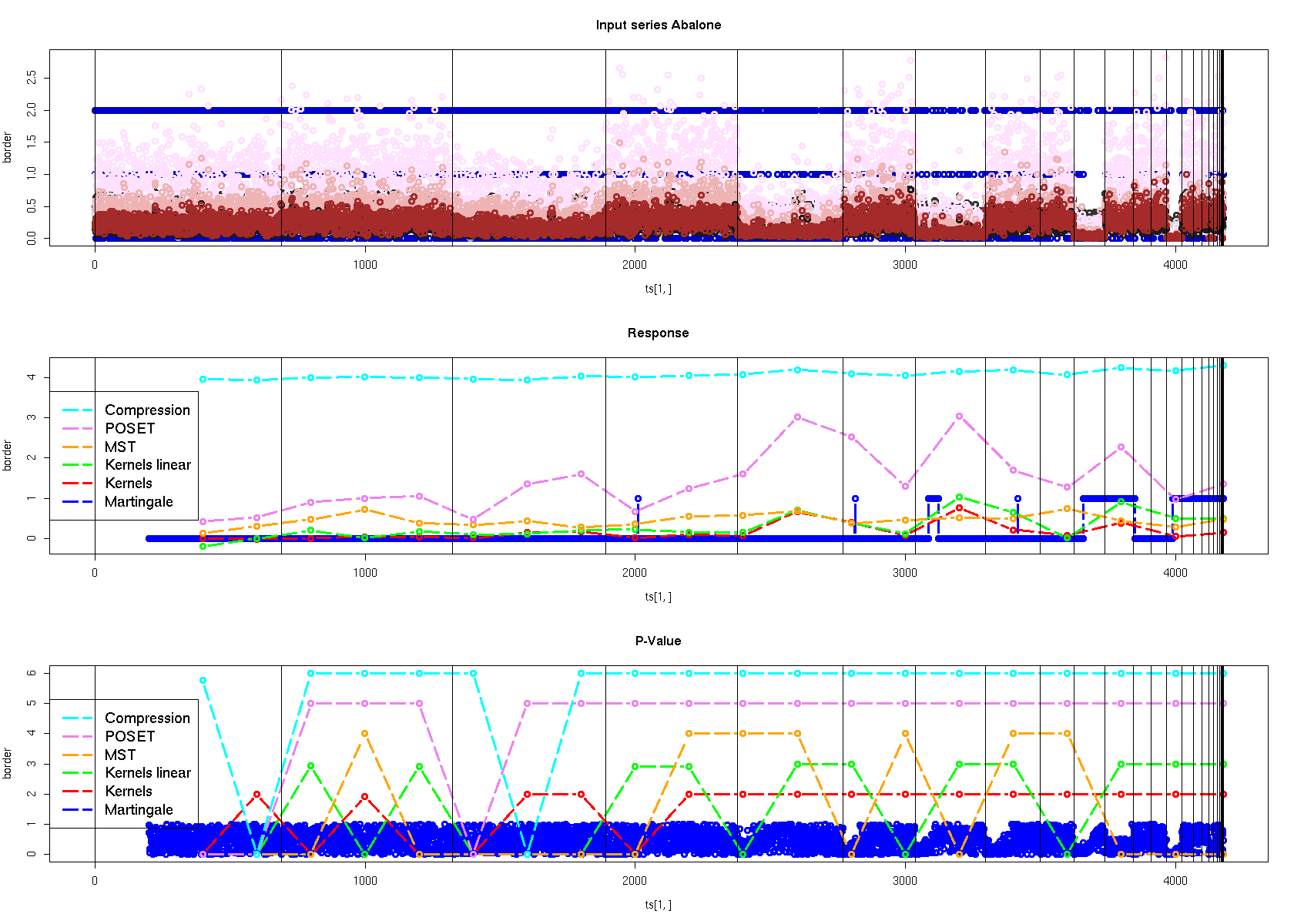} \\
\includegraphics[width=0.5\linewidth]{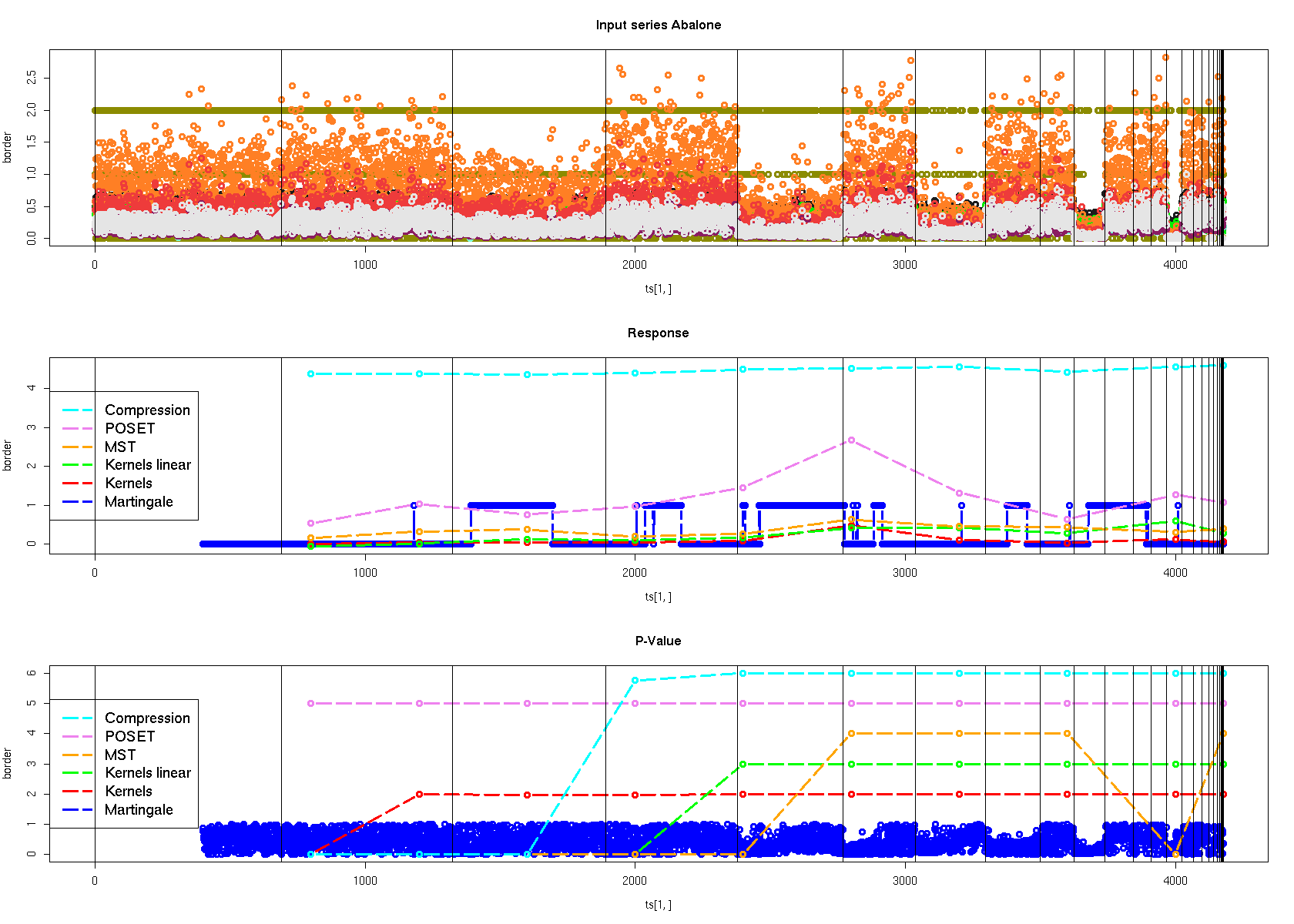} \includegraphics[width=0.5\linewidth]{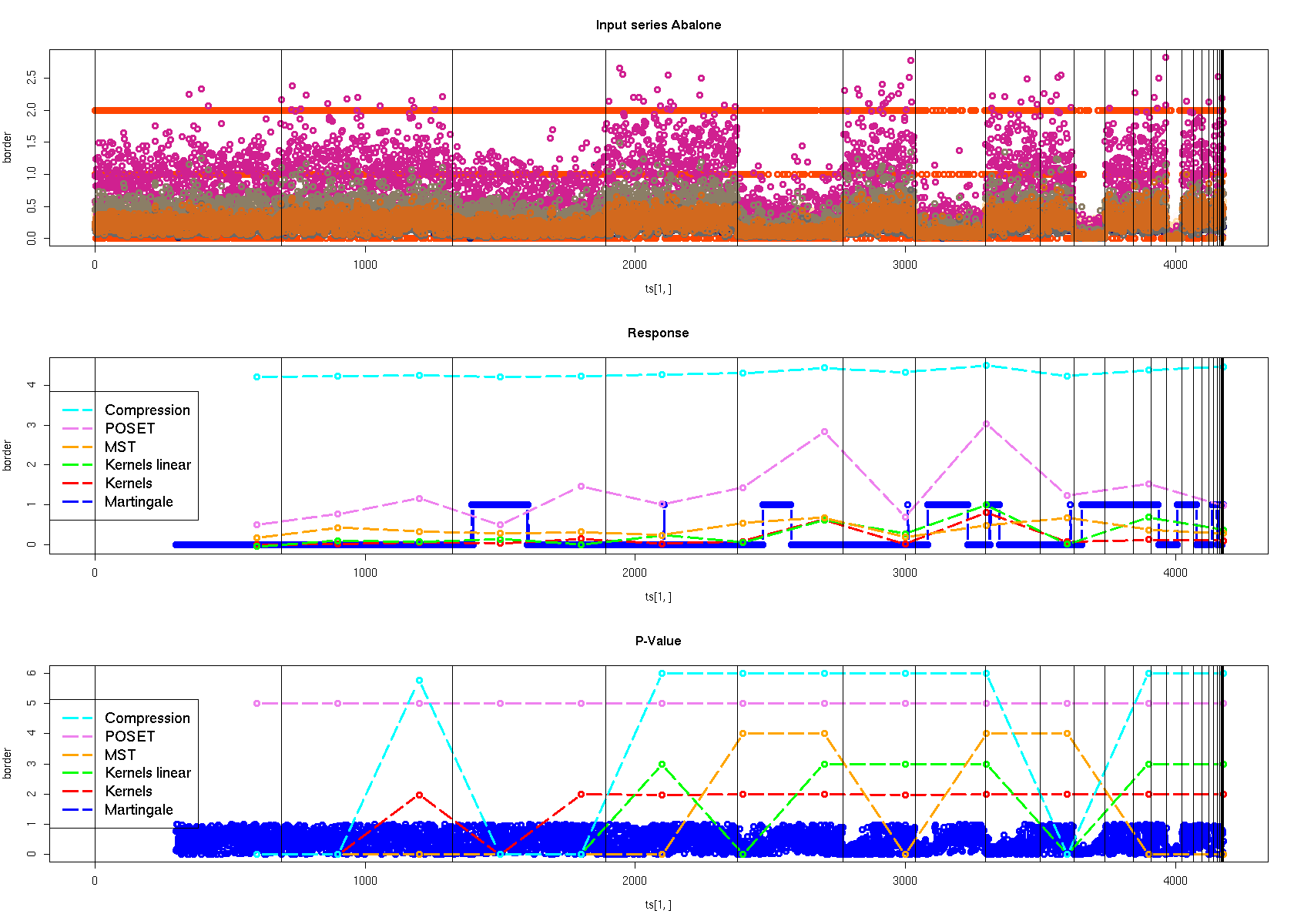} 
\caption{Abalone data set with window size 100, 200, 300, and 400 (clock-wise from top left).}
\label{fig:abalone}
\end{figure*}

\begin{figure*}
\includegraphics[width=0.5\linewidth]{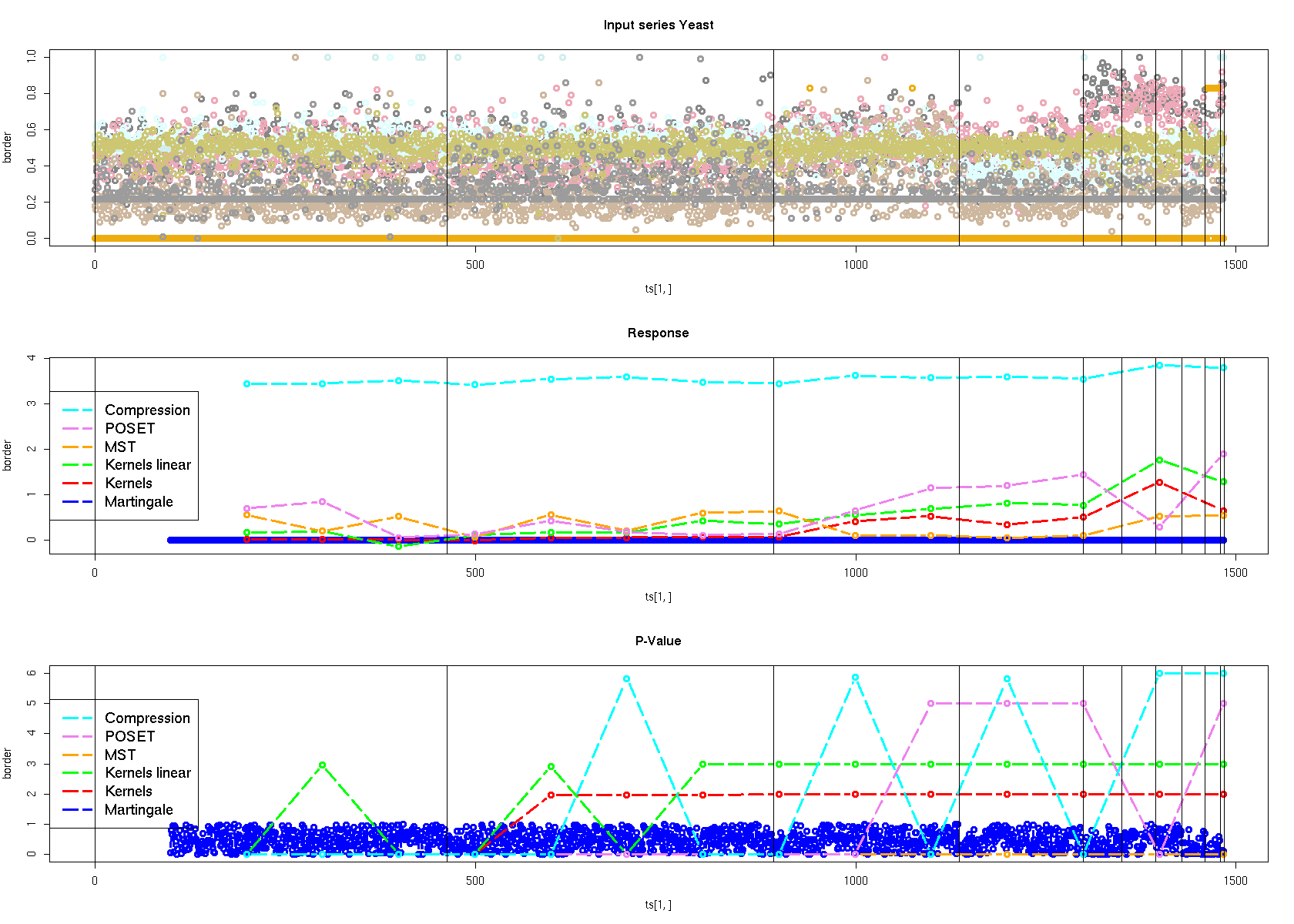} \includegraphics[width=0.5\linewidth]{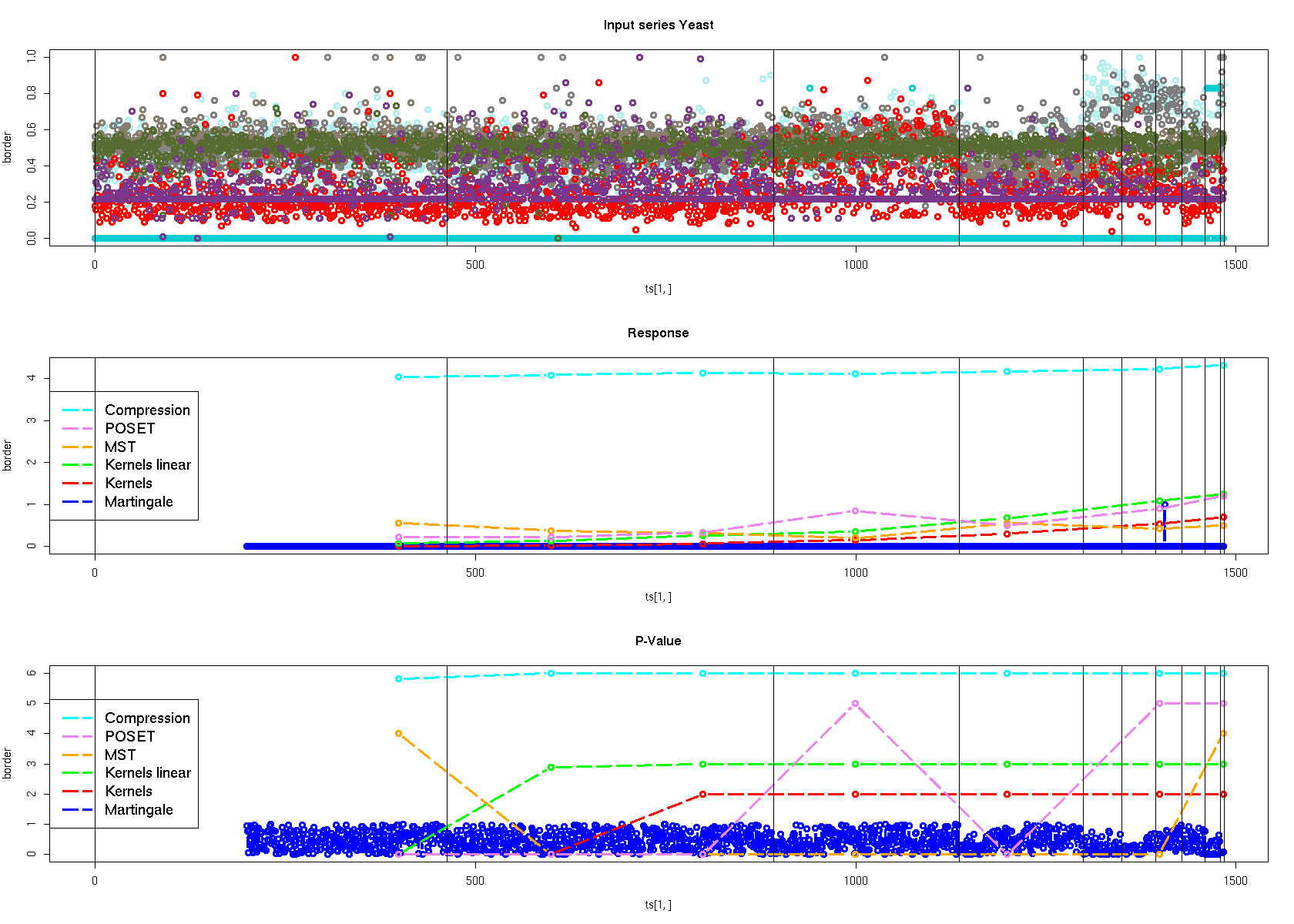} \\
\includegraphics[width=0.5\linewidth]{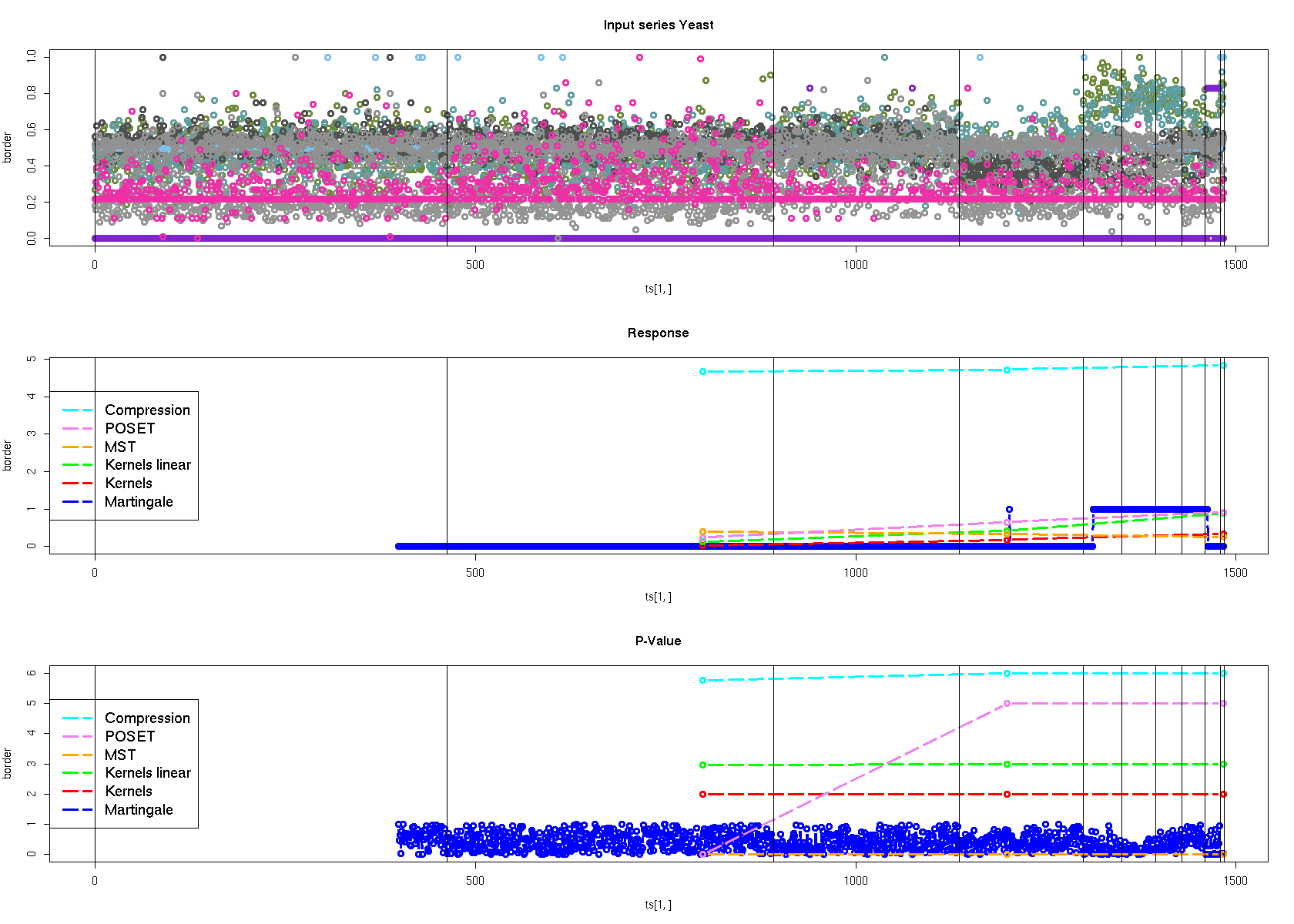} \includegraphics[width=0.5\linewidth]{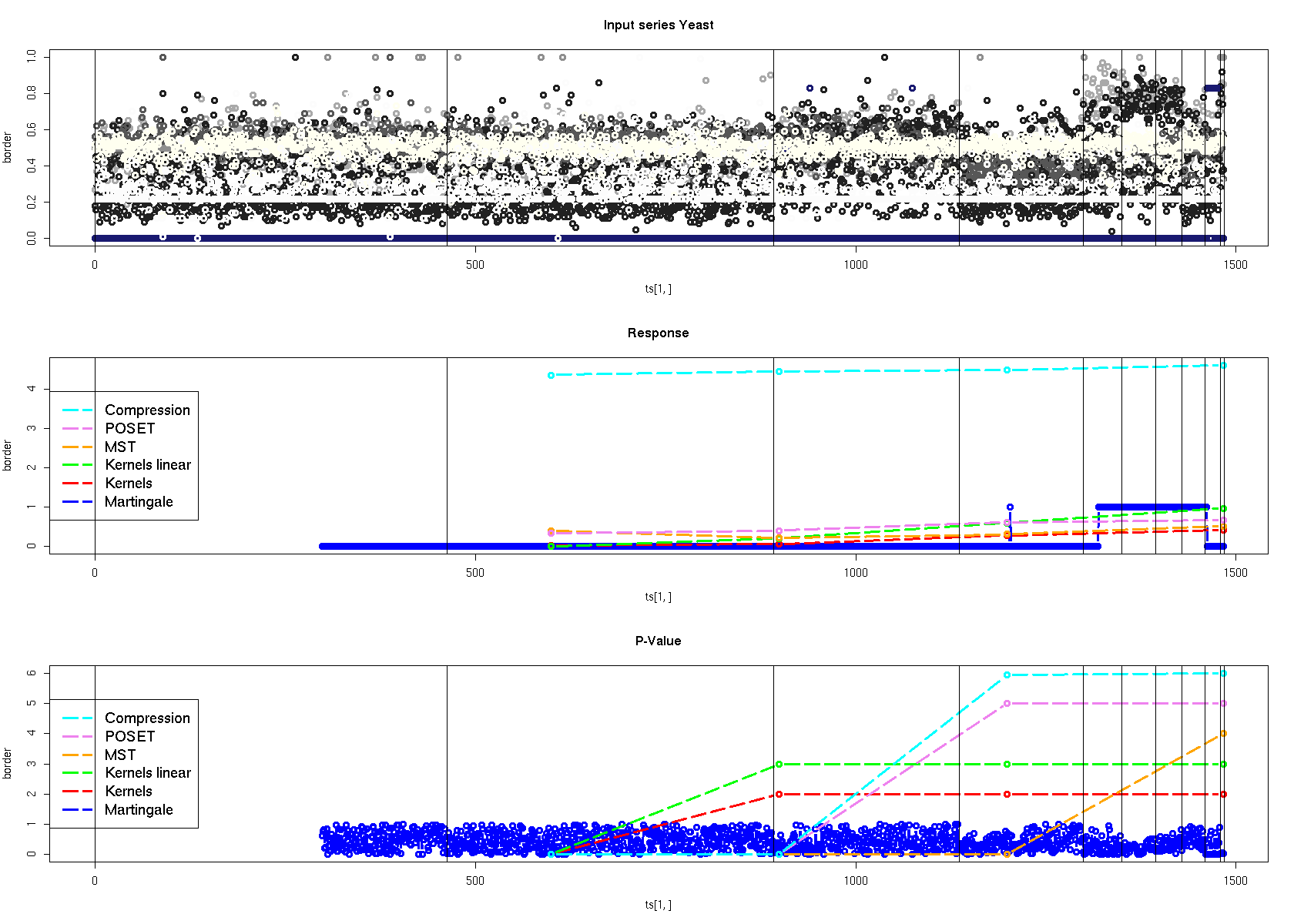} 
\caption{Yeast data set with window size 100, 200, 300, and 400 (clock-wise from top left).}
\label{fig:yeast}
\end{figure*}

\begin{figure*}
\includegraphics[width=0.5\linewidth]{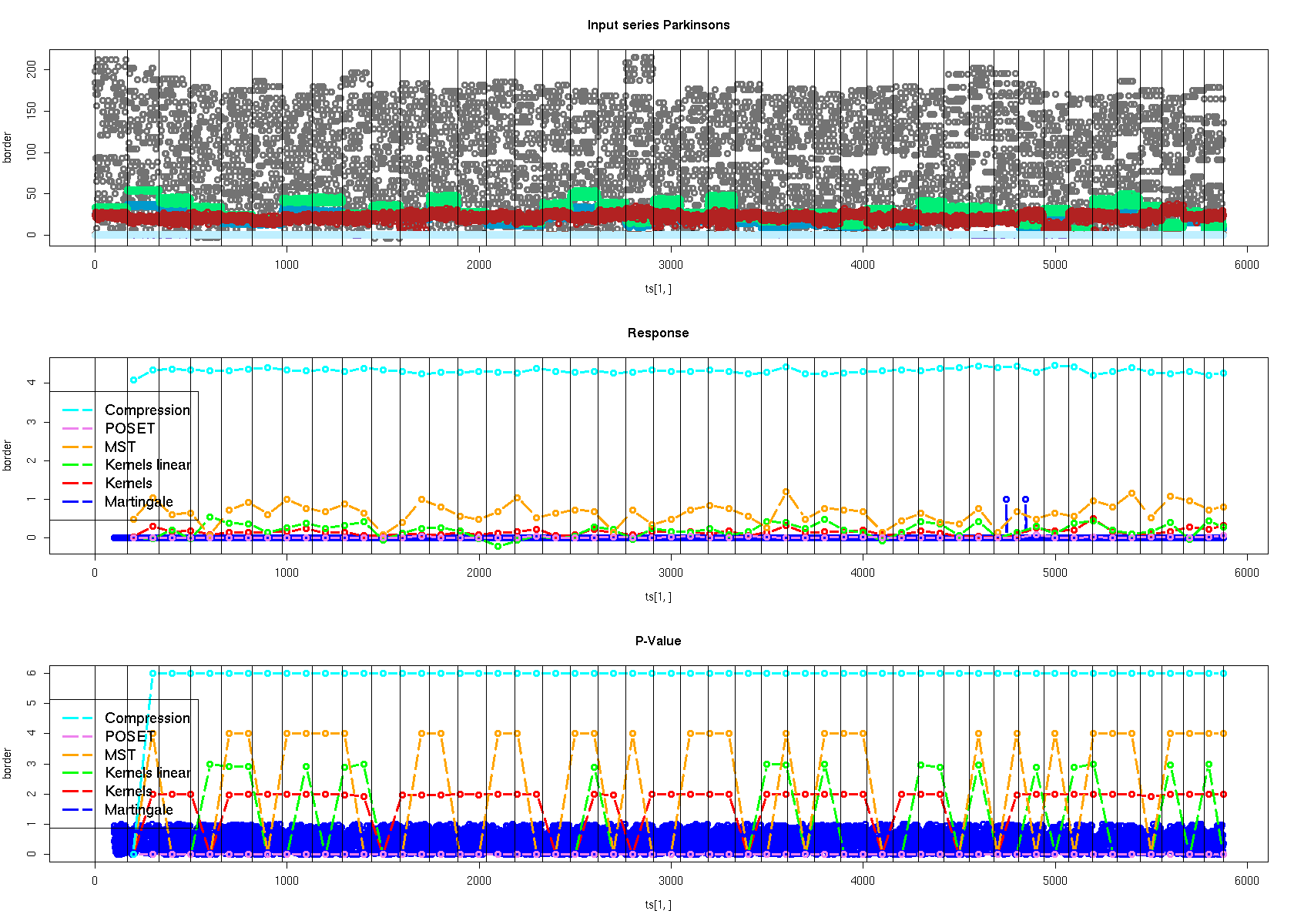} \includegraphics[width=0.5\linewidth]{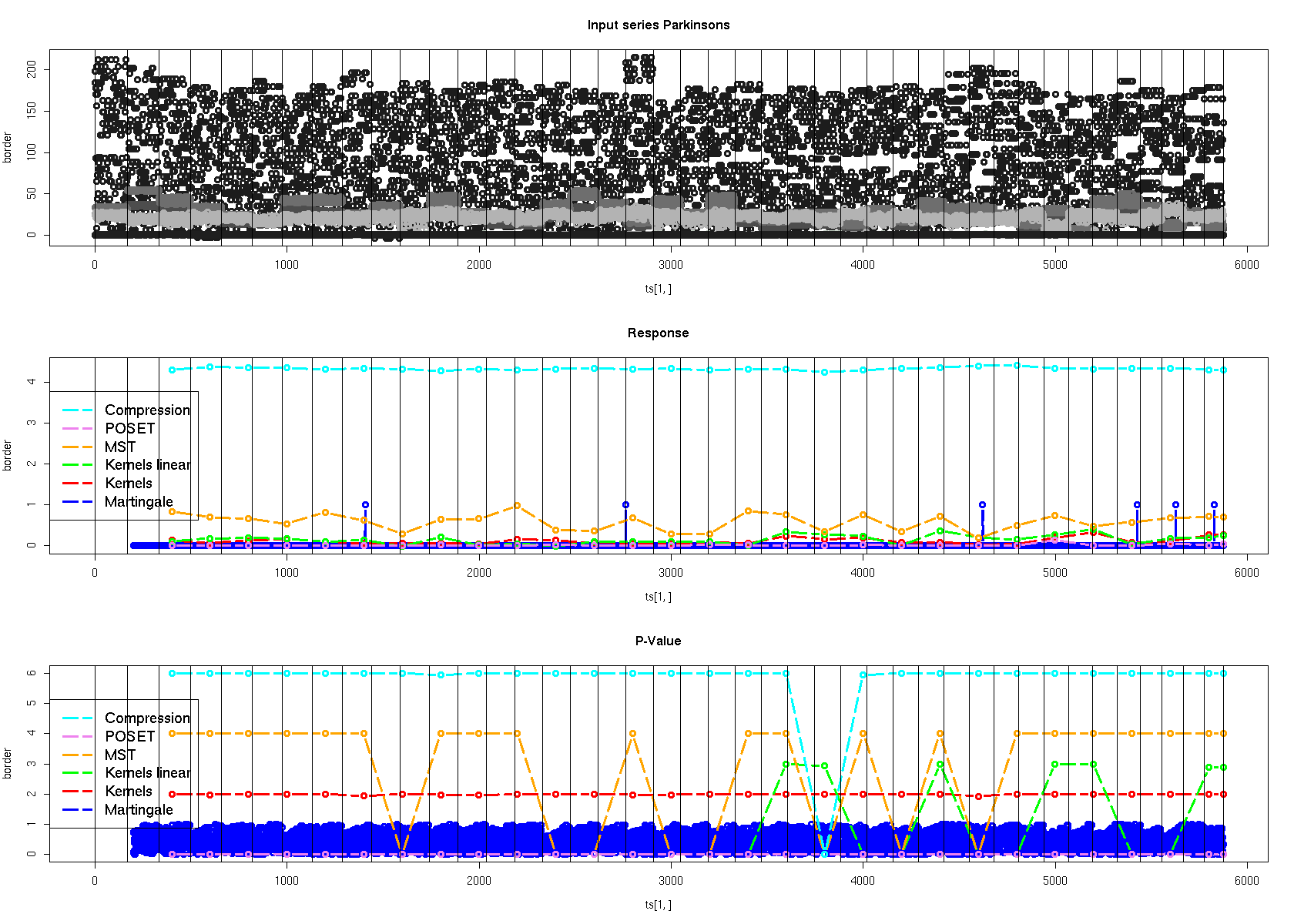} \\
\includegraphics[width=0.5\linewidth]{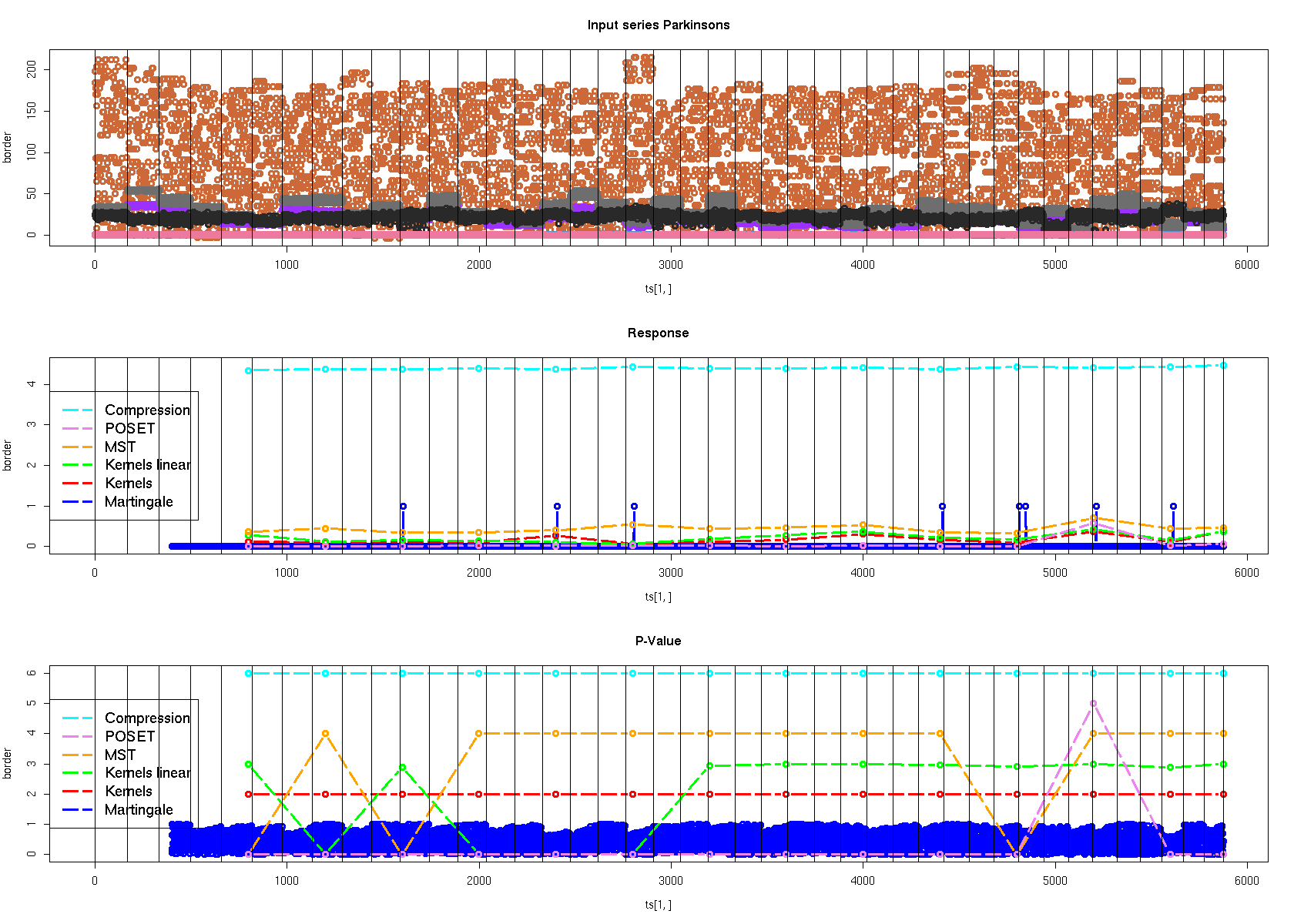} \includegraphics[width=0.5\linewidth]{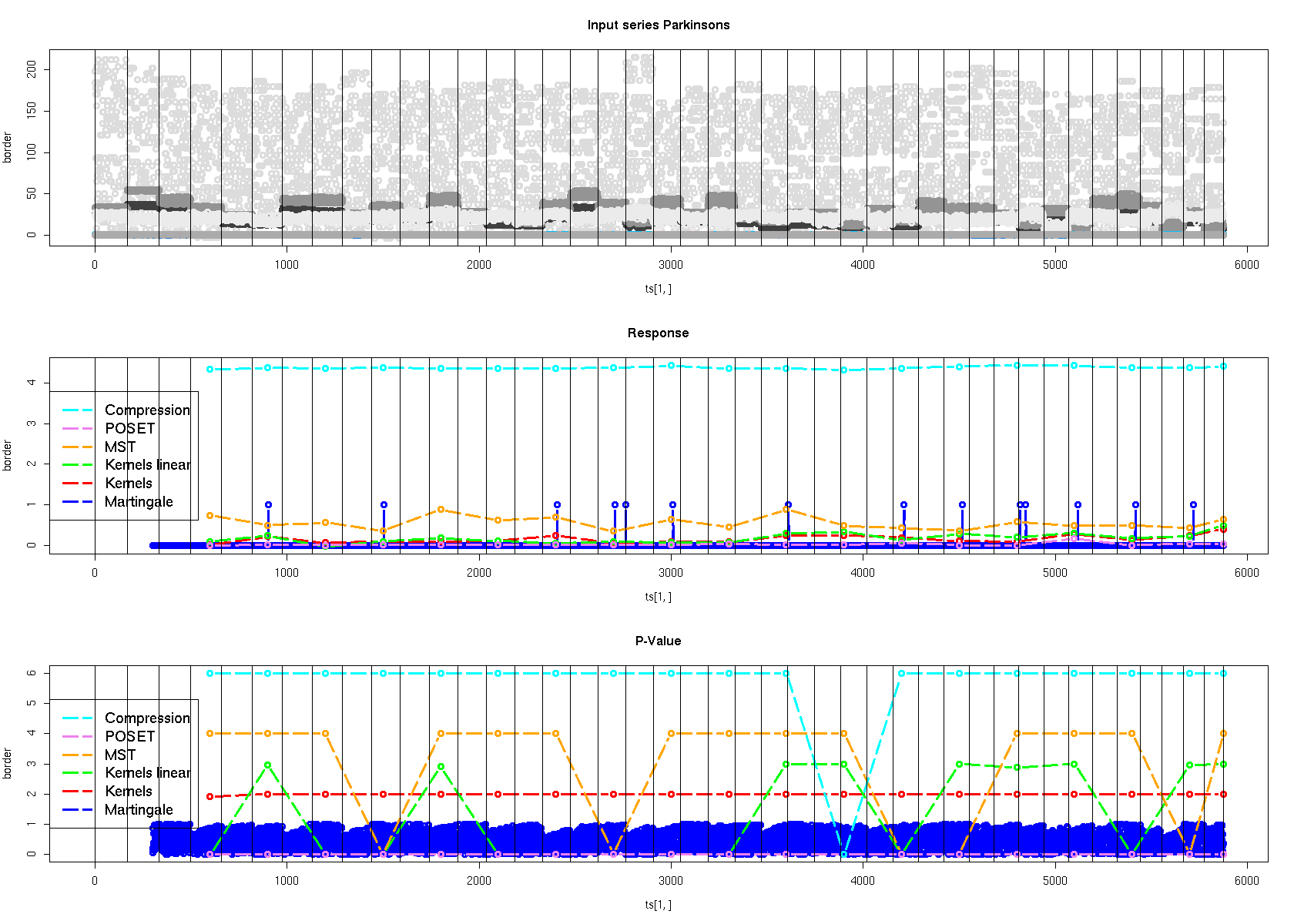} 
\caption{Parkinson's disease data set with window size 100, 200, 300, and 400 (clock-wise from top left).}
\label{fig:parkinsons}
\end{figure*}

\begin{figure*}
\includegraphics[width=0.5\linewidth]{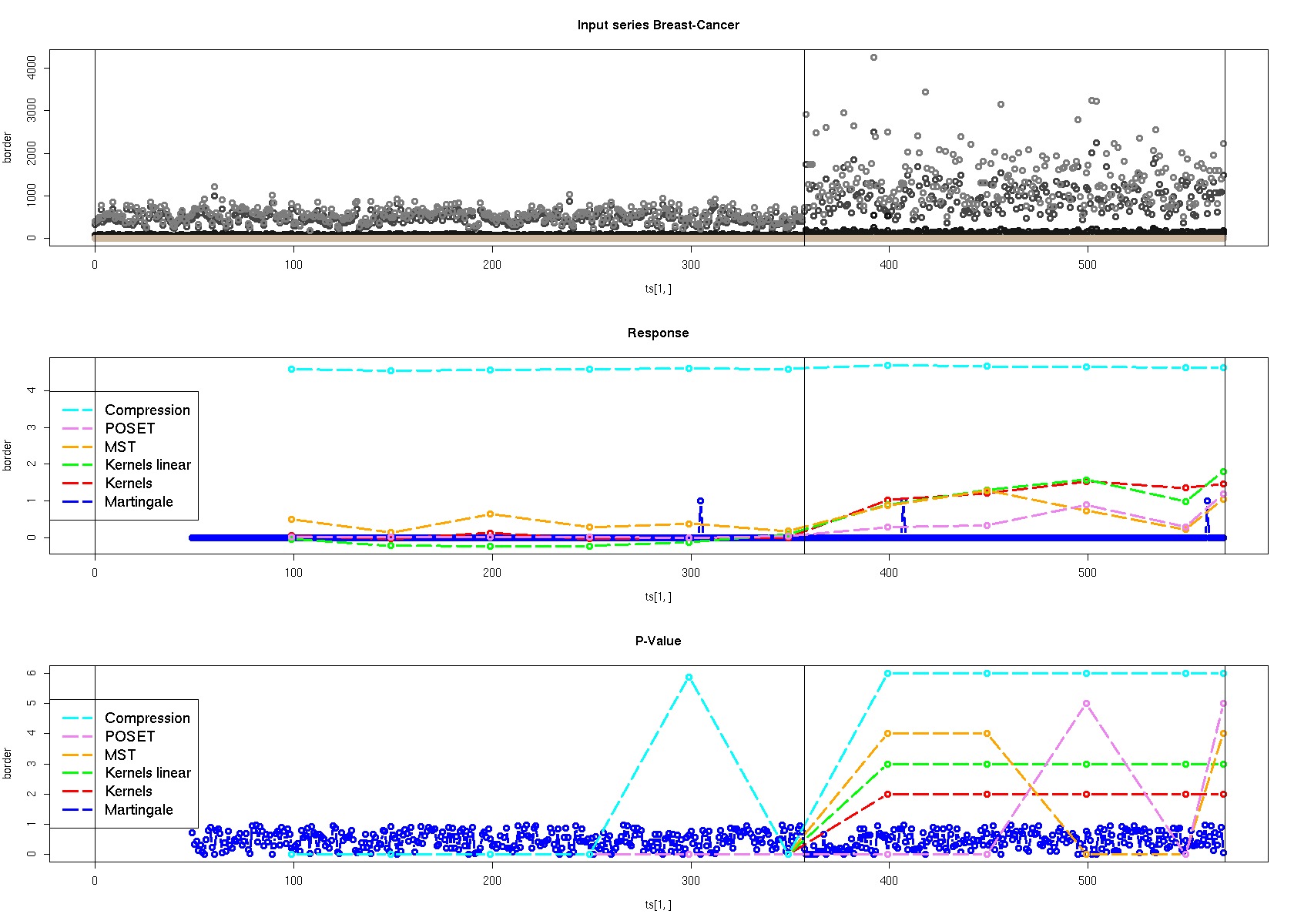}  \includegraphics[width=0.5\linewidth]{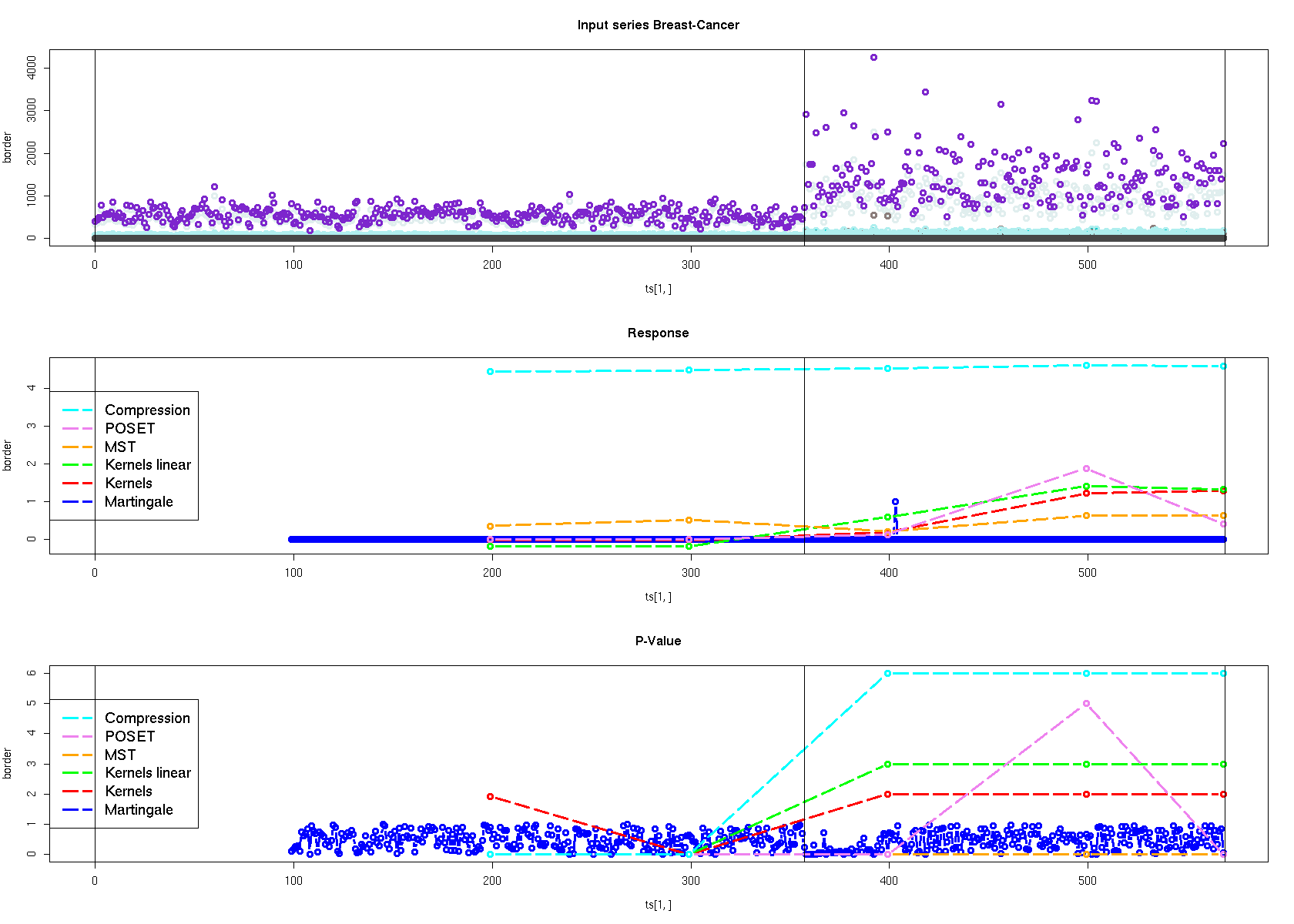} \\
\includegraphics[width=0.5\linewidth]{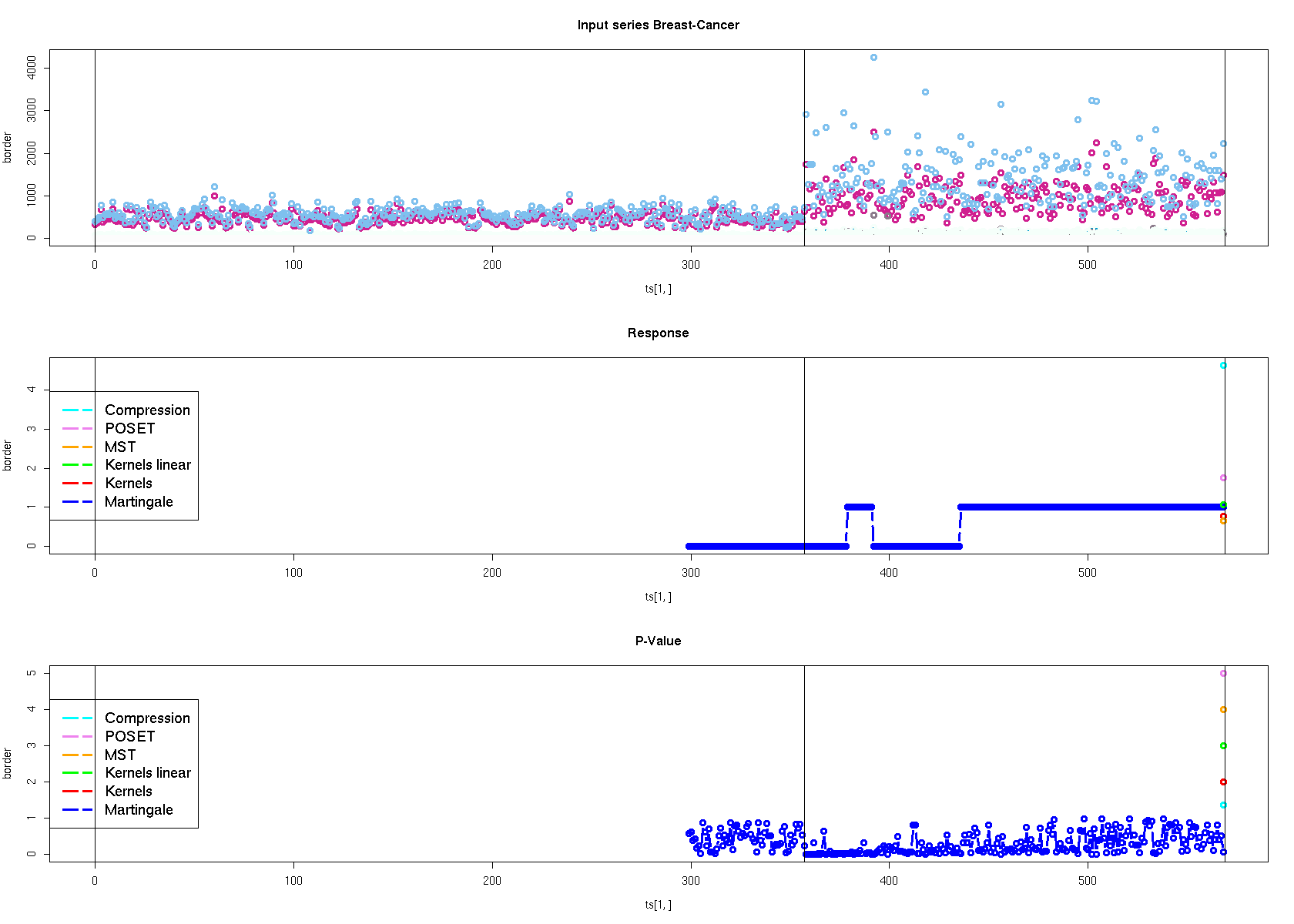} \includegraphics[width=0.5\linewidth]{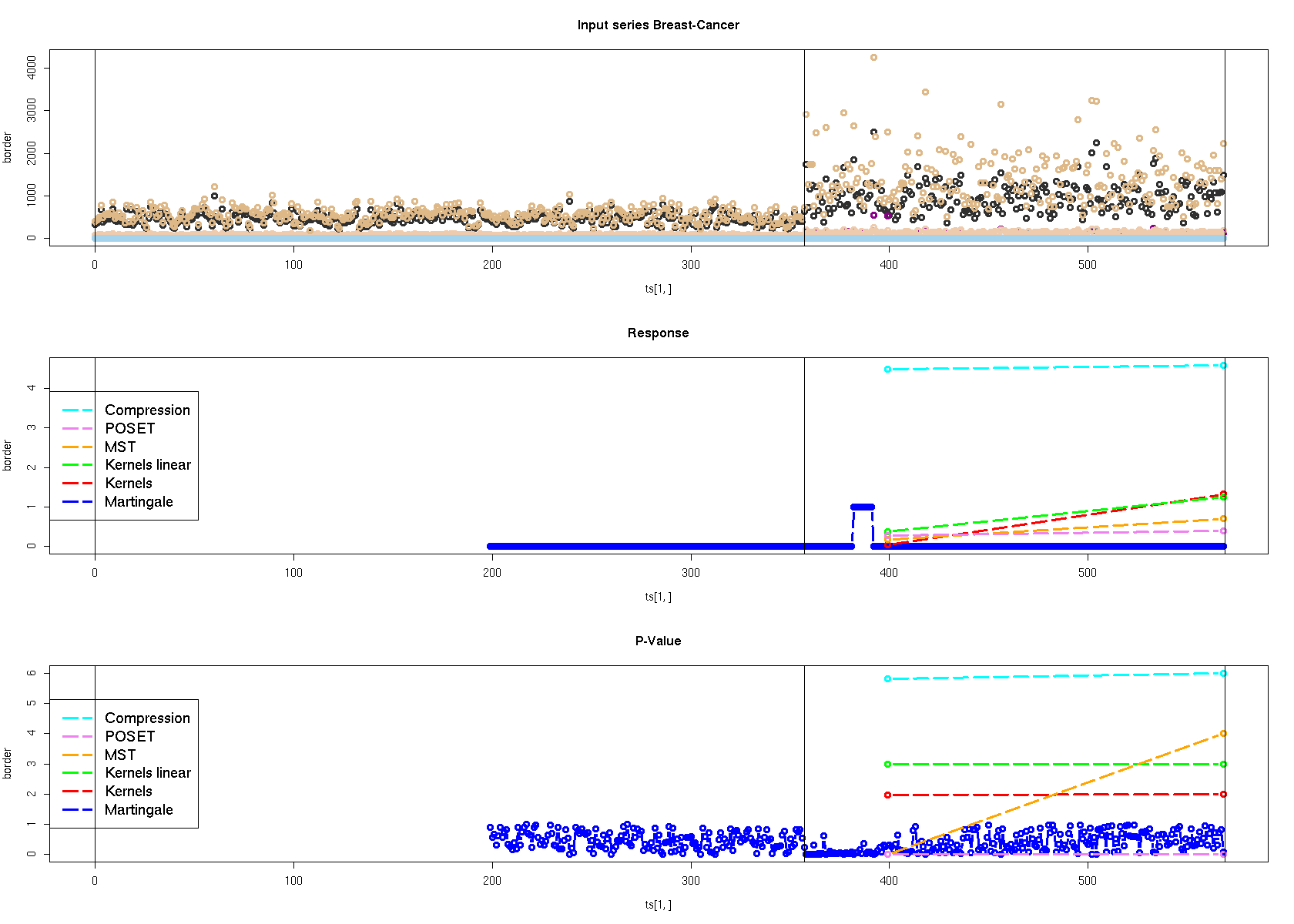} 
\caption{Breast cancer data set with window size 50, 100, 200, and 300 (clock-wise from top left).}
\label{fig:breastcancer}
\end{figure*}

We tested our six methods using four different moving window sizes.
For the breast cancer data set we used window sizes of 50, 100, 200,
and 300; and for the other data sets we used window sizes of 100, 200,
300, and 400.  The results are presented in Figure \ref{fig:abalone},
\ref{fig:yeast}, \ref{fig:parkinsons}, and \ref{fig:breastcancer}.

\singlefigure{1}{Abalone100.png}{Abalone data set with window size
  100}{fig:abalone-100}

To ease the introduction of the data sets and the application of the
methods, we turn our attention to the abalone data with window size
100, see Figure \ref{fig:abalone-100}. We shall take the same steps
for the other classified data set. 

First, we start describing how we represent the input series, which is
always the first scattered plot in each figure.  We use colors to plot
the different dimensions of the series in order to capture more facets
of the data.  We mark the borders between intervals by using a
vertical line, making it easy to distinguish the known
classifications. For example, for the abalone data, we have 21
different intervals with different lengths.

Second, we shall show the response of every method.  The response from
the Martingale method is encoded in the response row, where a value of
1 indicates a difference and 0 indicates no difference. For the other
methods, we plot a normalized distance measure as we shall explain
shortly for the $p$-values. For example in Figure
\ref{fig:abalone-100}, the Martingale method flags a difference at:
1100, 1600, 2100, 2200 and after 4100. Notice that the first four
differences are false differences because the moving window is
entirely contained into a classified interval. The last differences
are actually correct. In the following, we shall present performance
with a window size of 400 points: with a larger window we shall
capture better the variations of larger classified intervals but not
for smaller ones having the opposite situation that the one presented
here.

Third, we report the $p$-value. A $p$-value larger than 0.95 indicates
a significant change. To encapsulate all of the change responses for
the methods in a single figure, we draw $k*p$-value on a single plot,
where $k$ is an integer in $[1,6]$ and $k$ is uniquely associated with
a method (the response is computed similarly) and color coded for easy
consultation. A rejection is represented by a rise of the $p$-value
line to $k$.

Therefore, if all the methods reject all the intervals, then the
resulting plot will show six parallel lines on the interval 1--6.
Recall that the only exception is the Martingale method, for which we
use the $p$-value of the transducer in conjunction with the
response. 

For example in Figure \ref{fig:abalone-100}, The first interval of 100
points is taken as reference, then we let the moving window scan the
series with interval of length 100 points. The poset method (pink
line) detects a difference within the first interval, which is false;
however, the poset method detects correctly all other interval as
different. The compression method detects two false differences, but
overall it is capable to find always a small window in each classified
interval that is different from the reference. Notice that every
method detect a difference when the moving window is scanning the
interval 2400--2800: we can see five parallel lines.

Obviously, our goal is to show that we can distinguish between
different classified intervals. However, the differences in the
interval lengths cause problems to the scanning methodology, as the
moving window may be too large that it spans multiple intervals or may
be too small to completely cover other intervals. In this case, we
show that the reference window is from a single interval and is
different from a combination of intervals. Recall that the reference
window and the moving window are of the same size and we shall present
results for different sizes. 

Note this section is meant to provide a qualitative presentation of
the discriminative powers of the methods.

%\begin{comment}

%\end{comment}

\mypar{Abalone Data Set, 8 Dimensions} In Figure \ref{fig:abalone}, we
identify an interval based on the age of the abalone. The series is
generated by concatenating the intervals in decreasing order of the
number of points. The sex attribute is transformed into an integer
value of 0, 1, or 2, where the mapping is arbitrary but fixed across
the tests. The first interval contains enough points to perform
bootstrapping for the Compression method. We present the responses for
different window lengths of $|W|$: 100, 200, 300 and 400. When
$|R|{=}|W|{=}100$, the moving window is able to cover almost every
interval. We notice that, aside from a single false negative, the
poset method always successfully detects the changes.  The Martingale
method performs well on the larger intervals, for which the moving
window covers the interval, because it is able to capture the
$p$-value distribution change. The Kernel method performs well and the
Compression method is consistent as well.  As the $|R|$ and $|W|$
increase to 400 points, the discriminative powers of the methods also
increase, particularly the Kernel method.

We discuss further the results we observed by putting them in the
context of the properties of the methods that we presented in previous
sections: this series presents a change of variance.

\mypar{Yeast Data Set, 9 Dimensions} We created the intervals in the
same manner as described previously; however, we excluded the database
identification number and used the yeast type as the partitioning key.
Notice that, for any moving window size, the linear kernel method is
the fastest at identifying changes, even faster than the regular
kernel method. The Compression method does well, while the MST method
did not perform well, and the poset method required a window size of
at least 300 to be able to consistently identify changes.

This series presents an average change.

\mypar{Parkinson's Disease Data Set, 26 Dimensions} We kept only the
telemetry and motor-skill data, and excluded the age and sex
attributes when using this data set. Each patient has 200 data points,
see Figure \ref{fig:parkinsons}. Note: A hierarchical classification
of the patients into classes of the disease stage would provide a
better/more-appropriate test.

Due to the number of dimensions, the poset method did not work and it
does not find any distinction; however, the Compression, MST and
Kernel methods worked well. Notice that the Compression response
should be taken literally only for the interval which contains 100
points (the top left figure), because the bootstrap is built within a
single class. For larger window sizes we considered a reference window
of at most two patients. The Kernel method and the Compression method
are the most consistent measures. The MST method's rejection rate
increases as the number of points in each interval increases. The
Martingale method does not work well on this data for two reasons; the
series is built on intervals that are relatively short, and the
Martingale moving window $W$ will encounter $p$-value changes often
enough that it will become a normal behavior. In other words, we found
that the $p$-value distribution was indistinguishable from the
stochastic measure on the $p$-value, resulting in the Martingale value
being reset to 1 at regular intervals. In this case, a more {\em time
  sensitive} approach should be deployed for the $p$-value change to
let the Martingale value fluctuate naturally.

One final note about the Parkinson's disease data set. If we were
interested in finding similarity among patients, such as identifying
progression groups for the disease, such as early, advanced, and
final, we would not use a method that always finds differences between
the intervals above ---i.e., because each interval represent a
patient.

\mypar{Breast Cancer Data Set, 30 Dimensions} We consider two types of
breast cancer, benign and malignant, and sort the data by the number
of cases in each class. Visual inspection of the data in Figure
\ref{fig:breastcancer} shows that the series contains changes in both
the average and variance. We should expect all methods to catch one
type of change or the other. Despite the number of dimensions, the
poset method was able to identify the two classes with as few points
as 50. Notice that the Compression method has a single false negative
for the smallest window (of 50 points), because there was not enough
data for bootstrapping to allow the method to identify a difference in
the largest window.

\mypar{Summary} For series built from pre-classified data, we showed
that our methods provide a set of powerful tools. The changes in the
series are captured by the methods in a consistent manner for a different
number of window sizes and a different number of dimensions. Furthermore,
we can classify the changes of the series by noticing which methods
find the changes.

\subsection{Application to Hardware/Software Performance.}
\label{sec:hwcounters}
In this and the following section, we present examples of series
for which we want to find similarities, when there is neither
{\em a priori} classification nor known properties for the data.

Let us consider an application $A$ and an architecture $H$. Assuming
that we are interested in collecting a set of measures, such as processor
stalls due to cache misses or cycles per instruction, that are aimed
at measuring the execution times of $A$. If we sample these measures
during the execution of $A$ on $H$, then we are generating a
multi-dimensional time series. In this paper, we consider the six
measures described in \cite{CammarotaKDPVN2011}, and generate
6-dimensional series. These measures account for the cycles spent
waiting for resources, and directly account for performance losses.

Consider an application $A$ which is a property of Yahoo! We would
like to find another application, which exhibits similar performance
characteristics on the hardware architecture $H$. In this experiment
we used SPEC INT 2006, {\bf SPECINT}, which is part of a larger
application set \cite{Henning2006}. We generate multiple series by
concatenating the series generated by each application
$B\in SPECINT$ onto $A$. We reduce the problem to one of verifying
that a change exists in the series.

\mypar{Designing a Series} The set of applications that we run consist
of the Yahoo! application $A$ and the set of SPECINT applications
$SPECINT$. We run each application on the architecture $H$ with a
representative set of inputs, and collect performance statistics by
sampling the hardware counters. This process generates a series where
each point is an average result in an interval that is close to the
sample point. The series is the result of a deterministic, albeit
complex, process. The fact that there is a random component to the
series means that, although two series should be similar, it is
extremely unlikely that two series will be the same.

\begin{table*}%
  \tbl{A search property $C$ and a comparison with SPECINT: h264ref and perlbench are always different.
    \label{tab:spec2006}}{%
    \footnotesize
  \begin{tabular}{lrrrrrrrrrrrr}
    \hline
    {\bf   Method/App}& astar       & bzip2       & gcc         & gobmk       & hmmer       & libquantum  & mcf         & omnetpp    & sjeng       & xalancbmk \\ \hline\hline
    Compression       & diff        & diff        & diff        & diff        & diff        & diff        & diff        & diff       & diff        & diff  \\ 
    MST               & diff        & {\bf equiv} & {\bf equiv} & {\bf equiv} & {\bf equiv} & diff        & {\bf equiv} & diff       & {\bf equiv} & diff  \\ 
    POSET             & {\bf equiv} & diff        & {\bf equiv} & {\bf equiv} & {\bf equiv} & diff        & diff        & {\bf equiv}& diff        & {\bf equiv}  \\ 
    Martingale        & diff        & diff        & {\bf equiv} & {\bf equiv} & diff        & {\bf equiv} & {\bf equiv} & diff       & {\bf equiv} & {\bf equiv}  \\ 
    Kernel Lin.       & diff        & diff        & diff        & diff        & diff        & diff        & diff        & diff       & diff        & diff  \\ 
    Kernel            & diff        & diff        & diff        & diff        & diff        & diff        & diff        & diff       & diff        & diff  \\ 
    \hline
  \end{tabular}}
\end{table*}

\mypar{Value of Identifying Similar Series} If we can find an application
$B\in SPECINT$ that is similar to the application $A$ for the architecture
$H$, we may consider $B$ as being representative of $A$ in general.
When another architecture $I$ is introduced that improves the
performance of $B$, we may conjecture that $I$ will also improve $A$,
and vice versa. Notice that it is not necessary to run either $A$
or $B\in SPECINT$ on a prospective new architecture, because
manufacturers will provide results for SPECINT.

As a pruning device, if $I$ does not improve $B$, then $I$ will not
improve $A$ either ---i.e., very likely $I$ will not improve $A$
either.

\mypar{Constructing the Series} To facilitate the tuning of the tools,
we followed the same process described in Section \ref{sec:synthetic}.
That is, $A$ and $B$ produce two series $T_A$ and $T_B$, we create a
new series $T_A+T_A+T_B$ (or $T_B+T_B+T_A$ in case $B$ has fewer points
than $A$), where $+$ signifies concatenation.

\singlefigure{0.95}{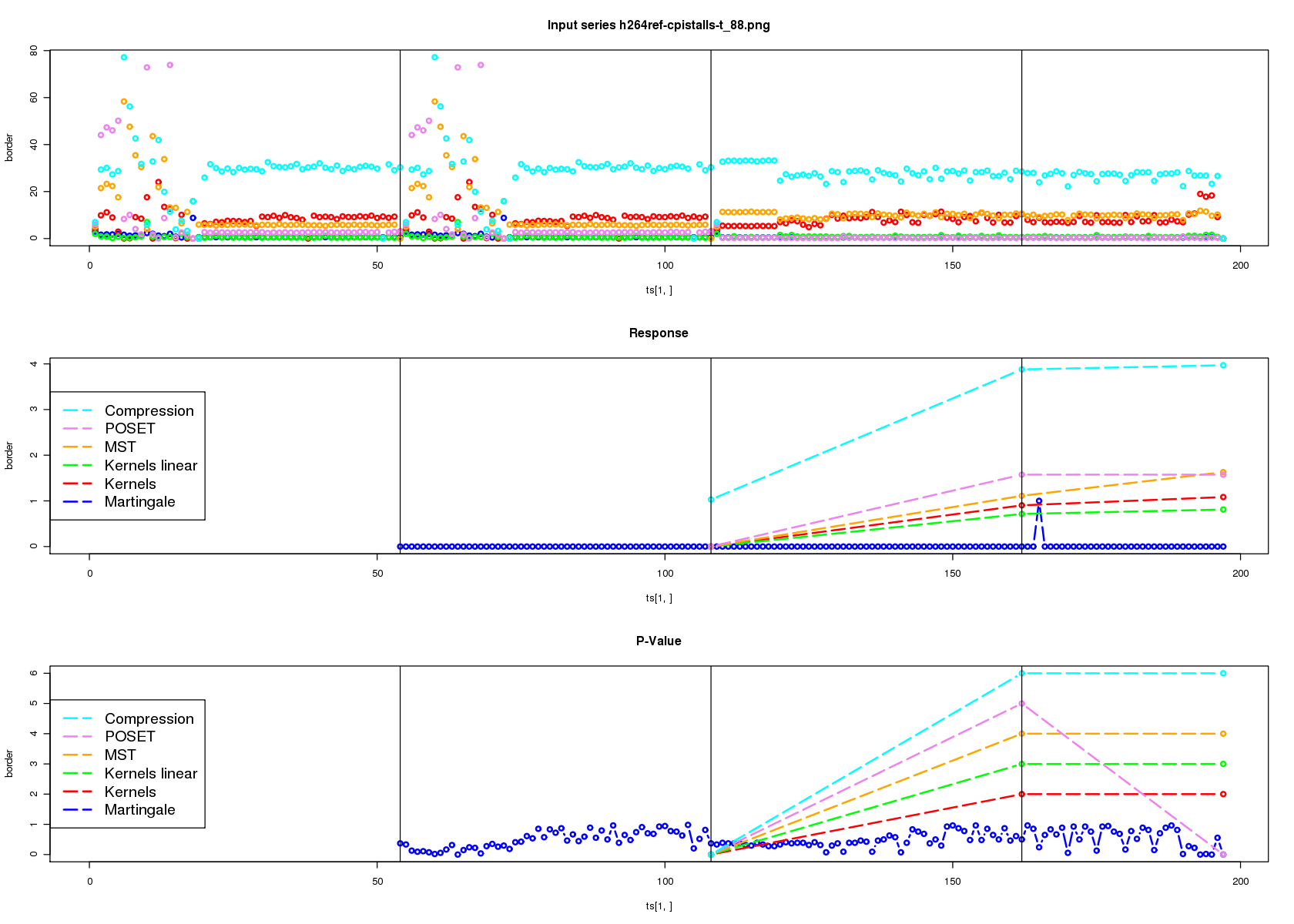}{Comparison with {\em h264ref}
	applications by all methods.}{fig:h264ref}

In Figure \ref{fig:h264ref}, we present an analysis using our methods
where we compare our application $A$ (the tail of the series) with the
$SPECINT$ applications (the head of the series), identified in the
literature as {\em h264ref}. All the methods identified the two series
as different.

In Table \ref{tab:spec2006}, we present the final comparison results
for all the benchmarks. For this paper, we attach no specific importance
to the application $A$, as long as it is not from $SPECINT$. In
principle, we know that a discriminative method would differentiate
$A$ from any application in SPECINT. In fact, the Kernel method and
the Compression method, which take into account the evolution of the
series over time, find no similarity (see Section \ref{sec:kernel-linear}
for a reminder of the linear Kernel method).

For the Martingale, MST, and poset methods, the evolution of the
series over time is relative; and by contrast these methods are more
sensitive to the frequency of the events. As a result, these methods
find similarities that may not be apparent in the series. From our
observations, we find that our application $A$ is similar to {\em
  gcc}, {\em gobmk}, and less similar to {\em mcf}, {\em sjeng}, and
{\em xalancbmk}, because at least two different methods suggests a
close similarity.

%\Doublefigure{0.70}{Historical_Data/nasdaq_csv_6000_250}{Historical_Data/nasdaq_csv_8750_250}{NASDAQ
%  IXIC quote series where the 24-th and 25-th year are similar (above)
%  and the 45-th and 50-th year are similar (below). The index series
%  is not normalized.}{fig:nasdaq}

\mypar{Summary} In this section, we introduced series that represent
the performance of software applications and the interaction with the
hardware that runs them. If we are interested in identifying the
evolution of the series over time, then we should consider the Kernel
method or the Compression method as candidates. If we are interested
in identifying the similarity of distributions, then we should consider
the MST or poset method. If we are interested in interchangeability,
then the Martingale method is more appropriate. Here we used all the
methods and a quorum to infer similarity among series.

\subsection{Stock Market Quotes} 
\label{sec:stockmarket}    
\begin{comment}
\Doublefigure{0.85}{Historical_Data/nasdaq_csv_6000_250}{Historical_Data/nasdaq_csv_8750_250}{NASDAQ
  IXIC quote series where the 24-th and 25-th year are similar (above)
  and the 45-th and 50-th year are similar (below). The index series
  is not normalized.}{fig:nasdaq}
\end{comment}
\Doublefigure{0.85}{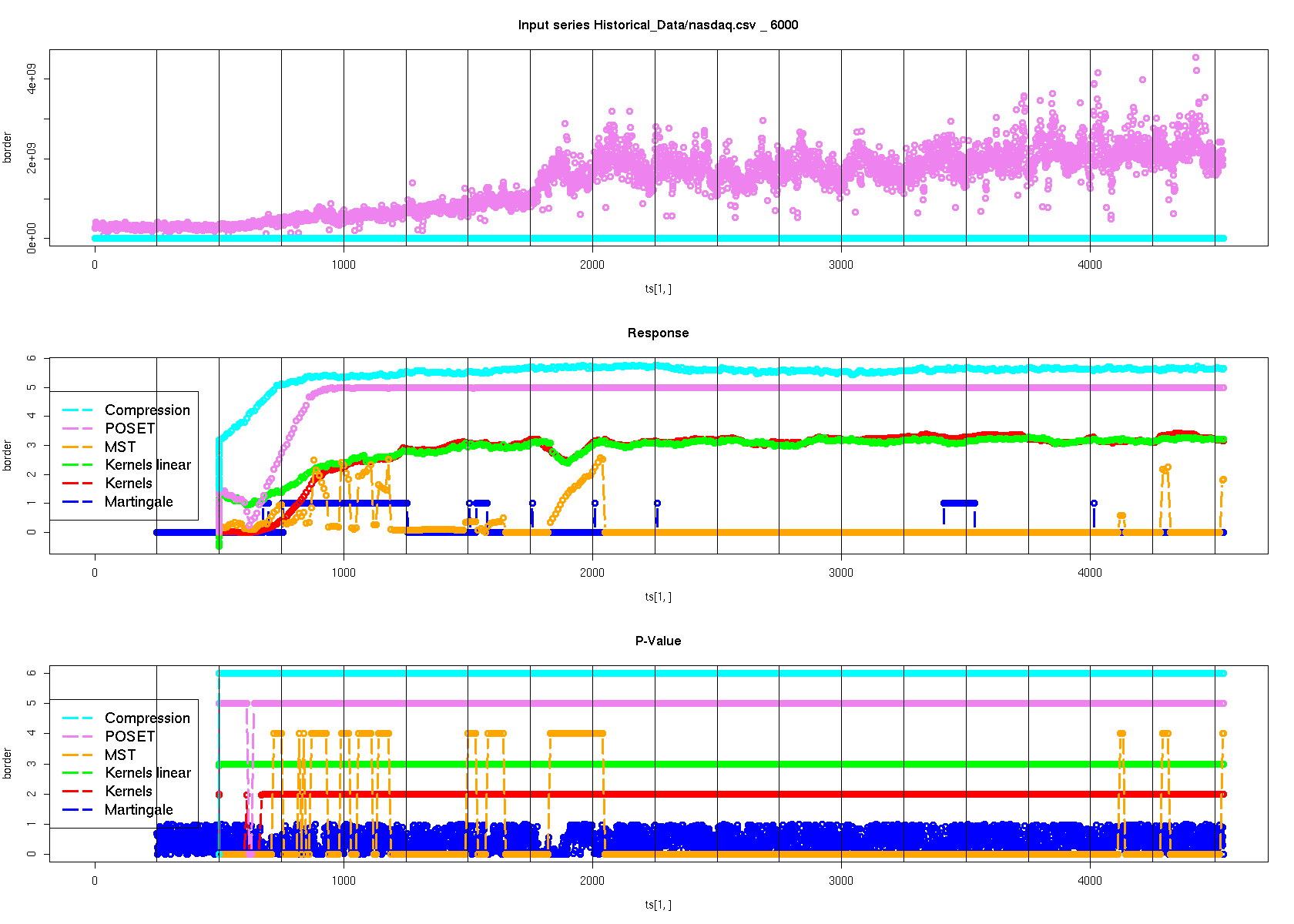}{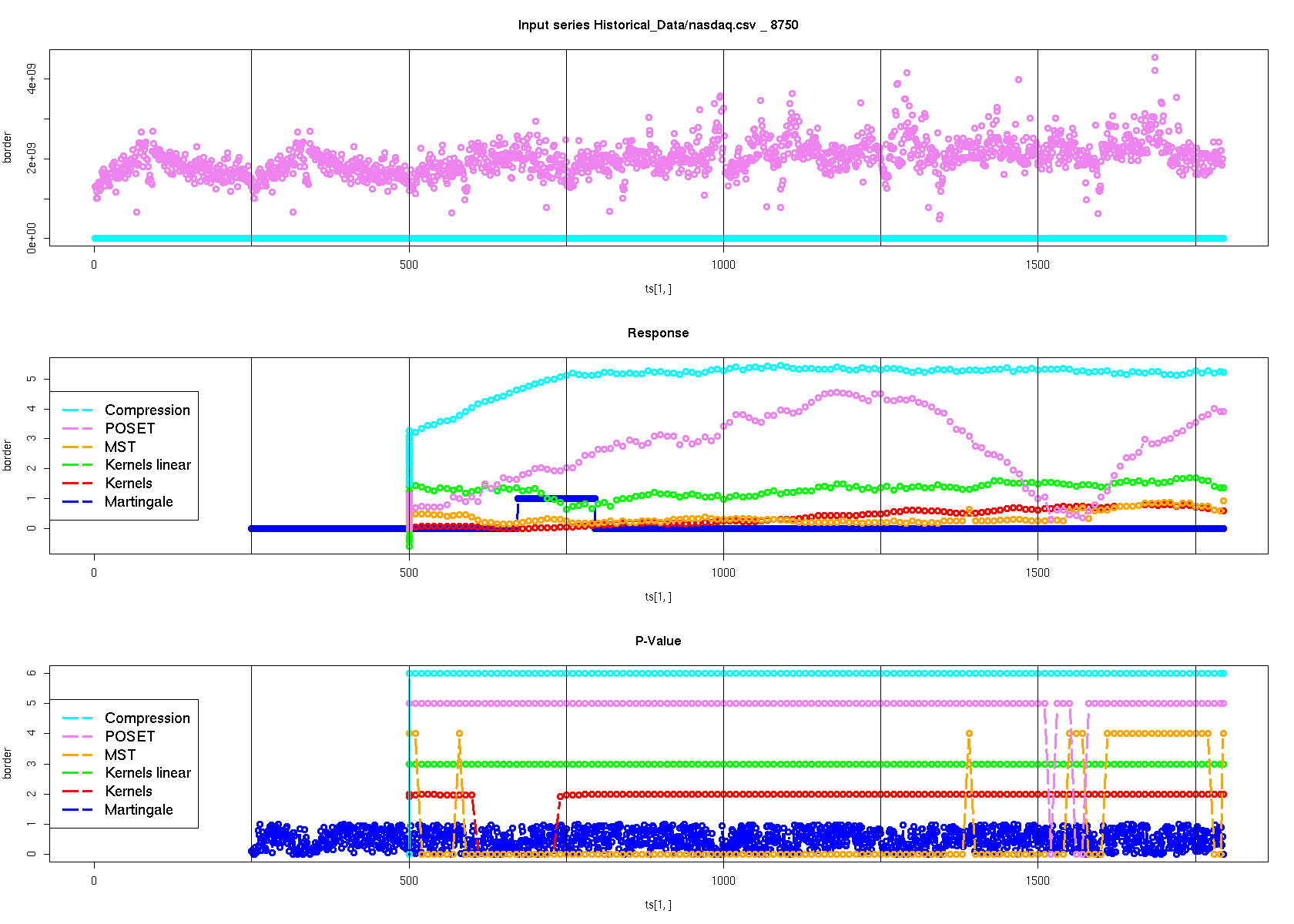}{NASDAQ
  IXIC quote series where the 24-th and 25-th year are similar (above)
  and the 45-th and 50-th year are similar (below). The index series
  is not normalized.}{fig:nasdaq}

In this section, we consider historical stock quote data as a
series generated by a stochastic process. A series of quotes is a
6-dimensional series with the following components: open price,
high price, low price, close price, volume, and adjusted price.
Each point in the series represents a single trading day (ignoring
holidays and weekends), and we consider it a continuous series.

We considered four ticker symbols: YHOO (Yahoo!), AAPL (Apple), GOOG
(Google), and NASDAQ (index). The first three provide a historical
picture of companies in the technology sector, and the last one is an
index, which {\em contains} these companies. We are interested in
exploring whether a stock's performance has a repeating pattern, and
then whether a year can repeat itself.

In the previous examples, we performed the interval analysis by
specifying the window length $|R|$ and $|W|$ and shifted the moving
window $W$ in steps of size $|W|$. In this experiment we performed a
scan analysis. We consider the full historical series, and use the
first year as the reference window $R$ and for tuning purposes. The
moving window $W$ has the same length as $R$, and is shifted by small
increments of ten points to scan the series. This results in $W$
scanning the series by two-week interval steps. Once the scan is
completed, we remove the first year and, repeat the process for the
cropped series.  In this way, we tested the entire series. Notice that
the Martingale method always performs a scan of the series one point
at a time.
\begin{comment}
\singlefigure{0.95}{Historical_Data/apple_csv_2250_250}{AAPL quote
  series: the 9-th year is similar to 17-th and 20-th year}{fig:appl}
\end{comment}
\singlefigure{0.95}{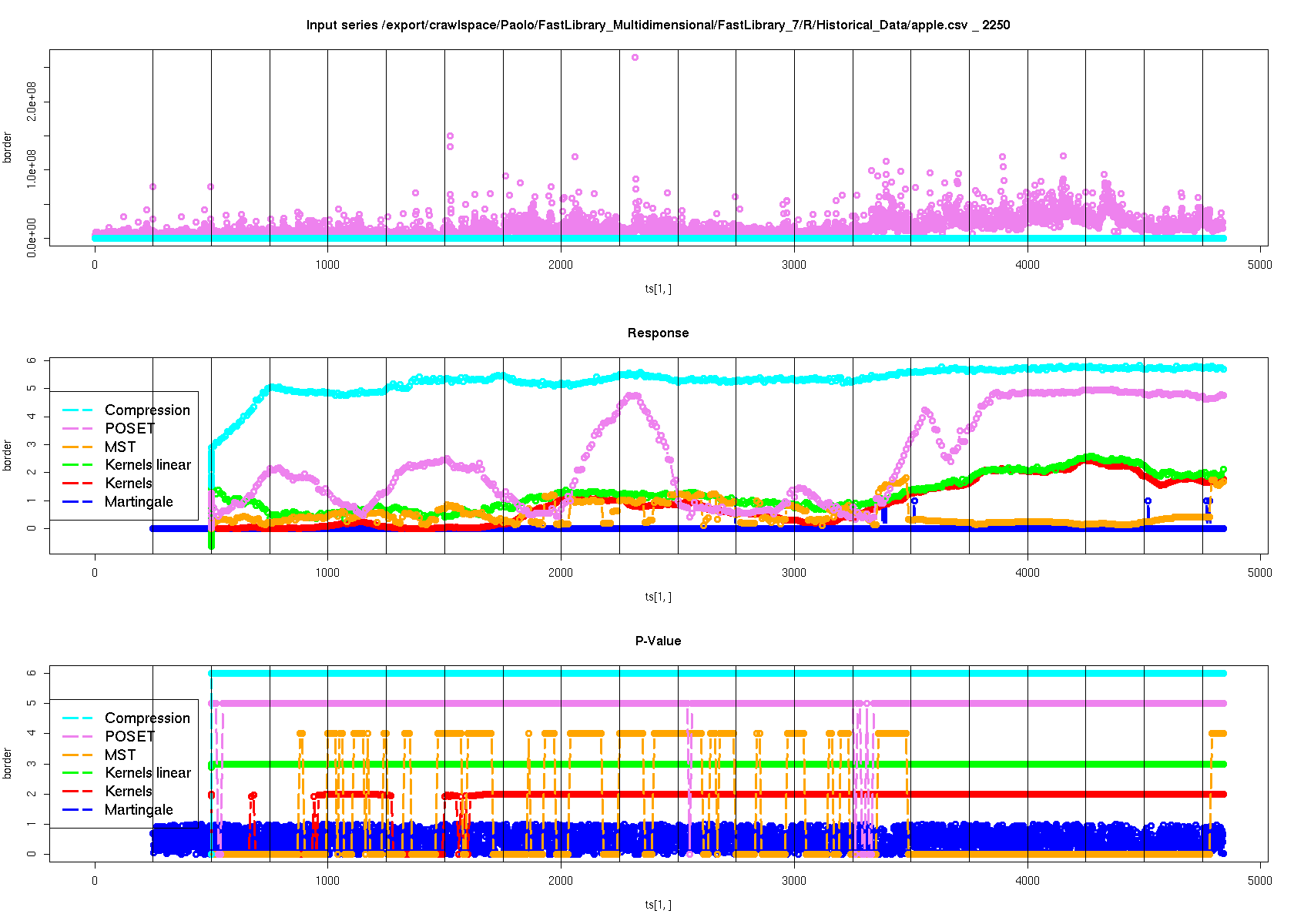}{AAPL quote
  series: the 9-th year is similar to 17-th and 20-th year}{fig:appl}

Our goal is to find similarity in order to prove that there is a
repeating pattern or at least {\em identical} years. We replace the
exact dates in favor of referring to the number of years since the
first stock offering was traded. Due to the small number of
dimensions, the poset method should provide a reasonable set of
similar years.  We do not normalize the series in order to take
account of inflation or other dollar valuations. Thus, we can identify
recession years, when a quote drops to previous values, and recovering
years. For example, an index such a NASDAQ is bound to increase during
its lifetime as a result of adding new stocks or accounting for the
value of the dollar. For such a quote, we are most likely to find only
recession years. By contrast, a quote such as AAPL may exhibit periods
of growth after a recession.

\mypar{IXIC NASDAQ} The NASDAQ index presents two pairs of years with
similarity that coincide with recessions. We have found one period of
no growth between the 24--25-th years and another similar recession
between the 45-th and the 50-th years. These two findings indeed match
two weak market periods. In fact, the 50-th interval coincides with
the 2008 fiscal year. In Figure \ref{fig:nasdaq}, we present the
experimental results.

\mypar{APPL} We found at least two similar years, both representing
a period before growth of the company (9-th and 21-th). In Figure
\ref{fig:appl}, we present the experimental results.

\begin{comment}
\Doublefigure{0.85}{Historical_Data/yahoo_csv_1500_250}{Historical_Data/yahoo_csv_2250_250}{YHOO
  quote series: 6-th and 14-th year are similar (top) 9-th and 10-th
  year are similar (bottom)}{fig:yhoo}
\end{comment}
\Doublefigure{0.85}{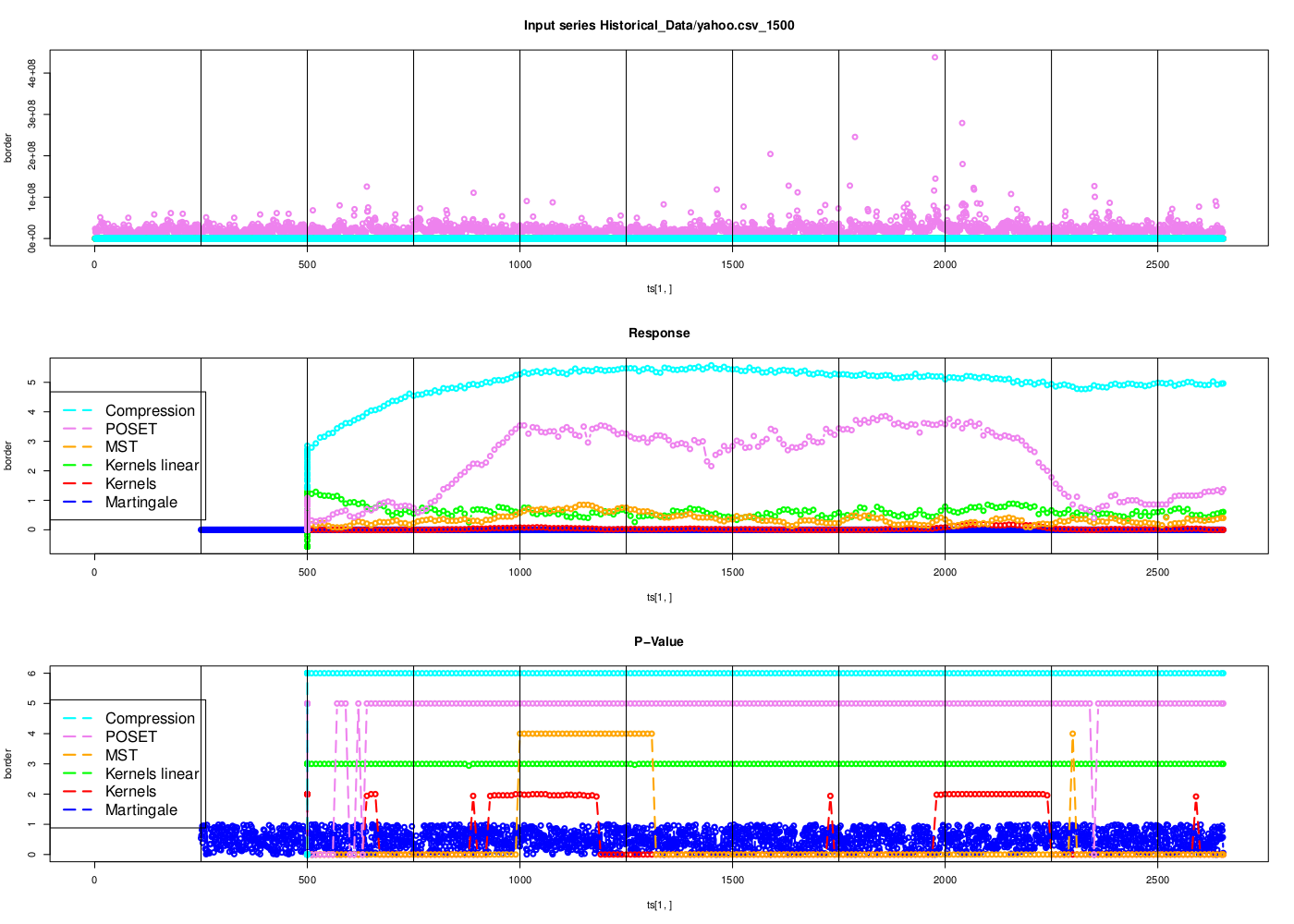}{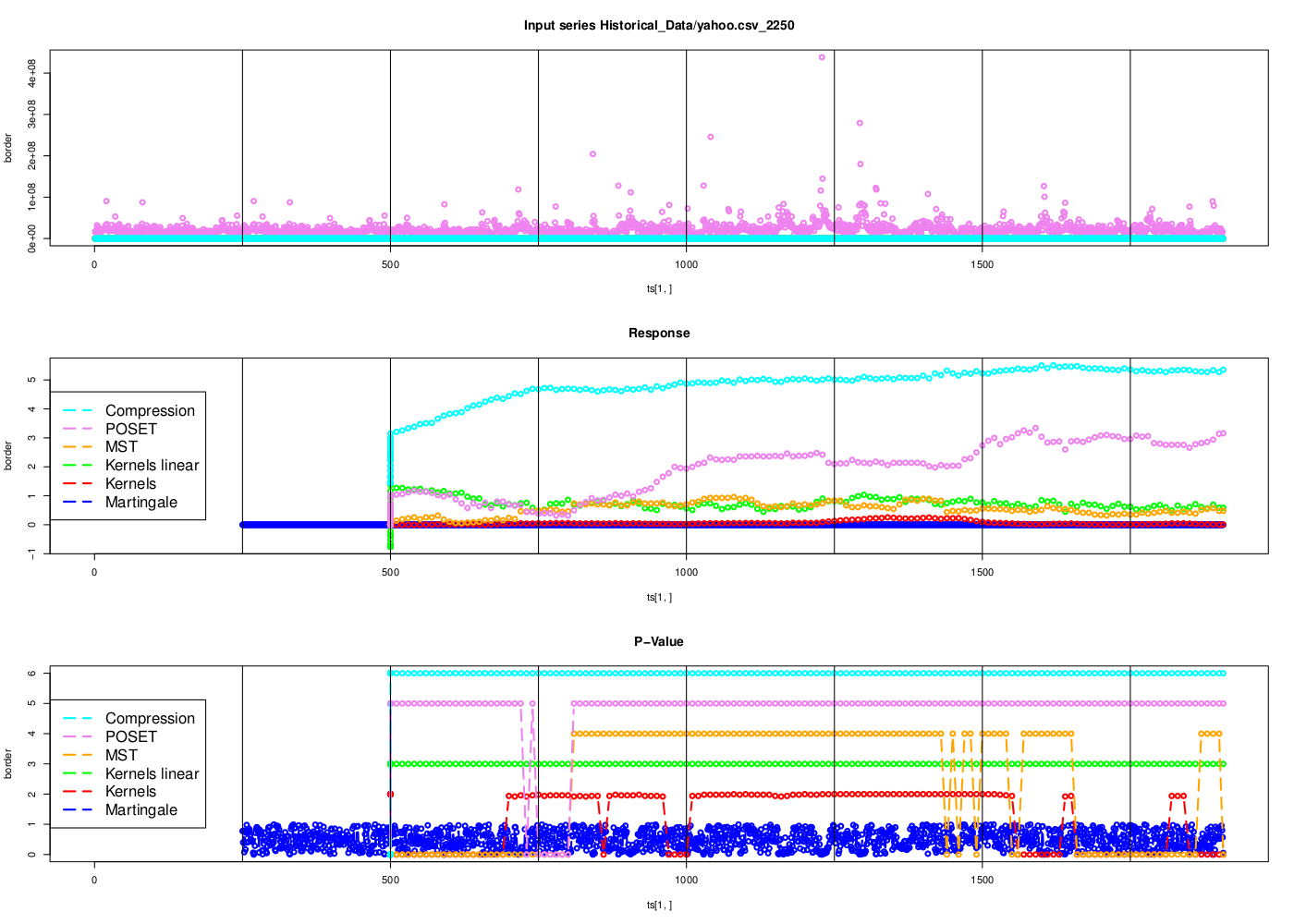}{YHOO
  quote series: 6-th and 14-th year are similar (top) 9-th and 10-th
  year are similar (bottom)}{fig:yhoo}

\mypar{YHOO} We found two similar years. The 6-th and the 14-th years
represent a rebound of the quote right after the so called {\em tech
  bubble} and {\em housing bubble} burst. The 9-th and the 10-th
represent a stable period for the company. We present the results for
both these time series in Figure \ref{fig:yhoo}.

\mypar{GOOG} As a result of a constant growth pattern, we were not able to
find similar years.

%\newpage

\newpage
\section{Single-Dimensional Experimental Results}
\label{sec:uniexperimental} 

% Paolo Sept 8 2011 10:54 pm

In this section, we validate the power of our CDF-based measures and
compare with the state of the art methods. The comparison is based
upon the performance on single-dimension series. For comparison
purpose, the measures are organized into two sets:

\begin{itemize}
\item[{\bf Standard}:] This set consists of the following five measures:
  Wilcoxon--Mann--Whitney, t--test, Kolmogorov--Smirnov, $\phi$, and
  $\Xi$.

\item[{\bf Extension}:] This set consists of the following nine measures:
  Kullback--Leibler (symmetric), Jin-L, Jensen--Shannon, $\chi^2$,
  Hellinger, Variational, Cram\'er--von Mises, Minkowsky, and
  Euclid. Among these methods, the Cram\'er--von Mises has been
  previously used in the literature; however, to the best of our
  knowledge, none of the methods have been used for series in the
  continuous domain $\R$.
\end{itemize}
We shall show that our extension measures are comparable to the
standard measures, and they may be used separately or together. The
overall measure will permit a more effective statistical test for
series with real values.

\subsection{Setup}
\label{sec:set-up} 

We apply the standard and extension measures, separately and together,
on a set of series. The series are generated by repeating the
following process 1000 times:
\begin{itemize}
\item The time series is composed by a set of consecutive windows. The
  number of windows is randomly chosen as ${\bf M}\in[2,20]$. The
  series is composed of at least two windows and at most twenty.

\item The window size is randomly chosen as $T\in [1,10]*100$, without
  any bias to a particular window size. That is, a series is composed
  of $M$ windows of equal size; where every windows has a minimum size
  of 100 points and a maximum of 1,000.

\item A window $E$ with the same distribution as the first window $R$
  is embedded into the series in $[2,M]$ randomly.

\end{itemize}
The goal of the tests is to find $E$ in the time series and to reject
every other window, by scanning the created series using a moving
window $W$. Three types of series are generated to reflect changes in
the average, changes in the variance, and changes in both the average
and the variance.

\mypar{Change in Average} Using a normal distribution generator ${\cal
  N}(0,10)$, we determine the reference average $m_0$ and variance
$v_0$. We generated two windows $R$ and $E$ using either a normal
distribution ${\cal N}(m_0,v_0)$ or a uniform distribution
$U(m_0-v_0,m_0+v0)$.  For every other window, we selected at random
$m_i = m_0+r*\frac{m_0}{i}$ where $r=\pm 1$, switching its sign with
equal probability. Therefore, as $M$ increases, the series becomes
longer, the tailing window of the series tends to be closer to the
reference $R$. Using a 20\% disagreement threshold, the system
recognized the similarity with a sharp positive pulse, correctly
flagging all the other intervals as different.

\mypar{Change in Variance} We generated two windows $R$ and $E$ using
either a normal distribution or a uniform distribution, as described
previously, using $m_0$ and $v_0$. For every other window, we selected
at random $v_i = v_0+r*\frac{v_0}{i}$ where $r=\pm 1$, switching its
sign with equal probability. Therefore, as $M$ increases, the series
becomes longer, the tailing window of the series tends to be closer to
the reference $R$. Using a 20\% disagreement threshold, the system
recognized the similarity with a rather slow positive pulse; however,
it also recognized other intervals as similar, resulting in false
positives.

\mypar{Changing Both Average and Variance} We generated two windows
$R$ and $E$ using either a normal distribution ${\cal N}(m_0,v_0)$ or
a uniform distribution $U(m_0-v_0,m_0+v0)$. For every other window, we
selected at random $m_i$ and $v_i$ from ${\cal N}(0,10)$. Using a 20\%
disagreement threshold, the system recognized the similarity with a
sharp positive pulse, correctly flagging all the other intervals as
different.

For the experiments where we changed only the average or the variance,
the test series converged to the reference window. For large enough
$M$, the measures would find it harder and harder to distinguish them.
Yet, there was only one window in the series that had the same
attributes as the reference.

\mypar{Moving Window} The moving window $W$ scans the series. The $W$
moves by a fixed size step of 100 epochs/points. Note that 100 is the
smallest windows size possible, thus the sliding window $W$ is always
contained into a classified window with a well defined statistical
properties.

\mypar{Disagreement} We quantified the number of times the system
recognized two windows as {\em the same} for different level of
disagreement (0.1, 0.2, 0.3, 0.4., 0.5., 0.6., 0.7, 0.8, 0.9, 1). That
is, with a disagreement of 0.1, two windows are recognized as {\em
  equal} if, at most, 1 out of 10 measures does not say so, while the
remaining measures do.

\mypar{Matching and Found} We define a {\em match} as having occurred
when the system produces a positive response for two windows $R$ and
$E$, which we know are equal. The golden standard for matches is 1000,
which is equal to the number of windows that were drawn from the same
distribution when the series was generated. We define {\em found} as
having occurred when the system produces a positive response
independent of the position in the series, because the sliding window
has a fixed step of 100 epochs instead of the effective window
size. For example, if $|W| = 1,000$, thus the series is composed of
widows of size 1,000, there could be one match and 19 founds, because
$W$ will lie on top of $E$ once, but it will overlap with $E$ about 19
times.

\subsection{Summary Results}
\label{sec:summary} 

We define the number of times an error was committed as
$(found-1000)+(1000-matches)$. That is, the combination of false positives
and false negatives, where false positives are when the system identifies
two windows as being equal minus the golden standard, and false negatives
are the number of misses made by the system.

\doublefigure{0.80}{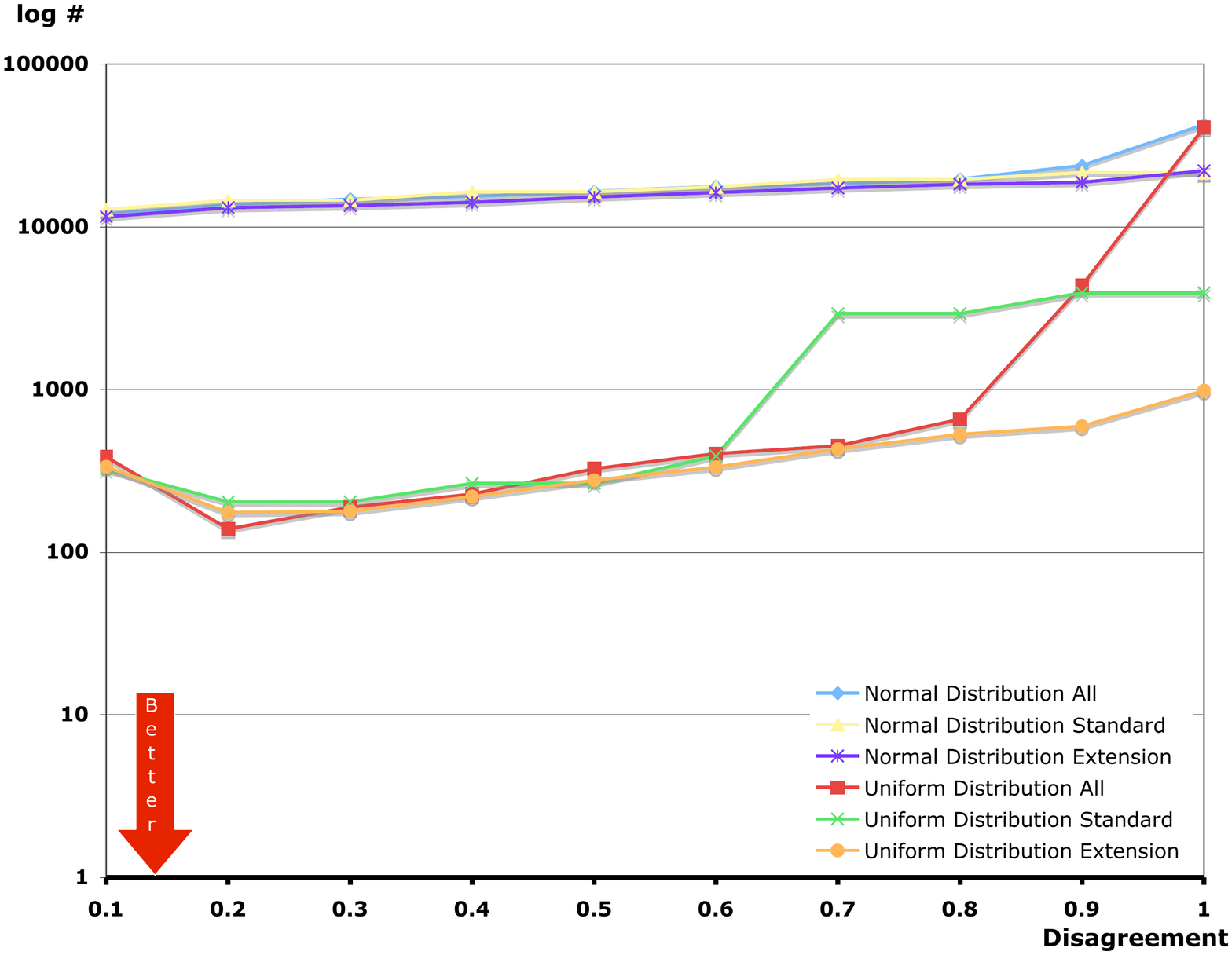}{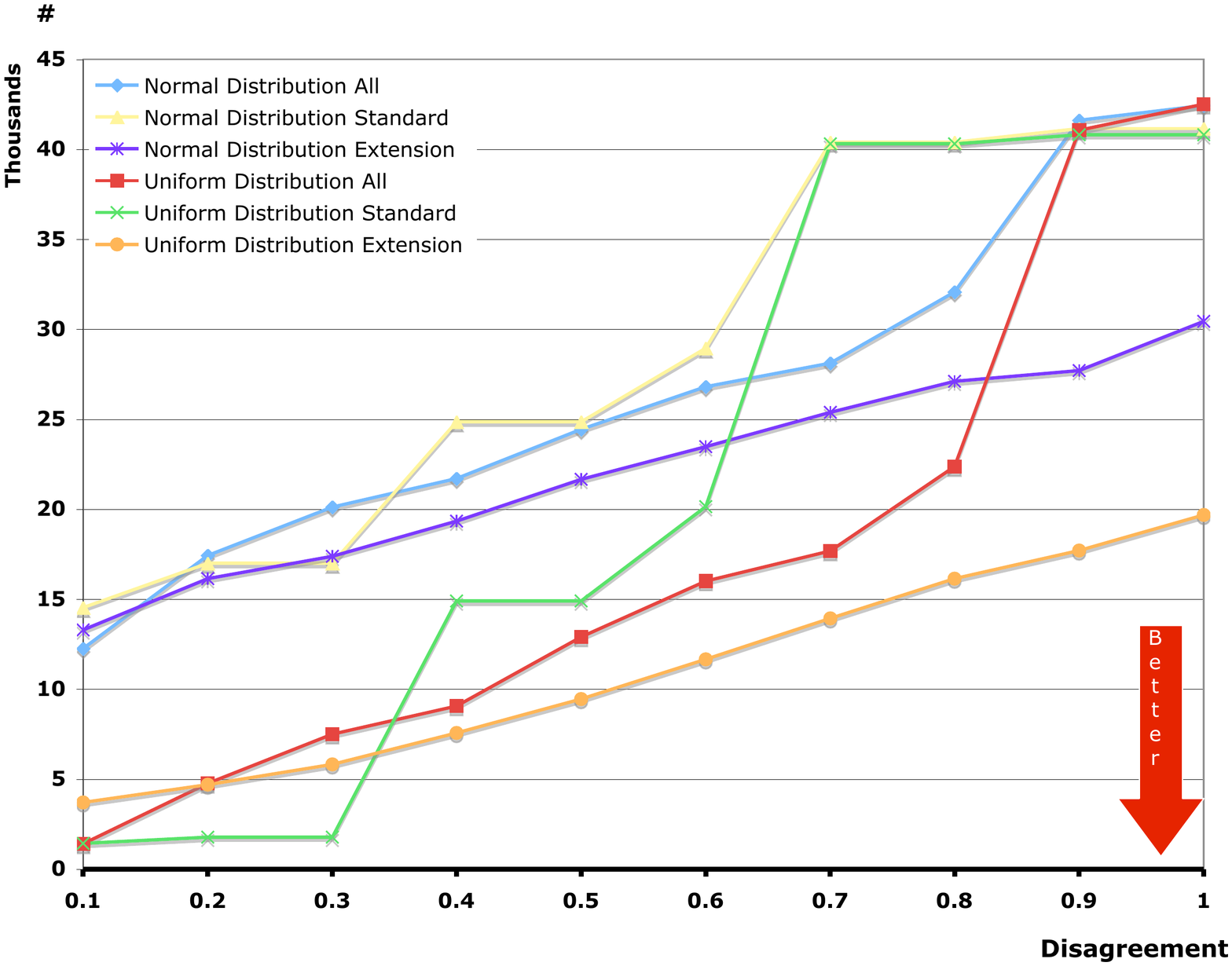}{Sum of false positives and false
negatives for series with change in average only (above) and change in
variance only (below).}{fig:mverror} 

In Figure~\ref{fig:mverror}, we present the error for change in
average only and change in variance only. Notice that for a change in
average only, the combination of both standard and extension measures
produces the smallest error, resulting in a Gestalt effect. This
effect alone justifies the addition of our methods.

\singlefigure{0.80}{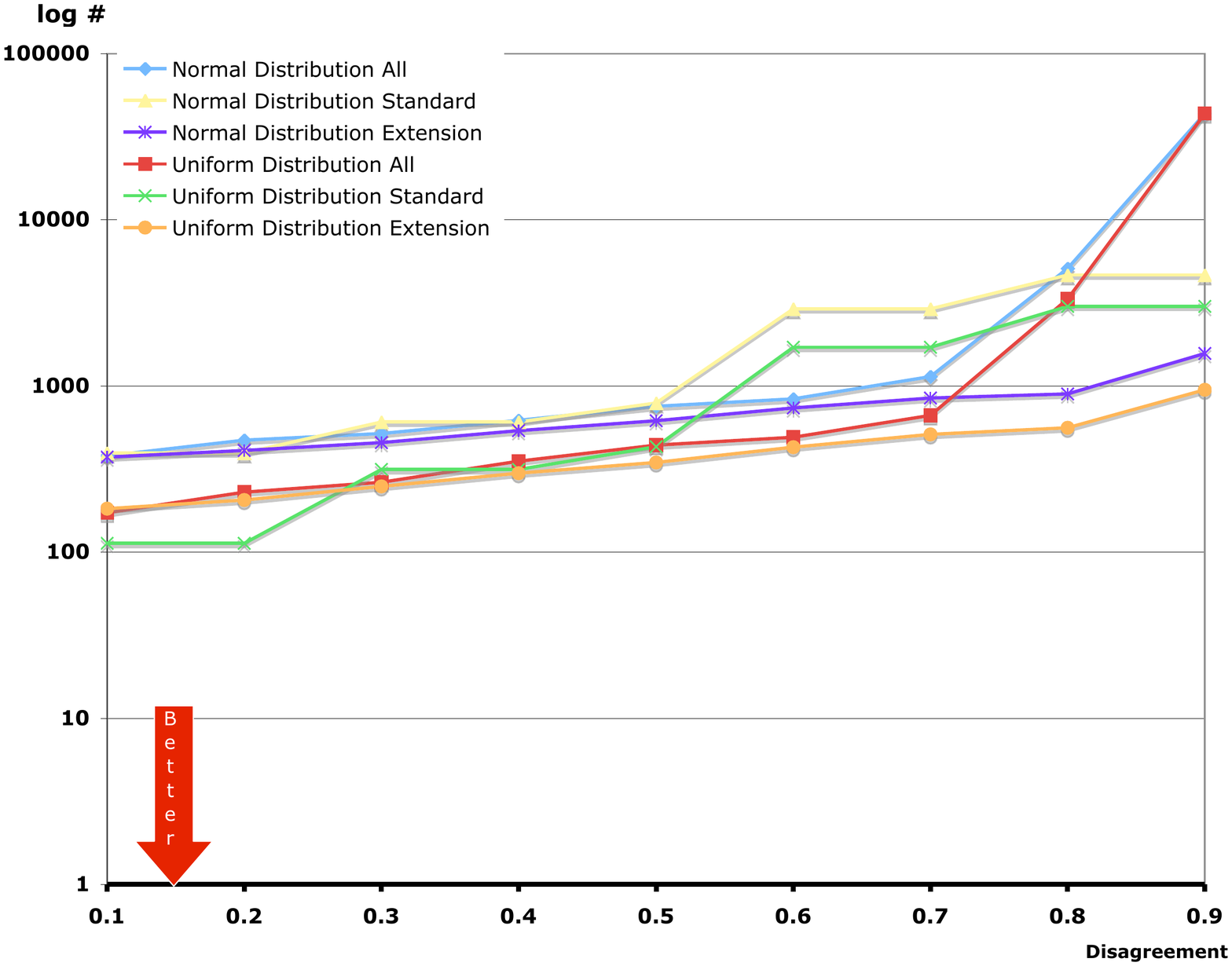}{Sum of false positives and false
negatives for series with change in average and variance}{fig:mandverror}

In Figure \ref{fig:mandverror}, we present the error for changes in
both the average and variance. Notice that the standard approach has a
smaller error over the range of 0.1 and 0.2 for uniformly distributed
series. However it has the same error as our extension methods for
series using normal distributions. Overall, our extensions work well,
performing consistently as a function of the disagreement factor.

Note: our new CDF based methods are a simple to deploy and a good
addition to the standard measures. As we showed for multidimensional
series, we present methods that add on top of the existing methods
without replacing them or be overshadowed by them. These methods are a
contribution to the field.

\subsection{Results for Change in Average}
\label{sec:M}

In Figure~\ref{fig:mstatistics}, we show that our extension methods
perform better than the standard methods, which means our methods
are more sensitive to detecting changes in the average.

\doublefigure{0.80}{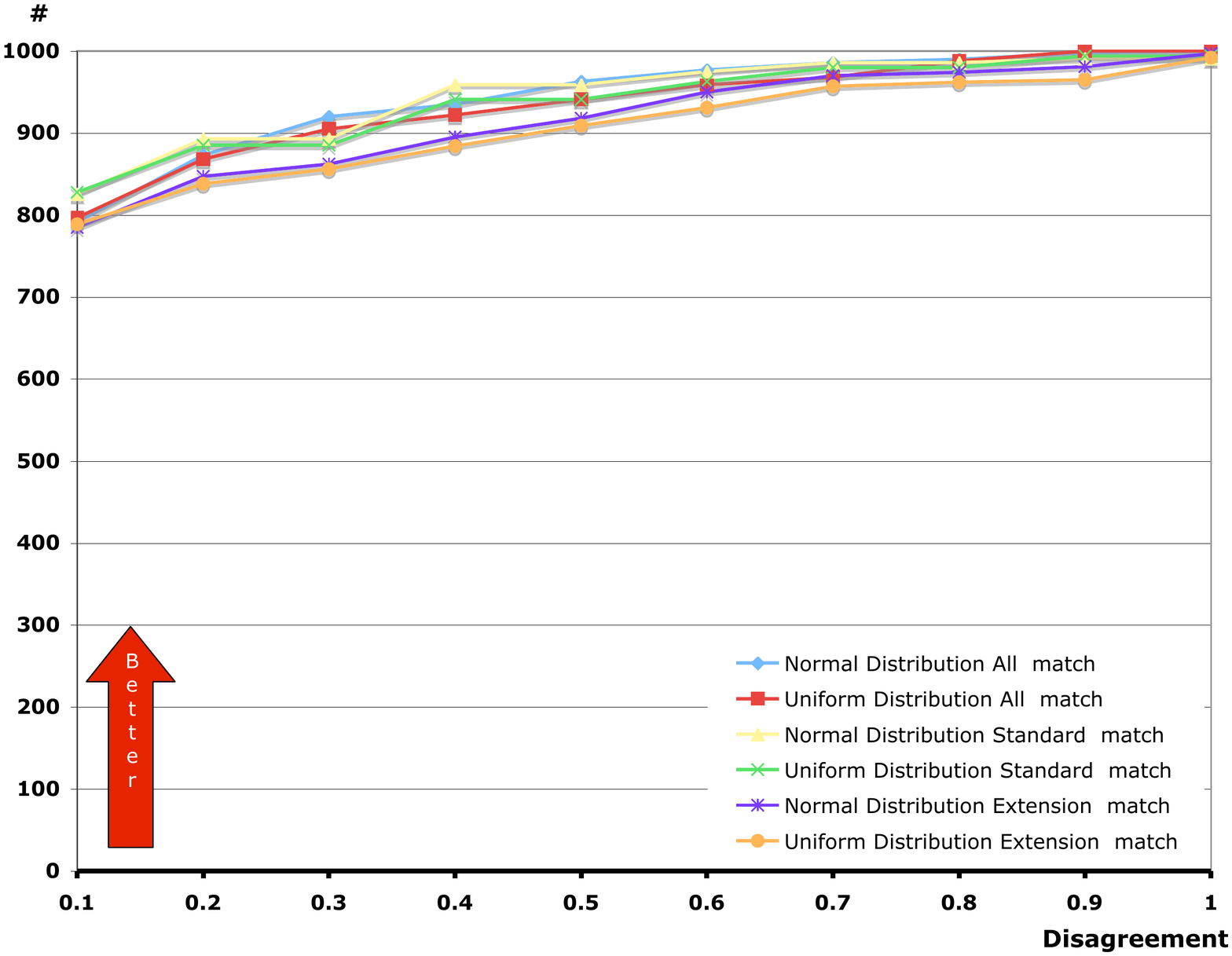}{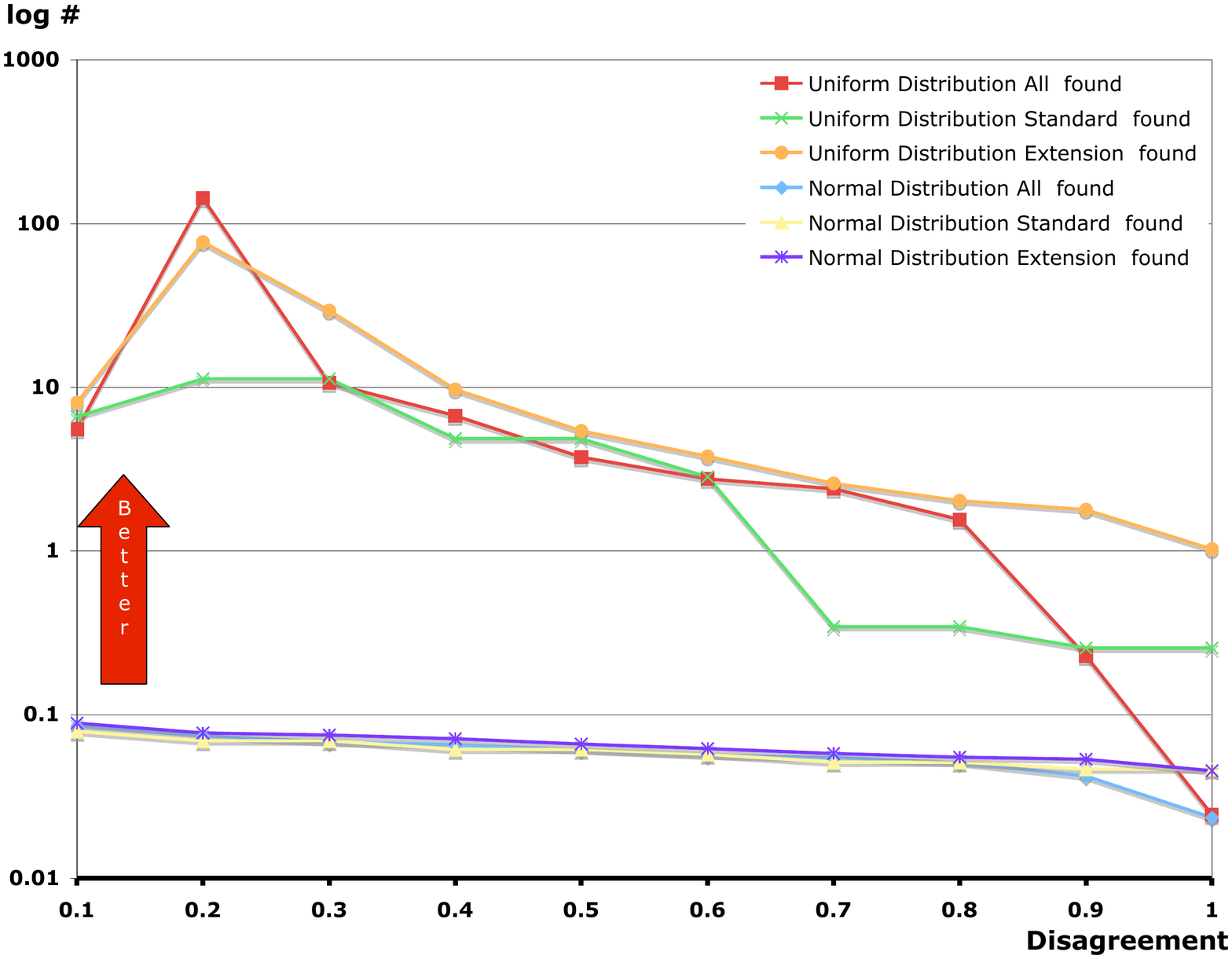}{Performance on series
	with changes in average: matches (above) and
	$\log\big(\frac{1}{|1000-Found|}\big)$ (below).}{fig:mstatistics}

\subsection{Results for Change in Variance}
\label{sec:V}

In Figure~\ref{fig:vstatistics}, we find that the standard measures
offer a more sensitive tool for detecting true similarity (or
conversely an anomaly).

\doublefigure{0.80}{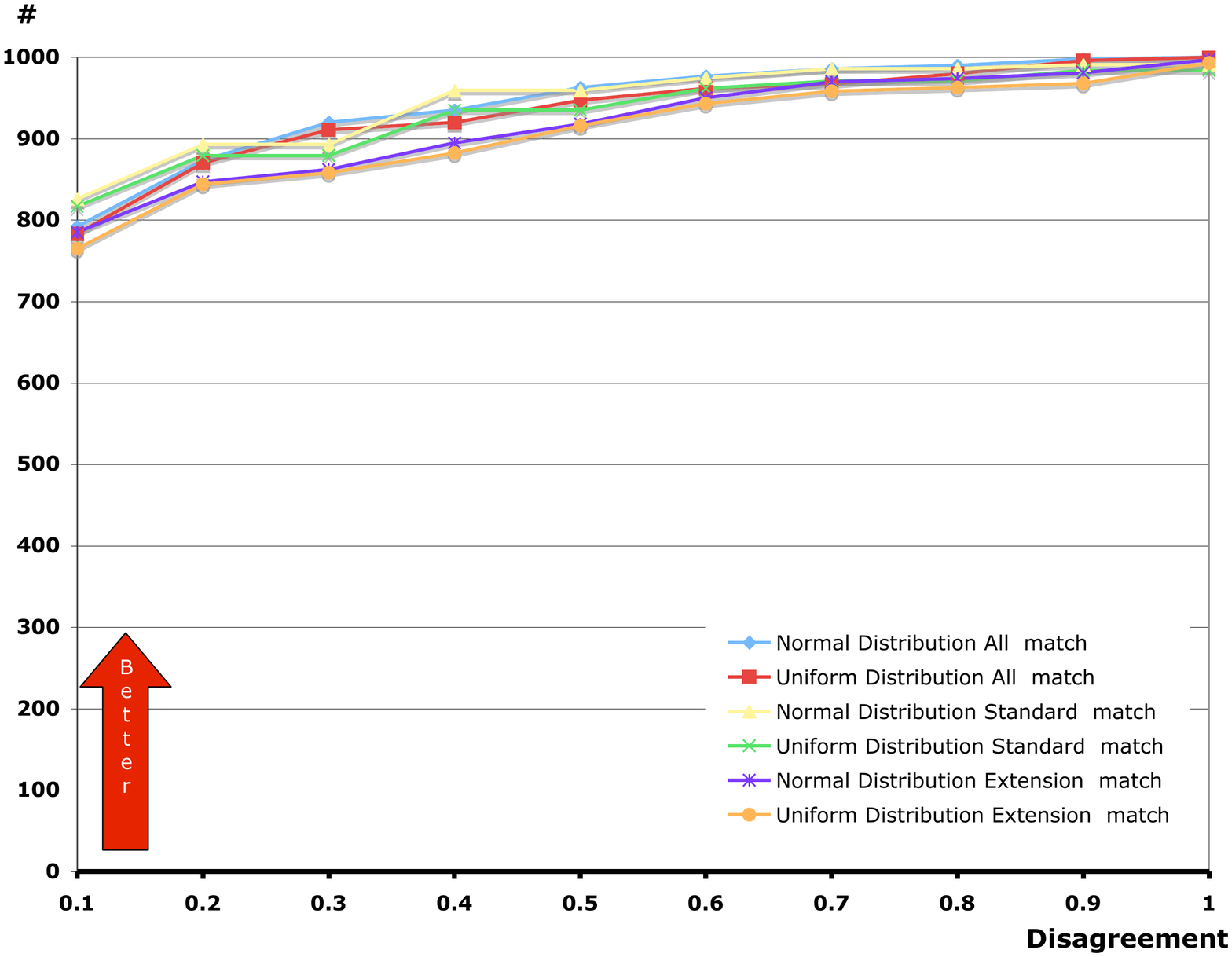}{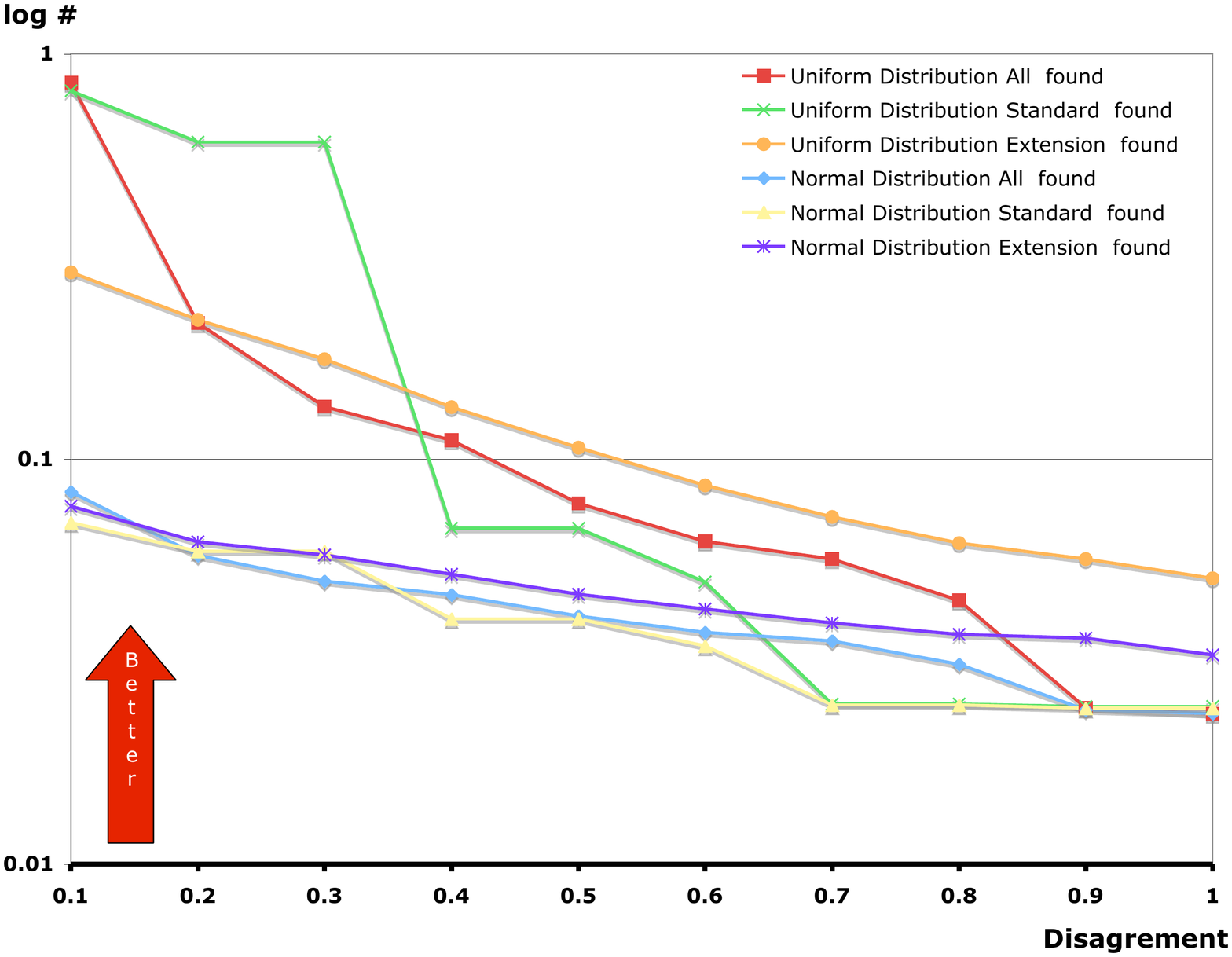}{Performance on series
	with changes in variance: matches (above) and
	$\log\big(\frac{1}{|1000-Found|}\big)$ (below)}{fig:vstatistics}

\subsection{Results for Change in Average and Variance}
\label{sec:mandv}

In Figure~\ref{fig:mandvstatistics}, we present the number of perfect
matches and the number of found matches by all approaches. The
standard methods appear to out-perform the extensions when using the
right balance of consensus and disagreement for series built using
a uniform distribution. However, our extension methods deliver a
predictable and ultimately better performance for inputs drawn from
a normal stochastic process.

\doublefigure{0.80}{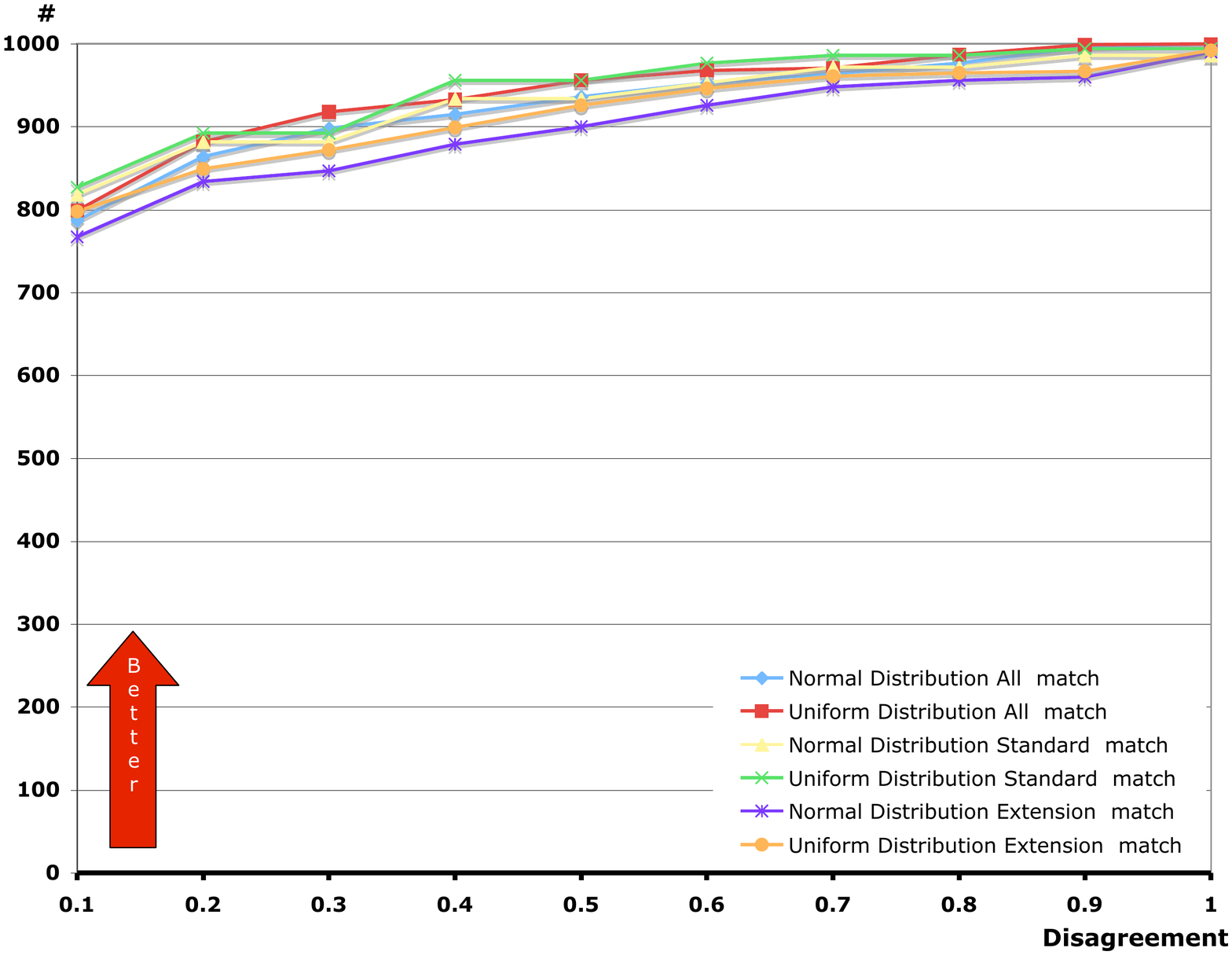}{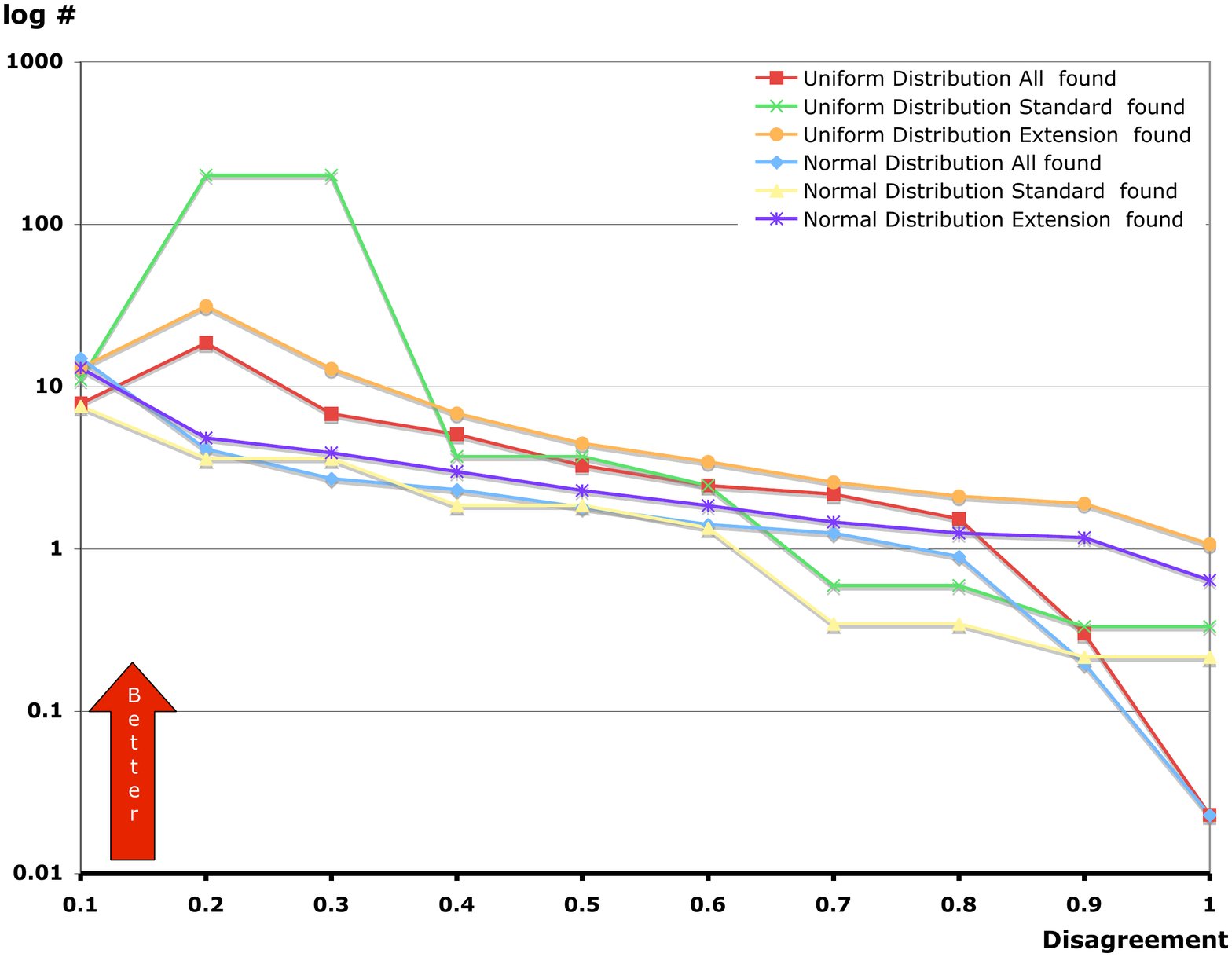}{Performance on
	series with changes in average and variance: matches (above) and
	$\log\big(\frac{1}{|1000-Found|}\big)$ (below).}{fig:mandvstatistics}

\section{Conclusions}
\label{sec:conclusions}

We have presented a survey of the different methods for detecting
stochastic change in multi-dimensional series. This included a
detailed examination of four types of promising methods: Kolmogorov's
information measure, the Martingale measure in conjunction with
conformal prediction, kernel methods for the computation of the
maximum mean discrepancy measure, and topological-order based methods
that are built on the comparison of empirical distribution
functions. To complete this work, we proposed new measures for the
comparison of empirical distributions previously applied only to
single-dimension series, and applied them to multi-dimensional series.

We have shown that each measure is different and exploits different
properties of the data. As a result, each measure provides valuable
information about the data. Although the measures range in
computational complexity, there are measures which are relatively
fast, such as the linear kernel method at $O(N)$, the poset-based
method and compression method at $O(N\log_2N)$, and finally the
Martingale, kernel, and MST methods at $O(N^2)$. Furthermore, we show
that this variety of methods and their inherent capabilities is also
apparent for one--dimensional-series methods, which have a much longer
history.

This paper is intended to be a self-contained work that includes a
survey of existing methods and an original contribution. The authors
would like to emphasize that although this is long standing research
field, mature and widely used--applied, it is far from being solved
having a never ending, always interesting research depth and still
offering surprising new results.

\begin{acks}
The authors would like to thank: Prof. Daniel Kifer of the Department
of Computer Science and Engineering Penn State University for the
discussions about single-dimension series stochastic distances,
Prof. Vladimir Vovk of the Department of Computer Science University
of London for the discussions about conformal prediction, Prof. Paul
Vitanyi of University of Amsterdam and Rudi Cilibrasi Ph.D. for the
discussions about compression distance, Alexander Smola
Ph.D. principal research scientist Yahoo! and Prof. Australian
National University for discussions about kernel methods, Arun
Kejariwal Ph.D. (now at Netflix) and Rosario Cammarota
(Ph.D. Candidate ICS UCI) for the discussions and data about the SPEC
hardware/software performance.

The authors thank Yahoo! Inc. for the resources, the support given,
and the understanding for this research that spanned about 4
years. The final result of this work is a stochastic library that the
authors will provide to interested affiliations.
\end{acks}

\bibliographystyle{acmsmall}
\bibliography{datamining}

\begin{thebibliography}{}

\bibitem[\protect\citeauthoryear{Ali and Silvey}{Ali and
  Silvey}{1966}]{AliS1966}
{\sc Ali, S.} {\sc and} {\sc Silvey, S.} 1966.
\newblock A general class of coefficients of divergence of one distribution
  from another.
\newblock {\em Journal of the Royal Statistical Society. Series B\/}~{\em
  28,\/}~1, 131--142.

\bibitem[\protect\citeauthoryear{Anderson}{Anderson}{1962}]{Anderson1962}
{\sc Anderson, T.} 1962.
\newblock On the distribution of the two-sample {C}ramer--von {M}ises
  criterion.
\newblock {\em Annals of Mathematics Statistics\/}~{\em 33,\/}~3, 1148--1159.

\bibitem[\protect\citeauthoryear{Aronszajn}{Aronszajn}{1950}]{Aronszajn1950}
{\sc Aronszajn, N.} 1950.
\newblock Theory of reproducing kernels.
\newblock {\em Transactions of the American Mathematical Society\/}~{\em
  68,\/}~3, 337--404.

\bibitem[\protect\citeauthoryear{Batchelor}{Batchelor}{1978}]{Batchelor1978}
{\sc Batchelor, B.} 1978.
\newblock {\em Pattern Recognition: {I}dea and Practice}.
\newblock New York: Plenum Press.

\bibitem[\protect\citeauthoryear{Bennett, G{\'a}cs, Li, Vit{\'a}nyi, and
  Zurek}{Bennett et~al\mbox{.}}{1998}]{BennetGLVZ1998}
{\sc Bennett, C.~H.}, {\sc G{\'a}cs, P.}, {\sc Li, M.}, {\sc Vit{\'a}nyi, P.
  M.~B.}, {\sc and} {\sc Zurek, W.~H.} 1998.
\newblock Information distance.
\newblock {\em IEEE Transactions on Information Theory\/}~{\em 44,\/}~4,
  1407--1423.

\bibitem[\protect\citeauthoryear{Bhattacharyya}{Bhattacharyya}{1943}]{Bhattach%
aryya1943}
{\sc Bhattacharyya, A.} 1943.
\newblock On a measure of divergence between two statistical populations
  defined by probability distributions.
\newblock {\em Bulletin Calcutta of Mathematics Society\/}~{\em 35}, 99--109.

\bibitem[\protect\citeauthoryear{Biau and Gyorfi}{Biau and
  Gyorfi}{2005}]{BiauG2005}
{\sc Biau, G.} {\sc and} {\sc Gyorfi, L.} 2005.
\newblock On the asymptotic properties of a nonparametric l1-test statistic of
  homogeneity.
\newblock {\em IEEE Transactions on Information Theory\/}~{\em 51,\/}~11, 3965
  -- 3973.

\bibitem[\protect\citeauthoryear{Bickel, Ritov, and Stoker}{Bickel
  et~al\mbox{.}}{2006}]{BickelRS2006}
{\sc Bickel, P.}, {\sc Ritov, Y.}, {\sc and} {\sc Stoker, T.} 2006.
\newblock Tailor-made tests for goodness-of-fit to semiparametric hypotheses.
\newblock {\em Annals Of Statistics\/}~{\em 34,\/}~2, 721--741.

\bibitem[\protect\citeauthoryear{Bickel}{Bickel}{1969}]{Bickel1969}
{\sc Bickel, P.~J.} 1969.
\newblock A distribution free version of the smirnov two sample test in the
  p-variate case.
\newblock {\em The Annals of Mathematical Statistics\/}~{\em 40,\/}~1, 1--23.

\bibitem[\protect\citeauthoryear{Borgwardt, Gretton, Rasch, Kriegel,
  Sch{\"o}lkopf, and Smola}{Borgwardt et~al\mbox{.}}{2006}]{BorgwardtGRKSS06}
{\sc Borgwardt, K.~M.}, {\sc Gretton, A.}, {\sc Rasch, M.~J.}, {\sc Kriegel,
  H.-P.}, {\sc Sch{\"o}lkopf, B.}, {\sc and} {\sc Smola, A.~J.} 2006.
\newblock Integrating structured biological data by kernel maximum mean
  discrepancy.
\newblock In {\em ISMB (Supplement of Bioinformatics)}. 49--57.

\bibitem[\protect\citeauthoryear{Cammarota, Kejariwal, D'Alberto, Panigrahi,
  Veidenbaum, and A.Nicolau}{Cammarota
  et~al\mbox{.}}{2011}]{CammarotaKDPVN2011}
{\sc Cammarota, R.}, {\sc Kejariwal, A.}, {\sc D'Alberto, P.}, {\sc Panigrahi,
  S.}, {\sc Veidenbaum, A.}, {\sc and} {\sc A.Nicolau}. 2011.
\newblock Pruning hardaware evaluation space via causality-driven application
  similarity analysis.
\newblock In {\em ACM International Conference on Computer Frontiers}.

\bibitem[\protect\citeauthoryear{Chakrabarti, Sarawagi, and Dom}{Chakrabarti
  et~al\mbox{.}}{1998}]{ChakrabartiSD1998}
{\sc Chakrabarti, S.}, {\sc Sarawagi, S.}, {\sc and} {\sc Dom, B.} 1998.
\newblock Mining surprising patterns using temporal description length.
\newblock In {\em VLDB '98: Proceedings of the 24rd International Conference on
  Very Large Data Bases}. Morgan Kaufmann Publishers Inc., San Francisco, CA,
  USA, 606--617.

\bibitem[\protect\citeauthoryear{Chernoff}{Chernoff}{1952}]{Chernoff1952}
{\sc Chernoff, H.} 1952.
\newblock A measure of asymptotic efficiency for tests of a hypothesis based on
  the sum of observations.
\newblock {\em The Annals of Mathematical Statistics\/}~{\em 23,\/}~4,
  493--507.

\bibitem[\protect\citeauthoryear{Cilibrasi and Vit{\'a}nyi}{Cilibrasi and
  Vit{\'a}nyi}{2005}]{CilibrasiV2005}
{\sc Cilibrasi, R.} {\sc and} {\sc Vit{\'a}nyi, P. M.~B.} 2005.
\newblock Clustering by compression.
\newblock {\em IEEE Transactions on Information Theory\/}~{\em 51,\/}~4,
  1523--1545.

\bibitem[\protect\citeauthoryear{D'Alberto and Dasdan}{D'Alberto and
  Dasdan}{2009}]{DAlbertoA2009}
{\sc D'Alberto, P.} {\sc and} {\sc Dasdan, A.} 2009.
\newblock Non-parametric information-theoretic measures of one-dimensional
  distribution functions from continuous time series.
\newblock In {\em Proceedings of the Ninth SIAM International Conference on
  Data Mining}, {SIAM}, Ed. Sparks, NV.

\bibitem[\protect\citeauthoryear{Daskalakis, Karp, Mossel, Riesenfeld, and
  Verbin}{Daskalakis et~al\mbox{.}}{2009}]{DaskalakisKMRV2009}
{\sc Daskalakis, C.}, {\sc Karp, R.~M.}, {\sc Mossel, E.}, {\sc Riesenfeld,
  S.}, {\sc and} {\sc Verbin, E.} 2009.
\newblock Sorting and selection in posets.
\newblock In {\em SODA '09: Proceedings of the twentieth Annual ACM-SIAM
  Symposium on Discrete Algorithms}. Society for Industrial and Applied
  Mathematics, Philadelphia, PA, USA, 392--401.

\bibitem[\protect\citeauthoryear{Dasu, Krishnan, Venkatasubramanian, and
  Yi}{Dasu et~al\mbox{.}}{2006}]{DasuKVY06}
{\sc Dasu, T.}, {\sc Krishnan, S.}, {\sc Venkatasubramanian, S.}, {\sc and}
  {\sc Yi, K.} 2006.
\newblock An information-theoretic approach to detecting changes in
  multi-dimensional data streams.
\newblock In {\em In Proc. Symp. on the Interface of Statistics, Computing
  Science, and Applications}.

\bibitem[\protect\citeauthoryear{Diday}{Diday}{1974}]{Diday1974}
{\sc Diday, E.} 1974.
\newblock Recent progress in distance and similarity measures in pattern
  recognition.
\newblock In {\em Second International Joint Conference on Pattern
  Recognition}. 534--539.

\bibitem[\protect\citeauthoryear{Einmahl and Khmaladze}{Einmahl and
  Khmaladze}{2001}]{EinmahlK2001}
{\sc Einmahl, J.} {\sc and} {\sc Khmaladze, E.} 2001.
\newblock The two-sample problem in rm and measure-valued martingales.
\newblock Open access publications from tilburg university, Tilburg University.

\bibitem[\protect\citeauthoryear{Faigle and Tur\'{a}n}{Faigle and
  Tur\'{a}n}{1988}]{FaigleT1988}
{\sc Faigle, U.} {\sc and} {\sc Tur\'{a}n, G.} 1988.
\newblock Sorting and recognition problems for ordered sets.
\newblock {\em SIAM Journal on Computing\/}~{\em 17,\/}~1, 100--113.

\bibitem[\protect\citeauthoryear{Feller}{Feller}{1948}]{Feller1948}
{\sc Feller, W.} 1948.
\newblock On the kolmogorov-smirnov limit theorems for empirical distributions.
\newblock {\em The Annals of Mathematical Statistics\/}~{\em 19,\/}~2,
  177--189.

\bibitem[\protect\citeauthoryear{Feller}{Feller}{1971}]{FellerII}
{\sc Feller, W.} 1971.
\newblock {\em An Introduction to Probability Theory and its Applications\/} 2
  Ed. Vol.~2.
\newblock John Wiley \& Sons.

\bibitem[\protect\citeauthoryear{Frank and Asuncion}{Frank and
  Asuncion}{2010}]{FrankAsuncion:2010}
{\sc Frank, A.} {\sc and} {\sc Asuncion, A.} 2010.
\newblock {UCI} machine learning repository.

\bibitem[\protect\citeauthoryear{Friedman and Rafsky}{Friedman and
  Rafsky}{1979}]{FriedmanR1979}
{\sc Friedman, J.} {\sc and} {\sc Rafsky, L.} 1979.
\newblock Multivariate generalizations of the wald-wolfowitz and smirnov
  two-sample tests.
\newblock {\em The Annals of Statistics\/}~{\em 7,\/}~4, 697--717.

\bibitem[\protect\citeauthoryear{Fuglede and Topsoe}{Fuglede and
  Topsoe}{2004}]{FugledeTopsoe}
{\sc Fuglede, B.} {\sc and} {\sc Topsoe, F.} 2004.
\newblock {Jensen-Shannon divergence and Hilbert space embedding}.
\newblock In {\em IEEE International Symposium on Information Theory}. 31--31.

\bibitem[\protect\citeauthoryear{Glaz, Naus, and Walleenstein}{Glaz
  et~al\mbox{.}}{2001}]{GlazNW2001}
{\sc Glaz, J.}, {\sc Naus, J.}, {\sc and} {\sc Walleenstein, S.} 2001.
\newblock {\em Scan Statistics}.
\newblock Springer series in statistics. Springer-Verlag.

\bibitem[\protect\citeauthoryear{Golub and Loan}{Golub and
  Loan}{1996}]{GolubVL1996}
{\sc Golub, G.} {\sc and} {\sc Loan, C.~V.} 1996.
\newblock {\em Matrix {C}omputations\/} 3 Ed.
\newblock The Johns Hopkins Univ. Press (Oct. 15, 1996).

\bibitem[\protect\citeauthoryear{Gretton, Borgwardt, Rasch, Sch{\"o}lkopf, and
  Smola}{Gretton et~al\mbox{.}}{2006}]{GrettonBRSS06}
{\sc Gretton, A.}, {\sc Borgwardt, K.~M.}, {\sc Rasch, M.~J.}, {\sc
  Sch{\"o}lkopf, B.}, {\sc and} {\sc Smola, A.~J.} 2006.
\newblock A kernel method for the two-sample-problem.
\newblock In {\em NIPS}, {B.~Sch{\"o}lkopf}, {J.~C. Platt}, {and} {T.~Hoffman},
  Eds. MIT Press, 513--520.

\bibitem[\protect\citeauthoryear{Gretton, Borgwardt, Rasch, Sch{\"o}lkopf, and
  Smola}{Gretton et~al\mbox{.}}{2008}]{GrettonBRSS08}
{\sc Gretton, A.}, {\sc Borgwardt, K.~M.}, {\sc Rasch, M.~J.}, {\sc
  Sch{\"o}lkopf, B.}, {\sc and} {\sc Smola, A.~J.} 2008.
\newblock A kernel method for the two-sample-problem.
\newblock Extended Version of the NIPS2006.

\bibitem[\protect\citeauthoryear{Gretton, Fukumizu, Teo, Song, Sch{\"o}lkopf,
  and Smola}{Gretton et~al\mbox{.}}{2007}]{GrettonFTSSS2007}
{\sc Gretton, A.}, {\sc Fukumizu, K.}, {\sc Teo, C.~H.}, {\sc Song, L.}, {\sc
  Sch{\"o}lkopf, B.}, {\sc and} {\sc Smola, A.~J.} 2007.
\newblock A kernel statistical test of independence.
\newblock In {\em NIPS}, {J.~C. Platt}, {D.~Koller}, {Y.~Singer}, {and} {S.~T.
  Roweis}, Eds. MIT Press.

\bibitem[\protect\citeauthoryear{Hahn}{Hahn}{1912}]{Hahn1912}
{\sc Hahn, H.} 1912.
\newblock {\"U}ber die integrale des herrn {H}ellinger und die
  orthogonalinvarianten der quadratischen formen von unendlich vielen
  ver\"anderlichen.
\newblock {\em Journal Monatshefte f\"ur Mathematik\/}~{\em 23,\/}~1, 161--224.

\bibitem[\protect\citeauthoryear{Hall and Tajvidi}{Hall and
  Tajvidi}{2002}]{HallT2002}
{\sc Hall, P.} {\sc and} {\sc Tajvidi, N.} 2002.
\newblock Permutation tests for equality of distributions in high-dimensional
  settings.
\newblock {\em Biometrika\/}~{\em 89,\/}~2, 359--374.

\bibitem[\protect\citeauthoryear{Harel}{Harel}{1993}]{Harel1993}
{\sc Harel, A.} 1993.
\newblock Random walk and the area below its path.
\newblock {\em Mathematics of Operations Research\/}~{\em 18,\/}~3, 566--577.

\bibitem[\protect\citeauthoryear{Henning}{Henning}{2006}]{Henning2006}
{\sc Henning, J.~L.} 2006.
\newblock Spec cpu2006 benchmark descriptions.
\newblock {\em SIGARCH Comput. Archit. News\/}~{\em 34,\/}~4, 1--17.

\bibitem[\protect\citeauthoryear{Ho}{Ho}{2005}]{Ho2005}
{\sc Ho, S.-S.} 2005.
\newblock A martingale framework for concept change detection in time-varying
  data streams.
\newblock In {\em Proceedings International Conference on Machine Learning
  (ICML)}. Bonn, Germany.

\bibitem[\protect\citeauthoryear{Ho and Wechsler}{Ho and
  Wechsler}{2005}]{HoW2005}
{\sc Ho, S.-S.} {\sc and} {\sc Wechsler, H.} 2005.
\newblock On the detection of concept change in time-varying data streams by
  testing exchangeability.
\newblock In {\em Proceedings Conference on Uncertainty in Artificial
  Intelligence (UAI)}. Edinburgh, Scotland.

\bibitem[\protect\citeauthoryear{Ho and Wechsler}{Ho and
  Wechsler}{2010}]{HoW2010}
{\sc Ho, S.-S.} {\sc and} {\sc Wechsler, H.} 2010.
\newblock A martingale framework for detecting changes in data streams by
  testing exchangeability.
\newblock {\em IEEE Transactions on Pattern Analysis and Machine
  Intelligence\/}~{\em 99,\/}~PrePrints.

\bibitem[\protect\citeauthoryear{Hope}{Hope}{1968}]{Hope1968}
{\sc Hope, A.~A.} 1968.
\newblock A simplified {M}onte {C}arlo significance test procedure.
\newblock {\em Journal of the Royal Statistical Society. Series B
  (Methodological)\/}~{\em 30,\/}~3, 582--598.

\bibitem[\protect\citeauthoryear{Horton and Nakai}{Horton and
  Nakai}{1996}]{HortonK1996}
{\sc Horton, P.} {\sc and} {\sc Nakai, K.} 1996.
\newblock A probabilistic classification system for predicting the cellular
  localization sites of proteins.
\newblock In {\em Proceedings of the Fourth International Conference on
  Intelligent Systems for Molecular Biology}. AAAI Press, 109--115.

\bibitem[\protect\citeauthoryear{Jensen}{Jensen}{1906}]{Jensen1906}
{\sc Jensen, J.} 1906.
\newblock Sur les fonctions convexes et les inégalités entre les valeurs
  moyennes.
\newblock {\em Acta Mathematique\/}~{\em 30}, 175--193.

\bibitem[\protect\citeauthoryear{Johnson and Sinanovic}{Johnson and
  Sinanovic}{}]{JohnsonS2000}
{\sc Johnson, D.} {\sc and} {\sc Sinanovic, S.}
\newblock Symmetrizing the {K}ullback--{L}eibler distance.

\bibitem[\protect\citeauthoryear{Jones and Furnas}{Jones and
  Furnas}{1987}]{JonesF1987}
{\sc Jones, W.} {\sc and} {\sc Furnas, G.} 1987.
\newblock Pictures of relevance: {A} geometric analysis of similarity measures.
\newblock {\em Journal of American Society for Information Science\/}~{\em
  38,\/}~6, 420--442.

\bibitem[\protect\citeauthoryear{Kagan}{Kagan}{1963}]{Kagan1963}
{\sc Kagan, A.} 1963.
\newblock Towards the theory of {F}isher's amount of information.
\newblock {\em Doklady Akademii nauk SSSR.\/}~{\em 151}, 277--278.
\newblock (in Russian).

\bibitem[\protect\citeauthoryear{Kailath}{Kailath}{1967}]{Kailath1967}
{\sc Kailath, T.} 1967.
\newblock The divergence and {B}hattacharyya distance measures in signal
  selection.
\newblock {\em IEEE Transactions on Communications\/}~{\em 15,\/}~1, 52--60.

\bibitem[\protect\citeauthoryear{Kendall}{Kendall}{1991}]{Kendall1991}
{\sc Kendall, D.} 1991.
\newblock Andrei {N}ikolaevich {K}olmogorov. 25 april 1903-20 october 1987.
\newblock {\em Biographical Memoirs of Fellows of the Royal Society,\/}~{\em
  37}, 301--319.

\bibitem[\protect\citeauthoryear{Keogh, Lonardi, and Ratanamahatana}{Keogh
  et~al\mbox{.}}{2004}]{KeoghLR2004}
{\sc Keogh, E.}, {\sc Lonardi, S.}, {\sc and} {\sc Ratanamahatana, C.~A.} 2004.
\newblock Towards parameter-free data mining.
\newblock In {\em KDD '04: Proceedings of the tenth ACM SIGKDD international
  conference on Knowledge discovery and data mining}. ACM, New York, NY, USA,
  206--215.

\bibitem[\protect\citeauthoryear{Kifer, Ben-David, and Gehrke}{Kifer
  et~al\mbox{.}}{2004}]{KiferBG2004}
{\sc Kifer, D.}, {\sc Ben-David, S.}, {\sc and} {\sc Gehrke, J.} 2004.
\newblock Detecting change in data streams.
\newblock In {\em Proceedings International Conference on Very Large Data Bases
  (VLDB).} Morgan Kaufmann, Elsevier, Toronto, Canada, 180--191.

\bibitem[\protect\citeauthoryear{Kim and Foutz}{Kim and Foutz}{1987}]{KimF1987}
{\sc Kim, K.-K.} {\sc and} {\sc Foutz, R.~V.} 1987.
\newblock Tests for the multivariate two-sample problem based on empirical
  probability measures.
\newblock {\em The Canadian Journal of Statistic\/}~{\em 15,\/}~1, 41--51.

\bibitem[\protect\citeauthoryear{Kolmogorov}{Kolmogorov}{1933}]{Kolmogorov1933%
e}
{\sc Kolmogorov, A.} 1933.
\newblock Sulla determinazione empririca di una legge di distribuzione.
\newblock {\em Giornale Istituzioni Italiane Attuari\/}~{\em 4}.

\bibitem[\protect\citeauthoryear{Kolmogorov and Uspenskii}{Kolmogorov and
  Uspenskii}{1987}]{KolmogorovU1987}
{\sc Kolmogorov, A.~N.} {\sc and} {\sc Uspenskii, V.~A.} 1987.
\newblock Algorithms and randomness.
\newblock {\em Theory of Probability and its Applications\/}~{\em 32,\/}~3,
  389--412.

\bibitem[\protect\citeauthoryear{Kullback and Leibler}{Kullback and
  Leibler}{1951}]{KullbackL1951}
{\sc Kullback, S.} {\sc and} {\sc Leibler, R.~A.} 1951.
\newblock On information and sufficiency.
\newblock {\em The Annals of Mathematical Statistics\/}~{\em 22,\/}~1, 79--86.

\bibitem[\protect\citeauthoryear{Kulldorf}{Kulldorf}{1997}]{Kulldorf1997}
{\sc Kulldorf, M.} 1997.
\newblock A spatial scan statistic.
\newblock {\em Communications In Statistics Theory And Methods\/}~{\em
  26,\/}~6, 1481--1496.

\bibitem[\protect\citeauthoryear{Kulldorff, Mostashari, Duczmal, Yih, Kleinman,
  and Platt}{Kulldorff et~al\mbox{.}}{2007}]{KulldorffMDYKP2007}
{\sc Kulldorff, M.}, {\sc Mostashari, F.}, {\sc Duczmal, L.}, {\sc Yih, K.},
  {\sc Kleinman, K.}, {\sc and} {\sc Platt, R.} 2007.
\newblock Multivariate spatial scan statistics for disease surveillance.
\newblock {\em Statistics in Medicine\/}~{\em 26}, 1824--1833.

\bibitem[\protect\citeauthoryear{Lee}{Lee}{1999}]{Lee1999}
{\sc Lee, L.} 1999.
\newblock Measures of distributional similarity.
\newblock In {\em Proceedings of the 37th annual meeting of the Association for
  Computational Linguistics on Computational Linguistics}. Morristown, NJ, USA,
  25--32.

\bibitem[\protect\citeauthoryear{Li, Chen, Li, Ma, and Vit{\'a}nyi}{Li
  et~al\mbox{.}}{2004}]{LiCLMV2004}
{\sc Li, M.}, {\sc Chen, X.}, {\sc Li, X.}, {\sc Ma, B.}, {\sc and} {\sc
  Vit{\'a}nyi, P. M.~B.} 2004.
\newblock The similarity metric.
\newblock {\em IEEE Transactions on Information Theory\/}~{\em 50,\/}~12,
  3250--3264.

\bibitem[\protect\citeauthoryear{Lin}{Lin}{1991}]{Jin1991}
{\sc Lin, J.} 1991.
\newblock Divergence measures based on the {S}hannon entropy.
\newblock {\em IEEE Transactions on Information Theory\/}~{\em 37,\/}~1,
  145--151.

\bibitem[\protect\citeauthoryear{Magel and Wibowo}{Magel and
  Wibowo}{1997}]{MagelW1997}
{\sc Magel, R.~C.} {\sc and} {\sc Wibowo, S.~H.} 1997.
\newblock Comparing the powers of the wald-wolfowitz and kolmogorov-smirnov
  tests.
\newblock {\em Biometrical Journal\/}~{\em 39,\/}~6, 665--675.

\bibitem[\protect\citeauthoryear{Mann and Whitney}{Mann and
  Whitney}{1947}]{MannW1947}
{\sc Mann, H.~B.} {\sc and} {\sc Whitney, D.~R.} 1947.
\newblock On a test of whether one of two random variables is stochastically
  larger than the other.
\newblock {\em The Annals of Mathematical Statistics\/}~{\em 18,\/}~1, 50--60.

\bibitem[\protect\citeauthoryear{Martin-Lof}{Martin-Lof}{1969}]{Lof1969}
{\sc Martin-Lof, P.} 1969.
\newblock Algorithms and randomness.
\newblock {\em Review of the International Statistical Institute\/}~{\em
  37,\/}~3, 265--272.

\bibitem[\protect\citeauthoryear{Meintanis and Iliopoulos}{Meintanis and
  Iliopoulos}{2008}]{MeintanisI2008}
{\sc Meintanis, S.~G.} {\sc and} {\sc Iliopoulos, G.} 2008.
\newblock Fourier methods for testing multivariate independence.
\newblock {\em Computational Statistics and Data Analysis\/}~{\em 52,\/}~4,
  1884--1895.

\bibitem[\protect\citeauthoryear{Melucci}{Melucci}{2007}]{Melucci2007}
{\sc Melucci, M.} 2007.
\newblock On rank correlation in information retrieval evaluation.
\newblock {\em SIGIR Forum\/}~{\em 41,\/}~1, 18--33.

\bibitem[\protect\citeauthoryear{M{\"u}ller}{M{\"u}ller}{1997}]{Muller1997}
{\sc M{\"u}ller, A.} 1997.
\newblock Integral probability metrics and their generating classes of
  functions.
\newblock {\em Advances in Applied Probability\/}~{\em 29,\/}~2, 429--443.

\bibitem[\protect\citeauthoryear{Nash, Sellers, Cawthorn, and Ford}{Nash
  et~al\mbox{.}}{1994}]{NashSTCF1994}
{\sc Nash, W.~J.}, {\sc Sellers, T.~L.}, {\sc Cawthorn, S. R. T. A.~J.}, {\sc
  and} {\sc Ford, W.~B.} 1994.
\newblock The population biology of abalone (haliotis species) in tasmania i.
  blacklip abalone (h. rubra) from the north coast and islands of bass strait.
\newblock Tech. Rep.~48, Sea Fisheries Division.

\bibitem[\protect\citeauthoryear{Pinsker}{Pinsker}{1960}]{Pinsker1960}
{\sc Pinsker, M.} 1960.
\newblock Information and information stability of random variables and
  processes.
\newblock {\em Probl. Peredachi Inf.\/}~{\em 7}.

\bibitem[\protect\citeauthoryear{Sch{\"o}lkopf and Smola}{Sch{\"o}lkopf and
  Smola}{2002}]{ScholkopfS2002}
{\sc Sch{\"o}lkopf, B.} {\sc and} {\sc Smola, A.} 2002.
\newblock {\em Learning with {K}ernels}.
\newblock MIT Press.

\bibitem[\protect\citeauthoryear{Shafer and Vovk}{Shafer and
  Vovk}{2008}]{ShaferV2008}
{\sc Shafer, G.} {\sc and} {\sc Vovk, V.} 2008.
\newblock A tutorial on conformal prediction.
\newblock {\em J. Mach. Learn. Res.\/}~{\em 9}, 371--421.

\bibitem[\protect\citeauthoryear{Shannon}{Shannon}{1948}]{Shannon1948}
{\sc Shannon, C.} 1948.
\newblock A mathematical theory of communication.
\newblock {\em The Bell System Technical Journal\/}~{\em 27}, 379--423 and
  623--656.

\bibitem[\protect\citeauthoryear{Shivakumar and Garc\'{\i}a-Molina}{Shivakumar
  and Garc\'{\i}a-Molina}{1995}]{ShivakumarM95}
{\sc Shivakumar, N.} {\sc and} {\sc Garc\'{\i}a-Molina, H.} 1995.
\newblock {SCAM}: {A} copy detection mechanism for digital documents.
\newblock In {\em Proceedings 2nd Conference on the Theory and Practice of
  Digital Libraries}.

\bibitem[\protect\citeauthoryear{Siegel}{Siegel}{1959}]{Siegel1959}
{\sc Siegel, S.} 1959.
\newblock {\em Nonparametric statistics, for the behavioral sciences}.
\newblock Series in Psychology. McGraw-Hill book company.

\bibitem[\protect\citeauthoryear{Song, Wu, Jermaine, and Ranka}{Song
  et~al\mbox{.}}{2007}]{SongWCR2007}
{\sc Song, X.}, {\sc Wu, M.}, {\sc Jermaine, C.}, {\sc and} {\sc Ranka, S.}
  2007.
\newblock Statistical change detection for multi-dimensional data.
\newblock In {\em KDD '07: Proceedings of the 13th ACM SIGKDD international
  conference on Knowledge discovery and data mining}. ACM, New York, NY, USA,
  667--676.

\bibitem[\protect\citeauthoryear{Sriperumbudur, Gretton, Fukumizu, Lanckriet,
  and Sch{\"o}lkopf}{Sriperumbudur et~al\mbox{.}}{2009}]{SriperumbudurGFLS2009}
{\sc Sriperumbudur, B.~K.}, {\sc Gretton, A.}, {\sc Fukumizu, K.}, {\sc
  Lanckriet, G. R.~G.}, {\sc and} {\sc Sch{\"o}lkopf, B.} 2009.
\newblock A note on integral probability metrics and $\phi$-divergences.
\newblock {\em CoRR\/}~{\em abs/0901.2698}.

\bibitem[\protect\citeauthoryear{Street, Wolberg, and Mangasarian}{Street
  et~al\mbox{.}}{1993}]{StreetWM1993}
{\sc Street, W.}, {\sc Wolberg, W.}, {\sc and} {\sc Mangasarian, O.} 1993.
\newblock Nuclear feature extraction for breast tumor diagnosis.
\newblock In {\em IS\&T/SPIE 1993 International Symposium on Electronic
  Imaging: Science and Technology}. Vol. 1905. San Jose, CA, 861--870.

\bibitem[\protect\citeauthoryear{Tak\'acs}{Tak\'acs}{1991}]{Takas1991}
{\sc Tak\'acs, L.} 1991.
\newblock A bernoulli excursion and its various applications.
\newblock {\em Advances in Applied Probability\/}~{\em 23,\/}~3, 557--585.

\bibitem[\protect\citeauthoryear{Taneja and Kumar}{Taneja and
  Kumar}{2004}]{TanejaK2004}
{\sc Taneja, I.~J.} {\sc and} {\sc Kumar, P.} 2004.
\newblock Relative information of type s, {C}sisz\'ar's f-divergence, and
  information inequalities.
\newblock {\em Information Sciences\/}~{\em 166}, 105--125.

\bibitem[\protect\citeauthoryear{Terwijn, Torenvliet, and Vit{\'a}nyi}{Terwijn
  et~al\mbox{.}}{2010}]{TerwijnTV2010}
{\sc Terwijn, S.~A.}, {\sc Torenvliet, L.}, {\sc and} {\sc Vit{\'a}nyi, P.~M.}
  2010.
\newblock Nonapproximability of the normalized information distance.
\newblock {\em Journal of Computer and System Sciences\/}~{\em In Press,
  Corrected Proof}, --.

\bibitem[\protect\citeauthoryear{Tsanas, Little, McSharry, and Ramig}{Tsanas
  et~al\mbox{.}}{2010}]{TsanasLMR2010}
{\sc Tsanas, A.}, {\sc Little, M.}, {\sc McSharry, P.}, {\sc and} {\sc Ramig,
  L.} 2010.
\newblock Accurate telemonitoring of parkinson.s disease progression by
  non-invasive speech tests.
\newblock {\em Biomedical Engineering, IEEE Transactions on\/}.

\bibitem[\protect\citeauthoryear{Vajda}{Vajda}{1972}]{Vadja1972}
{\sc Vajda, I.} 1972.
\newblock On the $f$-divergence and singularity of probability measures.
\newblock {\em Journal Periodica Mathematica Hungarica\/}~{\em 2,\/}~1--4,
  223--234.

\bibitem[\protect\citeauthoryear{Vovk}{Vovk}{1993}]{Vovk1993}
{\sc Vovk, V.} 1993.
\newblock A logic of probability, with application to the foundations of
  statistics.
\newblock {\em Journal of the Royal Statistics Society\/}~{\em 55,\/}~2,
  317--351.

\bibitem[\protect\citeauthoryear{Vovk, Gammerman, and Shafer}{Vovk
  et~al\mbox{.}}{2005}]{VovkGS2005}
{\sc Vovk, V.}, {\sc Gammerman, A.}, {\sc and} {\sc Shafer, G.} 2005.
\newblock {\em Vovk, Vladimir, Gammerman, Alex, Shafer, Glenn}.
\newblock Springer.

\bibitem[\protect\citeauthoryear{Vovk, Nouretdinov, and Gammerman}{Vovk
  et~al\mbox{.}}{2003}]{VovkNG2003}
{\sc Vovk, V.}, {\sc Nouretdinov, I.}, {\sc and} {\sc Gammerman, A.} 2003.
\newblock Testing exchangeability on-line.
\newblock In {\em Proceedings International Conference on Machine Learning
  (ICML)}.

\bibitem[\protect\citeauthoryear{Wald}{Wald}{1947}]{Wald1947}
{\sc Wald, A.} 1947.
\newblock {\em Sequential Analysis}.
\newblock Dover Publications Inc, N.Y.

\bibitem[\protect\citeauthoryear{Wang, Kulkarni, and Verdú}{Wang
  et~al\mbox{.}}{2005}]{WangKV2005}
{\sc Wang, Q.}, {\sc Kulkarni, S.~R.}, {\sc and} {\sc Verdú, S.} 2005.
\newblock Divergence estimation of continuous distributions based on
  data-dependent partitions.
\newblock {\em IEEE Transactions on Information Theory\/}~{\em 51,\/}~9, 3064
  -- 3074.

\bibitem[\protect\citeauthoryear{Wang, Wong, and Yao}{Wang
  et~al\mbox{.}}{1992}]{WangWY1992}
{\sc Wang, Z.}, {\sc Wong, S.}, {\sc and} {\sc Yao, Y.} 1992.
\newblock An analysis of vector space models based on computational geometry.
\newblock In {\em Proceedings International Conference on Research and
  Development in Information Retrieval (SIGIR)}. ACM, New York, NY, USA,
  152--160.

\bibitem[\protect\citeauthoryear{Wilcoxon}{Wilcoxon}{1945}]{Wilcoxon1945}
{\sc Wilcoxon, F.} 1945.
\newblock Individual comparisons by ranking methods.
\newblock {\em Biometrics Bulletin\/}~{\em 1}, 80--83.

\bibitem[\protect\citeauthoryear{Wilson and Martinez}{Wilson and
  Martinez}{1997}]{WilsonM1997}
{\sc Wilson, D.~R.} {\sc and} {\sc Martinez, T.~R.} 1997.
\newblock Improved heterogeneous distance functions.
\newblock {\em Journal of Artificial Intelligence Research\/}~{\em 6}, 1--34.

\bibitem[\protect\citeauthoryear{Zhang, Song, Gretton, and Smola}{Zhang
  et~al\mbox{.}}{2008}]{ZhangSGS08}
{\sc Zhang, X.}, {\sc Song, L.}, {\sc Gretton, A.}, {\sc and} {\sc Smola,
  A.~J.} 2008.
\newblock Kernel measures of independence for non-iid data.
\newblock In {\em NIPS}, {D.~Koller}, {D.~Schuurmans}, {Y.~Bengio}, {and}
  {L.~Bottou}, Eds. MIT Press, 1937--1944.

\end{thebibliography}

\elecappendix

\medskip
\small

\section{Review 1}
\begin{verbatim}
This paper reviews 4 different families of non-parametric methods to
 detect anomalies in time series, more specifically a change in the
 distribution of points which are sampled sequentially and
 independently. This task is usually referred to as change point
 detection. Of all 4 techniques, part of the 1st one and the 4th one
 are original contributions. The 4th method aims at comparing two
 sub-series R and W by estimating their respective Cumulative Density
 Functions (CDF) F_R and F_W, and then computing many distances
 D(F_R,F_W) to evaluate how dissimilar they are. A change is detected
 whenever a "quorum" of alarms/flags are raised, that is a sufficient
 number of distances go above a given threshold. These distances are
 presented in Section 4. Section 5 follows with experimental results
 with 3 simulated sets of time series (toy data, classification data,
 applications calls) and 1 real-life dataset (quote history for 4
 stocks).

Regardless of the interest of its contribution, I think the paper has
 many problems with its current form. I think this paper would have
 greatly benefitted from further polishing, rewriting and
 proofreading. In particular, I have the feeling that the authors have
 settled for a paper structure and notations that could looks sloppy,
 which contrasts with their self advocated goal of proposing new tools
 but also proposing a unified framework (footnote 1, p.2 or bottom of
 p.2). Here are few items that illustrate these impressions, and which
 the authors need to take care:

- bibliographic pointers are most often given at the end of the
 sections (e.g middle of p.15, bottom of p.22, p.30 etc.) instead of
 being provided right next to the introduction of important concepts
 as is common practice in ML. This makes it particularly difficult to
 understand on which literature the methods which are presented here
 are based upon.  Nothing is more important for a reader interested in
 a review paper than being able to frequently check references exactly
 at the moment when they are introduced, with page and section
 numbers.

- Overall, mathematical statements and notations are loose and lack
 clarity. Mathematical terms are overloaded. Overall, there are very
 few equations and a lot of text, which makes it difficult to check or
 understand exactly the authors' claims. Here are a few examples: **
 The title "Non-parametric methods applied to time series comparison"
 is not well chosen. The authors only handle time series of
 independent measurements. This is a *very* restrictive case when
 studying time series, which is known as change point detection. I
 would suggest something along the lines of "Non-parametric methods
 for Change Point Detection"

** The crux of the problem, is not well posed. In particular, the
 authors use the following sentence (p.5, before 2.2) to introduce
 change point detection: "A change occurs any time that W is different
 from R". What do you mean by "different"? if taken literally, then I
 guess that unless a series is constant there are always changes. Do
 the authors mean in distribution? Then why are not any of the
 standard definitions of change given here? The authors may want to
 refer to "Detection of Abrupt Changes: Theory and Application",
 M. Basseville, I. Nikiforov for such definitions.

**Since the authors kick out their methodological section without
 actually defining the problem, it is extremely difficult for the
 reader to understand their perspective. The introduction of Section 3
 is, for instance, quite disorganized. For instance, I do not
 understand what the authors mean by "What distinguishes our work from
 others is the focus on the computational aspects of implementing each
 method in a context of a set of statistical tools" nor think that the
 "we feel that we have taken the works of Bickel [8] and
 Friedman-Rafsky [24] and succeeded in extracting the common features"
 is particularly convincing, specially in the introduction of this
 section.

** Section 3.3 is the most problematic, mathematically speaking, and
 has to be entirely rewritten by following standard conventions,
 e.g. being careful with the notation for f, the function, and f(x), a
 number. Non-exhaustive list of examples

--- "Assume that F is a Hilbert space defined in E; that is f(x) with
 x \in E" ??

--- "First, for every fixed y=y0, then K(x,y0)\in F". K(x,y0) is a
 number, not a function. "x -> K(x,y0)" would be a function, or
 K(.,y0) using the dot notation.

--- "found the existence of a mapping phi(x)", again, the mapping is
 phi or x -> phi(x) but not phi(x).  Since this paragraph is really
 standard, it is annoying to stumble every other line on these
 confusions.


** The main contribution of the authors, Section 3.4, is extremely
 difficult to parse, with no formal definition/proposition/algorithm
 box nor figure. The structure is very sloppy here: there are 7
 (non-numbered) remarks over 4 pages which are provided one after the
 other with very little context. Since this should be the strongest /
 best motivated part of the paper, this section is disappointing.

** Results in the experimental section are not well presented. The
 axis and/or legends of figures 2,3, 5, 8, 9, 10, 11, 12, 13 of
 section 5 are impossible to read, which is to say most figures are
 hardly of any use at this point. When readable, most of the times the
 text on the axis is not clear (e.g. "Result[[XVAL]][x]")

** minor issues which break the reading flow:

--- the authors use the word "measure" without a formal explanation
 from the beginning of section 3. I think what the authors call a
 measure is widely known as a divergence, or possibly a dissimilarity
 or a distance in some cases. A measure is usually a map from a
 sigma-algebra to real numbers that satisfies sigma-additivity, as
 defined in "measure theory". The words "similarity measure" is
 sometimes used to denote a similarity, but "measure" in itself is not
 used as a synonym of divergence in the machine learning
 literature. The authors also use "measure" for both a distance or a
 similarities (or a kernel actually), e.g. p.36 Bhattacharyya section.

--- Why separate references to seminal works of Jensen [40] and
 Shannon [67] to mention the Jensen-Shannon divergence? I do not think
 either of these authors actually proposed it, and certainly not in
 either of these papers.

--- The statement that follows, on "commonality" of phi-divergences
 and RKHS as illustrated by the "embedability" of the Jensen-Shannon
 divergence is enigmatic. The Jensen-Shannon divergence is a negative
 definite distance between probability measures.  It can be shown that
 probability measures can thus be embedded in a Hilbert space where
 the regular norm of that RKHS coincides with the distance,
 i.e. JS(p,q)= || phi(p) - phi(q) ||_K . What is the relation with
 using phi-divergences between measures mapped in a RKHS (induced by
 any kernel) as advocated by [71]?

--- in p.5 the authors distinguish epochs and timestamps in a not so
 clear way but then resolve to only focus on timestamps in the
 paper. this discussion is not needed.

--- p.11 section 3.1.2: transducers are introduced as f_A, but the
 authors never use that notation again. a transducer is defined as a
 function (S^*) -> [0,1], but the authors use a conditional form to
 define it right below, that takes at least 2 parameters - m and N -
 that do not fit your definition of a function from (S^*).

--- p.13 section 3.1.3 has three properties. I am not sure what the
 authors mean by "Property".

--- p.15 the introduction of section 3.2 needs to be rewritten, and
 key notations like Y properly defined.

--- is the normalized compression distance a distance (definite,
 triangle inequality)?

--- the definition of completeness and cauchy sequences at the bottom
 of p. 18 is wrong.

--- p.20 footnote 3: referring to "choosing a kernel" as an "art" is
 not particularly convincing, given that there have been hundreds of
 papers that try to select kernels adaptively (MKL) and a few more for
 the MMS case in particular.

--- I do not understand the motivation of the authors to refer to "red
 points" and "white points" without illustrating their ideas with a
 figure.

--- p.32, bottom: the definition of the 0 norm is pretty much
 standardized in the literature, and is the natural limit of the p
 norm when p goes to 0. I have never seen it defined as the algebraic
 sum of all terms of a vector as proposed by the authors (which would
 violate the fact that a norm needs to be non-negative anyway).

--- p.34 the authors should use italics more sparingly in this
 section, specially on subjective calls like "*more* trustworthy"
 etc..

--- p.36 Hellinger: "component-wise comparison is less biased", biased
 in what sense? the statement "components near the extremes (0 or 1)
 are moved closer to 1/2 [by the square root]" is wrong.

--- section 4.1.3 mentions a few other "measures" but only quotes
 Wilcoxon-Mann-Whitney.

Now, on the content itself. First, I have to say that is has been
 relatively difficult for me to read the paper because of the problems
 highlighted above. Yet, on the content itself, I have been puzzled by
 the following points:

##### no reference to classics in the change point detection
 literature

"Detection of Abrupt Changes: Theory and Application", M. Basseville,
 I. Nikiforov

nor on the recent literature on change point detection with kernels,

"Kernel Change-point Analysis", Harchaoui, Bach, Moulines, NIPS 2008

which is, in the context of this paper, more relevant that the MMS
 family of papers [71 etc.], which is a more general tool. The
 reference above discusses specifically the problem considered by the
 authors. Multiple change points have also been considered for
 instance.  "Fast detection of multiple change-points shared by many
 signals using group LARS" Jean-Philippe Vert and Kevin Bleakley
 . NIPS 2010.


##### I do not understand section 4. Why apply divergences that have
 been explicitly designed for probability measures (or probability
 densities) to CDF's? The authors mention "In this spirit, we can
 extend the measures commonly used for vectors [...] and apply them to
 CDF's as inputs". I may have missed some additional motivation, but
 at this point the whole idea behind this section does not make sense,
 looks like a hack, and is not supported by any theory in statistics,
 information theory, or information geometry ( e.g. "Methods of
 Information Geometry" Amari Nagaoka).

##### Inadequacy of the datasets. None of the experiments is truly
 convincing for a machine learning application:

-- The experiments with stocks dataset makes very little sense from a
 financial perspective.  No one in the financial industry studies
 stock prices as i.i.d. datasets.  Financial econometrics is the
 discipline that studies financial time series. What do you mean by
 "For example an index such a(sic) Nasdaq is bound to increase during
 its lifetime as a result of adding new stocks". indices are simply
 reweighted when they contain new stocks, not increased.

-- Using (low dimensional) classification datasets to generate time
 series is a nice toy example but not really convincing.

-- The hardware/software application would deserve more information
 and quantitative (dimensions, sample size etc..) description of the
 dataset.

##### The techniques that are proposed (section 3.4) do not work with
 high dimensional data, because of the need for constructing a partial
 order to estimate CDF's. Isn't high dimensions one of the most
 obvious challenges in machine learning today?

To summarize, I think the authors need to spend considerably more time
 on this paper to make it fit for a submission. At the moment the
 contribution (3.4) is poorly presented, one of the key ideas (using
 divergences for probability densities directly on CDF's) makes no
 sense in my opinion, the experimental section needs significant
 rewriting and the datasets are not particularly interesting nor
 relevant to the problem.

\end{verbatim}

\section{Review 2}
\begin{verbatim}
A Review on Non-parametric Methods Applied to Series Comparison.


Contributions of the paper: The paper contains a comprehensive survey
 about non-parametric methods applied to series comparison, and two
 new non-parametric cumulative distribution function comparison
 methods, which are extensions of the work of Bickel [8] and
 Friedman-Rafksy [24].

The papers studies very interesting problems. In my opinion, however,
 the presentation needs considerable improvement before the paper can
 be published. I am especially worried about the technical details;
 the mathematical notations are confusing sometimes, and at many
 places the authors used verbose sentences instead of concise
 mathematical expressions.

The authors intended to write a self-contained review, but
 unfortunately I do not think that readers who are not familiar with
 the topic can fully understand the discussed ideas because the
 mathematical details are not presented with enough care. At many
 places in the paper the explanations are simply vague but verbose
 sentences instead of precisely formulated mathematical expressions.


It is not clear in the paper what those conditions are when the
 discussed methods can be used. Many of the methods in the paper are
 developed only for i.i.d. series, but the authors use them for more
 complex time series without discussing these issues.

My detailed comments are below.

Section 2.2. Examples of Series and Change: In some of these examples
 I would explain the roles of the mathematical terms introduced in
 Section 2.1 such as s_i, x_i, y_i, S, R, W, etc. Currently, the
 mathematical expressions are defined in section 2.1, and examples are
 given in Section 2.2, but Section 2.2 does not use the notations
 introduced in Section 2.1.

Section 3.

I found it very confusing that the authors talk about ?methods for
 multidimensional series? and later they talk about processes, while
 clearly many of the papers cited here are applicable only for
 i.i.d. series of random variables, but they cannot be used for more
 general time series and stochastic processes. Section 3.1 talks only
 about i.i.d. sequences. Section 3.2 is about Kolomogorov?s
 information and discusses general series again.  The author should
 help the readers and explain which tools can be used for
 i.i.d. series only, and how they are going to use these methods for
 more general series, e.g. stock prices.

The are a couple of other nonparametric divergence estimators that the
 authors might want to cite:

?A Nearest-Neighbor Approach to Estimating Divergence between
 Continuous Random Vectors, Qing Wang, Sanjeev R. Kulkarni, Sergio
 Verdu, IEEE International Symposium on Information Theory (ISIT),
 2006.?

?Estimating divergence functionals and the likelihood ratio by convex
 risk minimization. X. Nguyen, M. J. Wainwright and M. I. Jordan. IEEE
 Transactions on Information Theory, 56, 5847-5861, 2010.?


?On the Estimation of alpha-divergences, B. Poczos and J. Schneider,
 International Conference on AI and Statistics (AISTATS), JMLR
 Workshop and Conference Proceedings (15), 609-617,2011.?
 
Please cite the PhD thesis of Bharath K Sriperumbudur too:
 ?Reproducing kernel space embeddings and metrics on probability
 measures B. K. Sriperumbudur Ph. D. Dissertation, UC San Diego,
 2010?. This work contains many important and interesting results on
 RKHS based divergences estimators.


Page 11: Could you provide a specific example for A_j?

Page 12: ?Although we have implemented a few transducers?. Which ones?

Page 13, Equation 3.5: Please do not use * for multiplication in
 equations (just simply omit the ?*?). Similarly fix this problem in
 the other equations of the paper.

Page 13: Please define p_i again, or reference where it has been
 defined earlier.

Page 13: Section 3.1.3 is called Martingale methods, but it is not
 explained why and which random variables form a martingale.

Page 15: I do not like that authors simple list references at the end
 of sections instead of citing them in the sections where the results
 are used.

Page 15: Please cite those papers where ?similarity metric?,
 ?universal nature of the measure?, ?the algorithmic information
 distance?, and the ?information distance? have been introduced.

Section 3.2, Normalized Compression Distance: The notations should be
 improved here. For example, it is not explained formally with
 mathematical terms what this sentence means: ?y can be described by a
 chain of descriptions xs?.


Page 18: I could not find where the bound M has been defined.

Page 19: <,> is not a vector norm; it is a square of the vector norm
 <,>^{1/2}.

Page 19: The notation needs to be revised: For example, technically
 this sentence does not make sense, although it is clear what the
 authors wanted to write: ?Assume that $\mathcal{F}$ is a Hilbert
 space defined in $E$; that is, f(x) with $x \in E$.? Please change
 this sentence to something like ?Let $\mathcal{F}=\{f: E\to \Real\}$
 be a Hilbert space?.

Furthermore, $K$ kernel has not been defined, and $K(x,y_0)\in
 \mathcal{F}$ is not true either, since $K(x,y_0)\in \Real$! Please
 use the $K(\cdot,y_0)\in \mathcal{F}$ notation instead.

Similarly, use $f(y)=<f,K(\cdot,y)>$ instead of $f(y)=<f(x),K(x,y)>$!
  The not precise enough notation is especially confusing for the
 feature functions phi, because here $\phi(x)(\cdot)\in
 \mathcal{F}$. Therefore, $f(x)=<f,phi(x)>$ is correct, but
 $phi(x)=<phi(y),K(y,x)>$ is not correct. Please use
 $phi(x)=<phi,K(\cdot,x)>$ instead.

Page 19: ?This is a single dimensional space?. What is a single
 dimensional space?

Page 20: E.q. 3.15: Note that ?U? stands for the U statistic and for
 the unbiased estimation of MMD^2. Sometimes capital U, sometimes
 lower case u is used for the same quantity. Please fix it.

Page 21: ?matrix $\Sigma$ is a semi-definite ...? -> ?matrix $\Sigma$
 is a positive semi-definite ...?

Page 22: Section 3.3.2. Please emphasize more that MMD has been
 defined only for i.i.d. series.

Page 23: The description of the statistical tests is very confusing
 because the authors use the same notations for both the empirical and
 the true cumulative distribution functions! So from the text
 currently it seems the authors want to build a test to decide if the
 two empirical distribution functions are the same, which does not
 make sense, since the empirical distribution functions can be
 computed. The goal of the tests should be to decide if the true
 distribution functions are the same or not.

Page 25: A figure would help to show the meaning of $vdash$ and
 $dashv$.

Page 26: $\vdash_i=\min x_i$, $\dashv_i=\max x_i$. This notation is
 again a bit confusing, because one might think that in the $\max x_i$
 the maximum is taken over $i$, so the r.h.s. does not depend on $i$,
 but the l.h.s. still depends on $i$.


Section 4:

Page 31: This section has the same problem as the previous. $H0: F_R
 \sim F_W$ means that we want to test if the empirical distributions
 F_R and F_W are the same. Instead, we should test if the true
 underlying distributions are the same.
 
Page 32: Please provide a better explanation of the quantities in
 E.q. (4.1).

Page 32: Definiton of \|x_0\|. This quantity can be negative? There is
 no absolute value here...
 
Page 35: Is it true that JS=J_{in}L/2 ?

Page 36: The authors might want to mention which measures are
 distances too, i.e. which ones are nonnegative where the triangle
 inequality also holds.

Page 37: It would be also important to mention Csiszar?s f divergence.
  http://en.wikipedia.org/wiki/F-divergence

Page 38: I do not like the ?*? symbols in Eq 4.15. Please omit them.

Page 39: Section 4.1.3. For completeness, the Wilcoxon test should be
 more detailed.

Page 39: Section 4.2. ?The distribution of the measure values is well
 studied? Please provide some references here.

Page 42 and the other figures in Section 5: Please increase the font
 size! I had very hard time trying to read the labels of the
 figures. Also, the colors cannot be distinguished when the paper is
 printed in black and white!

Page 42: Why are these experimental results presented here and not in
 section 5 among the other experimental results?

Page 48: It is a bit weird to read sentences like this ?stochastic
 process, which has the normal distribution?, because technically a
 stochastic process is a series of random variables. In this case I
 believe the authors wanted to write that this stochastic process is a
 series of i.i.d. N(0,1) random variables.

Page 50: ?W slides through the series?, ?W shifts by a full interval
 window size?, etc

In my opinion, the process generation should be explained with
 mathematical expressions too. This could help readers understand the
 paper better. Similarly, the text on Page 54: ?With successive
 intervals we mix points? could also be improved by extending it with
 mathematical formulae.


The paper also needs a complete English grammar checking. I collected
 a few typos and grammatical errors below.


Some Typos, Grammar, Wording problems:

page 4: ?We present out? [our?]
page 4: ?notations is used?
page 5: ?occurs any time that? [that -> when?]
page 9: ?the the?
page 11: ?... and a new data point, for which a measure of strangeness, and it returns??
page 12: ?has not effect? 
page 14: ?Lets? [Let us]
page 14: ?and, in the work by Vovk? [fix the commas]
page 23: ?these pairs allows? [allow]
page 24: ?introduced the definition distribution functions? [definition of]
page 27: ?X_i, in the topological ordering, should? [fix the commas]
page 31: ?In the literature are available methods? [there are available methods?] 
page 31: ?a similarity measures?
page 59: ?every methods?
\end{verbatim}
%% COMPLETED %%

\end{document}